\documentclass[10pt,journal,compsoc]{IEEEtran}

\ifCLASSOPTIONcompsoc
  \usepackage[nocompress]{cite}
\else
  \usepackage{cite}
\fi

\ifCLASSINFOpdf
\else
\fi
\usepackage{ragged2e}
\usepackage[table]{xcolor}
\usepackage[inline]{enumitem}
\usepackage{tikz}
\newcommand*\circled[1]{\tikz[baseline=(char.base)]{
            \node[shape=circle,fill,inner sep=0.5pt] (char) {\textcolor{white}{#1}};}}

\usepackage{afterpage}
\usepackage{psfrag} 
\usepackage{lscape}
\usepackage{booktabs}
\usepackage{subcaption}

\usepackage[most]{tcolorbox}
\usepackage{multirow}
\newcolumntype{P}[1]{>{\centering\arraybackslash}p{#1}}

\NewDocumentCommand\DownArrow{O{2.0ex} O{black}}{%
   \mathrel{\tikz[baseline] \draw [<-, line width=0.5pt, #2] (0,0) -- ++(0,#1);}
}

\usepackage{hyperref}
\usepackage{xfrac}

\newlist{myitemize}{itemize}{1}
\setlist[myitemize,1]{label=\textbullet,leftmargin=5mm}

\makeatletter
\let\old@lstKV@SwitchCases\lstKV@SwitchCases
\def\lstKV@SwitchCases#1#2#3{}
\makeatother

\usepackage{lstlinebgrd}
\makeatletter
\let\lstKV@SwitchCases\old@lstKV@SwitchCases
\lst@Key{numbers}{none}{%
    \def\lst@PlaceNumber{\lst@linebgrd}%
    \lstKV@SwitchCases{#1}%
    {none:\\%
     left:\def\lst@PlaceNumber{\llap{\normalfont
                \lst@numberstyle{\thelstnumber}\kern\lst@numbersep}\lst@linebgrd}\\%
     right:\def\lst@PlaceNumber{\rlap{\normalfont
                \kern\linewidth \kern\lst@numbersep
                \lst@numberstyle{\thelstnumber}}\lst@linebgrd}%
    }{\PackageError{Listings}{Numbers #1 unknown}\@ehc}}
\makeatother

\lstdefinestyle{base}{
  language=Java,
  basicstyle=\ttfamily\small,
  frame=lines,
  keywordstyle=\color{blue}\textbf,
  commentstyle=\color[rgb]{0.0,0.4,0.0}\scriptsize,
  extendedchars=true,         
  breaklines=true   
  showspaces=false,
  showstringspaces=false, 
   numbers=left,
    stepnumber=1,   
        tabsize=1,
    breaklines=true,      
    xleftmargin={0.5cm},
  moredelim=**[is][\color{green}]{!!}{!!},
  moredelim=**[is][\color{orange}]{^}{^},
  moredelim=**[is][\color{red}]{@}{@},
   breakindent=0pt,                
}

\usepackage{amssymb}
\usepackage{pdflscape}

\usepackage{float}
\usepackage{stfloats}

\usepackage{tablefootnote}
\usepackage{longtable}
\usepackage{balance}

\usepackage{multicol}

\begin{document}

\title{A Comprehensive Survey of Logging in Software: From Logging Statements Automation to Log Mining and Analysis}

\author{Sina~Gholamian,~\IEEEmembership{Student Member,~IEEE,}
        Paul~A.~S.~Ward,~\IEEEmembership{Member,~IEEE}
\IEEEcompsocitemizethanks{\IEEEcompsocthanksitem S. Gholamian and P. Ward are with University of Waterloo, Waterloo, Canada, N2L 3G1.\protect\\

E-mail: \{sgholamian, pasward\}@uwaterloo.ca }}


\IEEEtitleabstractindextext{%
\begin{abstract}
\justifying
Logs are widely used to record runtime information of software systems, such as the timestamp and the importance of an event, the unique ID of the source of the log, and a part of the state of a task's execution.
The rich information of logs enables system developers (and operators) to monitor the runtime behaviors of their systems and further track down system problems and perform analysis on log data in production settings. 
However, the prior research on utilizing logs is scattered and that limits the ability of new researchers in this field to quickly get to the speed and hampers currently active researchers to advance this field further.  
Therefore, this paper surveys and provides a systematic literature review and mapping of the contemporary logging practices and log statements' mining and monitoring techniques and their applications such as in system failure detection and diagnosis. 
We study a large number of conference and journal papers that appeared on top-level peer-reviewed venues. 
Additionally, we draw high-level trends of ongoing research and categorize publications into subdivisions. 
In the end, and based on our holistic observations during this survey, we provide a set of challenges and opportunities that will lead the researchers in academia and industry in moving the field forward.
\end{abstract}

}

\maketitle

\IEEEdisplaynontitleabstractindextext
\IEEEpeerreviewmaketitle

\section{Introduction}\label{introduction_section}
\IEEEPARstart{S}{oftware} systems are pervasive and play important and often critical roles in the society and economy such as in airplanes or surgery room patient monitoring systems.
Gathering feedback about software systems' states is a nontrivial task and plays a crucial role in system diagnosis in the case of a failure. 
In the interest of higher availability and reliability, software systems regularly generate \textit{log files} of their status and runtime information.
Developers insert logging statements into the source code which are then printed in the log files, also known as \textit{execution logs} and \textit{event logs}~\cite{cinque2012event}. 
Then, at a later time, while the system is running or postmortem, developers or operators would analyze the log files for various tasks. 
For example, the content of log files has been studied to achieve a variety of goals such as anomaly and fault detection~\cite{xu2009detecting,jiang2008automatic,fu2009execution,du2017deeplog}, online or postmortem performance and failure diagnosis~\cite{yuan2010sherlog,nagaraj2012structured,syer2013leveraging,zhao2016non,zhang2017pensieve}, pattern detection~\cite{aharon2009one,makanju2009clustering,vaarandi2015logcluster}, profile building~\cite{vaarandi2015logcluster}, business decision making~\cite{barik2016bones}, and user's behavior observation~\cite{lee2012unified}. 

Conventionally, software developers and practitioners apply testing and monitoring techniques to analyze the software systems. 
System testing happens during the development phase by developers, while practitioners utilize system monitoring techniques to understand the behavior of the system in the deployed environment~\cite{candido2019contemporary}. 
As such, it is a common practice to have running programs report on their internal state and variables, through log files that system administrators and operators can analyze~\cite{bertero2017experience} for different purposes. 
This continuous cycle of development and deployment of the software and looking at the system logs has also initiated and thrived adjacent fields of research such as DevOps~\cite{jabbari2016devops,cukier2013devops,dyck2015towards}. 
That is, the importance of log analysis and its computational intensity has also brought in other tools to scale up the effort. 
For example, considering the advancements of machine learning (ML) and artificial intelligence (AI) and the vast size of the log files, research has proposed the use of ML for automated operation analysis (AIOps)~\cite{urlaiops} of execution logs. 
From the commercialization perspective, the widespread need for log analysis has also contributed to the emergence of commercial products such as Splunk~\cite{urlsplunk} and Elastic Stack~\cite{urlelk}. 
Splunk makes the large-scale logs accessible by extracting patterns and correlating system metrics to diagnose problems and provide insight for business decisions. 
Elastic Stack~\cite{urlelk}, a.k.a. ELK, consists of three different subsystems of \textbf{E}lasticsearch~\cite{urlelastic}, \textbf{L}ogstash~\cite{urllogstash}, and \textbf{K}ibana~\cite{urlkibana}, works to ingest and process logs from different sources by Logstash, in a searchable format accomplished by Eleasticsearch, and Kibana lets users visualize data with charts and graphs. 

From the system's diagnosis perspective, the information provided through the logging statements combined with other system metrics, such as CPU, memory, and I/O utilization, serves an important role in anomaly detection and understanding and diagnosing the system's runtime behavior in the case of a failure. 
Despite the tremendous potential value hidden in execution logs, the inherent characteristics of logs, such as their heterogeneity and voluminosity~\cite{candido2019contemporary}, make the analysis of them difficult on a large scale and poses several challenges. 
Some of the associated challenges with logging statements and their analysis in software systems are as follows: 
\begin{enumerate}[label=\protect\circled{\arabic*}] 
\item Providing proper logging statements inside the source code remains a manual, \textit{ad-hoc}, and non-trivial task in many cases~\cite{li2018studying} due to the free-form text format of log statements and lack of a general guideline for logging.
\item As the size of computer systems increases and software becomes more complex and distributed, manual inspection of log files becomes cumbersome and impractical, and it calls for automated analysis of logs.  
\item Log data can be heterogeneous and voluminous, as within a large software system multiple subsystems may potentially generate a plethora of logs in different formats. Additionally, logging inefficiently introduces overhead on multiple subsystems such as I/O, network, and storage. 
\item Developers and automatic logging analysis tools that aim to automate the logging process always face challenging questions of \textit{``what, where, how, and whether to log?''}.
\end{enumerate}

\begin{figure}[h]
\centering
\includegraphics[scale=.14]{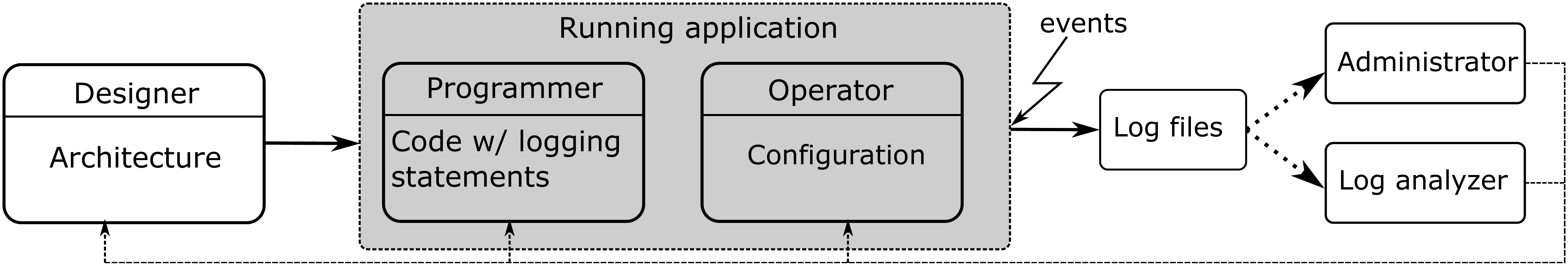}
\vspace*{-3mm}
\caption{Framework for creation and analysis of log files~\cite{salfner2004comprehensive}.}
\label{method_1}
\vspace*{-1mm}
\end{figure}

Considering the aforementioned challenges, prior and ongoing research has made numerous efforts to mine and understand log statements in the source code and execution log files to either gain more insights about logging practices, troubleshoot the software, or automate the logging process~\cite{yuan2012characterizing,chen2017characterizing,zhu2015learning}. 
Figure~\ref{method_1} depicts a framework in which the creation process and analysis of log files are illustrated. 
After the system's architecture is decided and programmers implement the source code with logging statements, the operators run the system by selecting proper runtime configuration parameters (\textit{e.g.}, logging verbosity level in Log4j~\cite{urllog4x}). 
While the software is running, events that are logged in the source code generate records within the log files. 
Next, administrators (, practitioners), and automated log analyzer tools may review the files and feedback the outcome to the designers, programmers, and operators to make adjustments to the architecture, source code, and system configuration if needed, respectively. 

As the size of computer systems increases, the manual process of developers placing logging statements into the source code and administrators reviewing the log files and detecting problems negatively affecting the system becomes less practical and less effective. 
Consequently, being able to automatically detect logging points and insert appropriate logging statements in the source code, as well as systematically analyze log files for detecting system issues are very beneficial and high-in-demand research topics~\cite{zhu2015learning,zhao2017log20}. 
Thus, researchers have dedicated a significant amount of studies in the area of software logging and log analysis techniques throughout the last decade~\cite{candido2019contemporary} to propose various approaches, including automated analysis and application of machine learning to process large-scale log files. 
However, after reviewing the prior work, we noticed what has not been addressed is a clear and comprehensive review of the current progress in software systems' logging and log analysis research. 
It is, therefore, hard for researchers to recognize how their current and future work will fit in the big picture of present-day logging research. 
Understanding where we are at the moment and creating a snapshot of the current research is a fundamental step towards understanding where we should go from here and what the necessary next steps of the research would be. 
Learning from our own experiences and the obstacles that we have to go through to holistically picturize the current research in the field, this survey aims to pave the road obstacles and provide a methodological review of logging, its practices, and its automation techniques and tools for software systems. 
Additionally, in this survey, we review the current state of logging in software to discover solutions for the aforementioned challenges and highlight the next steps for future research efforts. 
We review and study a vast number of peer-reviewed conference and journal papers from related research areas including software and distributed systems, dependability and reliability, and machine (and deep) learning. 
Moreover, we aim to build knowledge~\cite{basili1999building} and trends by connecting and combining findings from multiple research. 
Thus, we examine and categorize the prior research for \textit{logging costs}, \textit{logging practices}, \textit{automation of log analysis}, and \textit{efforts to automate the insertion and improvement of logging statements inside the source code}. 
Finally, we provide \textit{trends and opportunities for future work} based on the insights we gain during this survey. 
Before we dive deeper into the survey, we review the necessary vocabulary in the following. 

\subsection{Terminology}
\textbf{Log printing statement (LPS)}. LPSs are the log statements in the source code added by developers.  
We use \textit{``log printing statement''}, \textit{``log statement''}, and \textit{``logging statement''} interchangeably, as the prior work has used all of the variations~\cite{fu2009execution,zhao2017log20,fu2014developers,gholamian2020logging,li2020qualitative}. 

\textbf{Log message.} A log message, typically a single line, is the output of the LPS in the log file. 
Prior work also makes a subtle distinction between a log message and a \textbf{log entry}, and defines a \textit{``log entry''} or a \textit{``log record''} as a line in the log file composed of a log header and a log message~\cite{el2020systematic}. 
Log header contains timestamp, verbosity level, and source component, and its format is usually defined by the logging framework, \textit{e.g.}, \textit{log appenders}~\cite{urlslf4japp}, whereas a log message is written by the developer and consists of the amalgamation of the static part of the log message from the source code and the dynamic value of the variables during the runtime.
For our purpose in this survey, log message, log entry, log record, and log event are used interchangeably~\cite{pecchia2015industry,li2020qualitative,fronza2013failure,kubacki2017holistic}.

\textbf{Log file.} Log file(s) is a collection of log messages stored on a storage medium, also called \textit{``event logs''}, and \textit{``execution logs''}, or simply just \textit{``logs''}~\cite{fronza2013failure,kubacki2017holistic,fu2009execution}. 
In most cases, these terms can be used interchangeably, and we commonly use log file(s) as an umbrella term to cover the different naming variations throughout the survey. As a minor point, in special cases that we mean to refer to a set of log lines in general (\textit{i.e}., without binding them to specific files), it is more appropriate to refer to them as execution logs or log records (\textit{e.g.}, \textit{execution logs} are used for anomaly detection). 

Additionally, we might refer to computer (computing) systems and software systems interchangeably on some occasions throughout this survey with regards to logs, meaning that the log messages in the log files are generated from log printing statements within the source code of the software systems (for various software or hardware related events, concerns, or issues), which are also an artifact of computer systems as an umbrella for software, hardware, and anything in between.

\subsection{Survey Organization}
The rest of this survey is organized as follows. 
Section~\ref{log_file}, prior to explaining our study design, we provide the background for log statements, messages, and files. 
Section~\ref{sutdy_design} provides the details of our study design and research questions (RQs) based on the SLR and SM conventions in performing evidence-based surveys in software engineering~\cite{kitchenham2015evidence}. 
In Section~\ref{rq1_categorize}, we provide our findings for RQ1 and categorize the prior logging research. 
In Section~\ref{rq2_trends}, we present our findings for RQ2 and present the publication trends for different topics, years, and venues. 
Then, Section~\ref{rq3_breakdown} reviews the prior research in each category of logging in details and provides our findings for RQ3. 
In Section~\ref{opportunities}, we provide our findings for RQ4 and describe open problems and opportunities for future work, and Section~\ref{conclusions} concludes the survey.

\section{Log statements and Log Files}\label{log_file}
Logging is the process of recording and keeping track of the \textit{events of interest}, \textit{e.g.}, to developers, practitioners, system admins, and end users, while the software is running. 
As such, log messages aim to achieve this goal and record the events of interest that happen during the software system's execution and store them in the log files. 
Generally, the logging process starts with software developers (\textit{i.e.}, programmers) include logging statements with description, variables, and verbosity levels (Figures~\ref{log_example_c} and~\ref{log_example_lib}) into the source code. 
Then, while the software is running, the logging statements are \textit{\textbf{logged}}, if appropriate configurations (\textit{e.g.}, verbosity level) are enabled.
In the simplest case, log messages are written to a single log file. 
However, in a distributed system, there can be multiple log files in different formats.
The focus of our survey is on this type of logs which are also called execution logs or event logs~\cite{vaarandi2003data}. 
Event logs are the outcome of logging statements that software developers insert into the source code. 
Event logging and log files are playing an increasingly important role in computer systems and network management~\cite{vaarandi2003data,xu2009detecting,candido2019contemporary}, which we will review later in this survey.

\subsection{Transaction Logs}
It is also worth mentioning briefly the difference between the execution logs and transaction logs. 
A transaction log (also called journal) is a record file of the transactions between a system and the users of that system, or a data collection method that naturally captures the type, content, or time of transactions made by a user from interacting with the system. 
In a database server, a transaction log is a file in which the server stores a record of all the transactions performed on the database~\cite{davis2012sql}. 
The transaction log is an important component of database servers and cryptocurrency protocols (\textit{e.g.}, blockchain~\cite{beck2016blockchain})  when it comes to recovery. 
If there is a system failure, transaction logs are used to revert the database back to a consistent state. 
In summary, transaction logs act as a ledger to accurately record the transactions in the system which are agreed, shared, and synchronized among all the parties involved, whereas execution logs capture events of interests with different verbosity and severity levels, \textit{e.g.}, DEBUG, INFO, ERROR, \textit{etc.}, which do not necessarily require sharing, agreement (\textit{i.e.}, consensus), or synchronization between different software modules.   
  
\subsection{Log Example}
Software developers utilize logging statements inside the source code to gain insight into the internal state of applications amid their execution. 
In the simplest form, logging statements are \textit{print} statements utilized in different programming languages. 
In this case, the logging statement may contain a textual part indicating the context of the log, \textit{i.e.}, the \textit{description} of the log, and a \textit{variable} part providing contextual information about the event. 
Figure~\ref{log_example_c} shows an example of logging statements in \textit{C} programming language.  
\begin{figure}[h]
\small
\vspace{-2mm}
\begin{flushleft}
\centering
\fbox{\begin{minipage}{27em}
printf(``Cannot find BPService for bpid=\%d'', id);\\
\hspace*{8mm} $\vert$ \hspace{8mm} \textbf{description} \hspace{20mm} $\vert$ \textbf{variable}
\end{minipage}}
\end{flushleft}
\caption{Log example in \textit{C}.}
\label{log_example_c}
\vspace{-3mm}
\end{figure}

Logging statements may utilize logging libraries to improve the organization of the logged information. 
For example, in Java, libraries such as Log4j~\cite{urllog4x} and SLF4J~\cite{urlslf4j} provide a higher degree of flexibility to the developers. 
Logging libraries, also called logging utilities (LU)~\cite{chen2020studying} or logging libraries and utilities (LLUs), provide extra features, such as \textit{log level}, which indicate the verbosity and the severity of the logging statement. 
Log levels help to better distinguish the importance of runtime events and control the number of logs collected on the storage device~\cite{mizouchi2019padla}. 
For example, less verbose levels, \textit{i.e.}, \textit{FATAL}, \textit{ERROR}, and \textit{WARN}, are leveraged to alarm the user when a potential problem happens in the system, and more verbose levels such as \textit{INFO}, \textit{DEBUG}, and \textit{TRACE} are utilized to record more general system events and information or detailed debugging. 
In practice, \textit{INFO} and more verbose levels are utilized during the software development phase by programmers, and \textit{INFO} or less verbose levels are, by default, for the software deployment phase, as the end user observes. 
In case more insight about the internal state is needed, end users might enable more verbose logging.
An example of a logging statement with library usage for \textit{WARN} verbosity level is shown in Figure~\ref{log_example_lib}.
\begin{figure}[h]
\small
\vspace{-3mm}
\begin{flushleft}
\centering
\fbox{\begin{minipage}{27em}
log.warn("Cannot find BPService for bpid=" + id);\\
\hspace*{5mm}\textbf{level }$\vert$ \hspace{5mm}  \hspace{8mm} \textbf{description} \hspace{13mm} $\vert$ \textbf{variable}
\end{minipage}}
\end{flushleft}
\vspace{-3mm}
\caption{Log example with Log4j library.}
\label{log_example_lib}
\vspace{-1mm}
\end{figure}

\begin{figure}[h]
\centering
\includegraphics[scale=0.55]{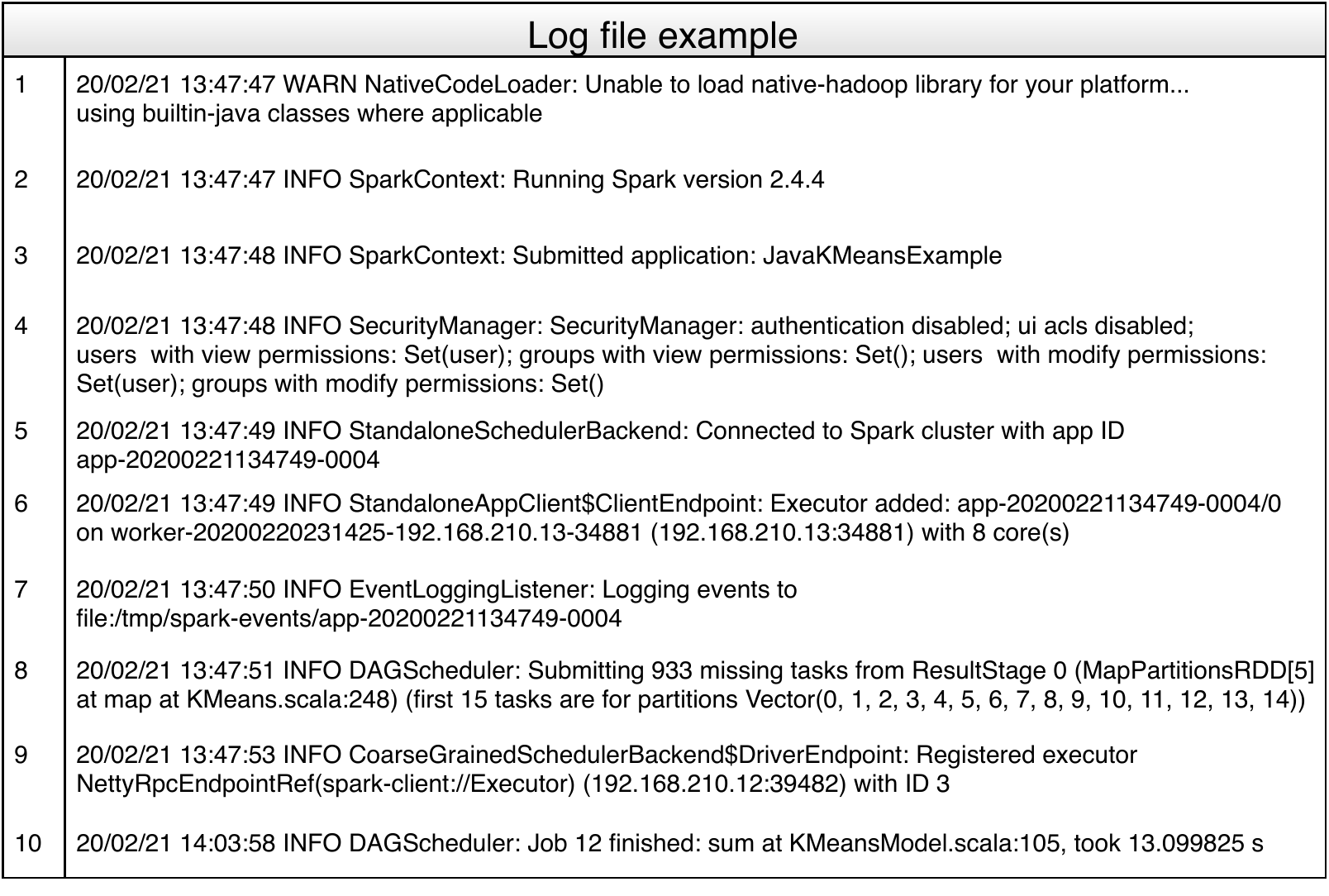}
\vspace*{-3mm}
\caption{A log file example from Apache Spark.}
\label{log_file_fig}
\vspace{-2mm}
\end{figure}

Logging statements are generally saved in log files. 
Figure~\ref{log_file_fig} shows ten lines of logs from Apache Spark~\cite{urlspark1} execution logs collected in our execution of \textit{k-means} clustering algorithm~\cite{macqueen1967some} on a standalone cluster. 
In real-world cases, a cloud computing system can generate millions of such log messages per minute~\cite{zhu2019tools}. 
For example, for an online store with millions of customers worldwide, it is common to generate tens of terabytes of logs in a single day~\cite{logothetis2011situ,liu2019logzip}.

As observed in Figure~\ref{log_file_fig}, a major part of logging messages is unstructured text. 
Thus, in order to make log files useful and avoid the hassle of manually processing a plethora of log files, the first and foremost step of log processing is the automatic \textit{``parsing''} of log messages, which transforms unstructured logs into structured events. 
Not to be confused with syntactic parsers in programming languages, which parses source code and confirm whether it follows the rules of the formal grammar, log parser, on the other hand, transforms unstructured raw log files into a sequence of structured events, to enable automated analysis of logs. 
We review the log parsing process later in Section~\ref{log_parsing}. 

\subsection{Program Traces vs. Logs}
The term log is often used to represent the way a program is used (such as security logs), while tracing (not to be confused with \textit{``TRACE''} log level in logging libraries such as Log4j) is used to capture the temporal sequence of events during a particular execution of a program~\cite{miranskyy2016operational}, in contrast to logs which are generally the consolidation of continuous execution of software systems. 
Tracing is typically performed by an external program/tool that instruments the runtime environment, such as network traffic traces, whereas logs are the direct output of logging statements' execution inside the software. 
Moreover, while traces are typically structured data, logs are free-form and unstructured text. 
For example, a trace can contain the software execution paths, the events triggered during the execution, and the value of variables, which are used for debugging and program understanding. 
Stack traces are common examples that are used for function call tree tracing during development and postmortem debugging. 
Some of the well-known tracing systems include Google's Dapper~\cite{sigelman2010dapper}, X-Trace~\cite{fonseca2007x}, Microsoft's MagPie\cite{barham2003magpie}, the \textit{black-box} approach~\cite{aguilera2003performance}, and Casper~\cite{wu2016casper}. 

\section{Study Design}\label{sutdy_design}
We perform a systematic literature review (SLR) and mapping study (MS) following Kitchenham \textit{et al.}~\cite{kitchenham2015evidence,keele2007guidelines} and Petersen \textit{et al.}~\cite{petersen2015guidelines} guidelines. 
In the following, we elaborate on the details of this process.

\subsection{Research Questions}\label{rq_subsection}
While we review the prior literature, we aim to pursue and answer the following research questions: 

\begin{itemize}
\item \textbf{RQ1:} How to systematically review and categorize prior logging research into different topics? 
\item \textbf{RQ2:} What are the publication trends based on venues, topics, and years? 
\item \textbf{RQ3:} How the research in each topic can be systematically compared with their \textit{approaches}, \textit{pros}, and \textit{cons}? 
\item \textbf{RQ4:} What open problems and future directions are foreseeable for logging research?
\end{itemize}

With the pursual of the aforementioned RQs, we ensure to follow the established evidence-based software engineering (EBSE) paradigm for our literature review~\cite{keele2007guidelines,kitchenham2015evidence}. 
As a contributing improvement, our survey combines and benefits from the advantages of both systematic literature review (SLR) and systematic mapping (SM) paradigms. 
Prior research indicates the main differences between SM and SLR are that SM methodology is broader and more based on qualitative measures, while SLR focuses on narrower research questions and quantitative measures~\cite{napoleao2017practical}. 
While being comprehensive, \textit{i.e.}, SM, our survey also provides details on the experimentation and results of each primary study, \textit{i.e.}, SLR.   
In summary, RQ1, RQ3, and RQ4 qualitatively assess the prior literature into different research categories base on different topics, \textit{i.e.}, systematic mapping (SM)~\cite{wohlin2012experimentation}, whereas RQ2 quantitatively measures the publications based on venues, topics, and years, and RQ3 provides details aligned with SLR data extraction methods for each study, \textit{e.g.,} \textit{the aim}, \textit{experiments}, \textit{results}, and \textit{findings}~\cite{napoleao2017practical}.

\afterpage{
{\renewcommand{\arraystretch}{1.4}
\begin{table}[h]
\fontsize{7.5}{11}\selectfont
\centering
\begin{tabular}{p{1cm} p{7cm}} 
 \toprule
 \cellcolor{blue!20}\textbf{Research databases} & \cellcolor{gray!15} ACM Digital Library, IEEE Xplore, ScienceDirect (Elsevier), Scopus, SpringerLink, and Google Scholar (snowballing).\\ [0.5ex] 

\midrule
 \cellcolor{blue!20}\textbf{Search query}& 
'software AND log AND (statement OR file OR record OR event)' AND 'Publication Date: (01/01/2010 TO 05/31/2021)' 
  
\\ [2ex] 

 \bottomrule
\end{tabular}
\caption{Literature review databases and keywords. We searched in various online research databases for different combinations of log-related keywords.
}
\label{table_lit_review}

\end{table}
}
}

\subsection{Search Strategy and Paper Selection}
\begin{figure}[h]
\centering 
\includegraphics[scale=.67]{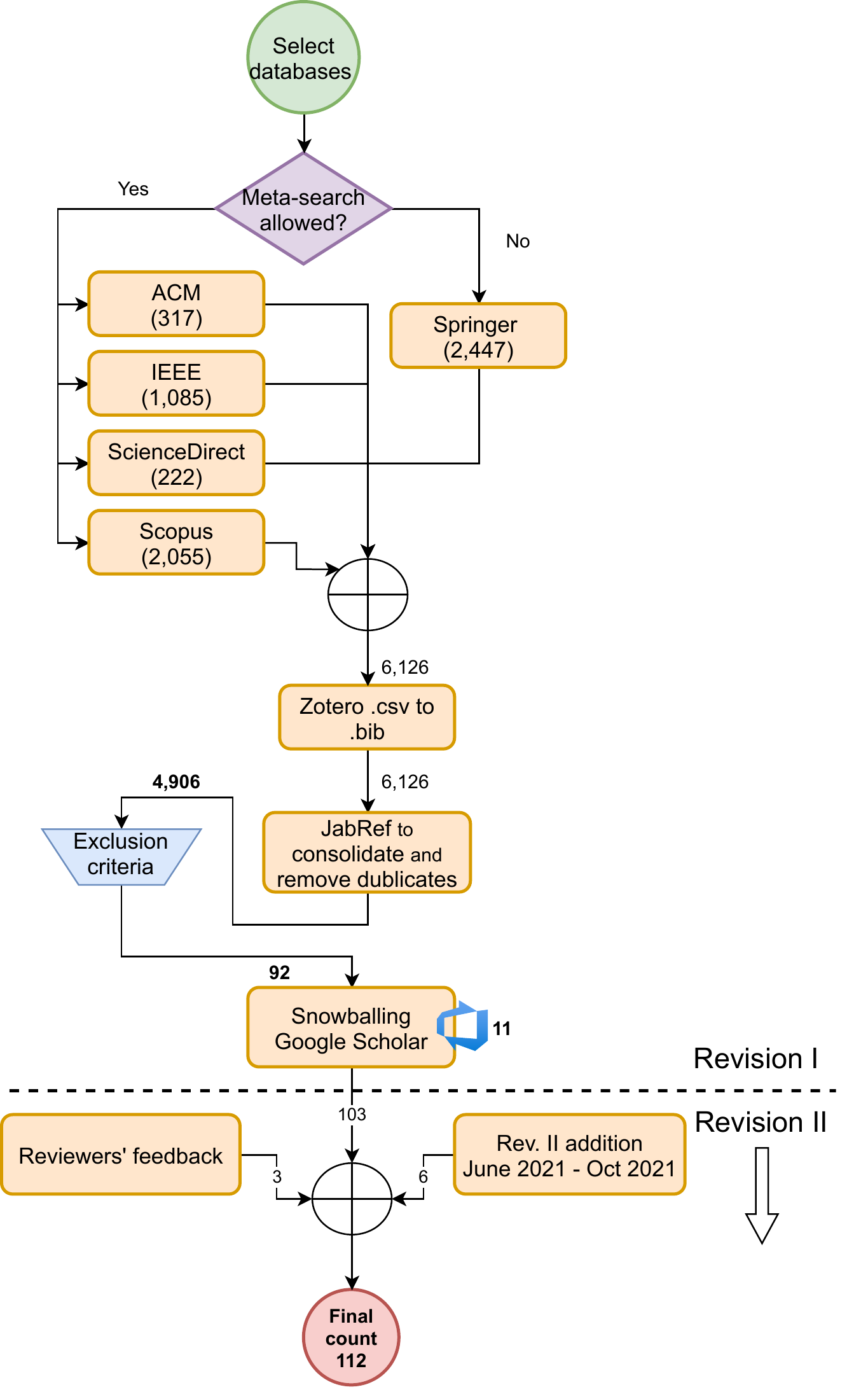}
\caption{This figure provides the steps involved in our survey methodology.}
\label{methodology_chart}
\end{figure}

Studying log mining and logging analysis techniques in software systems is a challenging and widespread topic, and there have been numerous prior studies that focus on log analysis. 
Table~\ref{table_lit_review} summarizes the databases and the search query used in our survey, and Figure~\ref{methodology_chart} provides the flowchart of the steps in the database selection and reference analysis. 
We start with the established research databases listed in the table and search for the list of log-related keywords. 
If the database allows for metadata search (\textit{i.e.}, title, abstract, and keywords), we limit our search to metadata, to avoid the inclusion of numerous unrelated research that has the keywords in their main text. 
However, for Springer, we could not use meta-search, and thus, the reason behind the high number of returned publications. 
We use reference management software, such as Zotero~\footnote{\url{https://www.zotero.org/}} and JabRef~\footnote{\url{https://www.jabref.org/}} to facilitate and automate our process of reference consolidation from different sources and duplicate removal. 
After duplicate removal, our process resulted in 4,906 papers in the first revision of this survey. 
In the second revision (June-October 2021), we gathered additional 633 publications. 

\subsection{Inclusion and Exclusion Criteria}

{\renewcommand{\arraystretch}{1.4}
\rowcolors[]{2}{gray!15}{white}
\begin{table}
\fontsize{7.5}{11}\selectfont
\centering
\begin{tabular}{p{.7cm} p{7cm}} 
 \toprule
 \rowcolor{blue!20}\textbf{Criteria} &  \textbf{Explanation}\\ [0.5ex] 
\midrule

C1& Primary study is in English language.\\
C2& Final (and long) version of an study is selected.\\  
C3& The date of publication is $\geq 2010$.\\
C4& It paper is published in very good to flagship venues.\\
C5& The search query exists in the title, abstract, or keywords of the publication.\\
C6& Additional studies that are included through single iteration of backward and forward \textit{snowballing} search.\\
 
\bottomrule
\end{tabular}
\caption{Selection criteria for primary studies.}
\label{table_criteria_extraction}
\end{table}
}

Table \ref{table_criteria_extraction} summarizes the selection criteria in our study. For \textbf{exclusion criteria}, our focus has been on the full (\textit{i.e.}, not short), recent papers in English in the last decade in the established venues, ranging from very good to flagship ones~\footnote{\url{https://www.core.edu.au/conference-portal}}. 
We focus on the last decade as our goal is to provide a fresh picture of the recent trends in logging research to enable relevant directions for future work. 
In addition, other surveys, such as Salfner et al.~\cite{salfner2010survey}, provided a comprehensive overview of failure detection approaches up to 2010, which also leverage log analysis. 
Moreover, Candido \textit{et al.}~\cite{candido2019contemporary} reviewed and found very few studies earlier than 2010 in their systematic literature review of software monitoring through logs. 
Laterally, we observe that another factor that has triggered a momentum in logging research is the recent rise in artificial intelligence (AI), machine learning (ML), and in particular deep learning (DL) approaches~\cite{greene20102019}. 
A significant portion of logging code automation and automated log analysis approaches leverage ML and DL algorithms.       

After we gather the publications with the search query, we manually investigated the results and included relevant publications that directly tackle issues associated with log statements or log files, and excluded the ones which had a weak association with logs. 
Due to the significant number of curated papers, for manually reading the papers, we follow the prior work's suggestion~\cite{wohlin2014guidelines} to read the title and abstract first and then other parts of the papers to decide on inclusion in an efficient way. 
A significant number of papers are excluded as we read the beginning of the papers.    
Also, if a research project has multiple variations, \textit{i.e.}, a conference paper followed by a more comprehensive journal paper, we only include the more comprehensive version. 
Additionally, once we find an influential paper (\textit{i.e.}, highly-cited based on the absolute number of citations $>100$~\cite{garousi2016highly}), we also check all of its references and its citations, \textit{i.e.}, a single round of backward and forward \textit{snowballing search}~\cite{wohlin2014guidelines} with Google Scholar. 
Followed by exclusion criteria, we are narrowed down to 92 publications, and at last, 11 additional references are added with snowballing, bringing the total count to 103 publications in the first revision of the paper. 
In the second revision, based on the reviewers' feedback, we added three additional primary studies~\cite{busany2016behavioral,amar2018using,bao2019statistical}. 
We also reviewed the databases and list of accepted papers in venues in Table~\ref{pub_venues_breakdown} for new publications, for June-October 2021 interval, and found six recent studies~\cite{yao2021improving,wang2021would,gujral2021exploratory,chen2021demystifying,gholamian2021distributed,le2021log}. 
This brings the total number of primary studies to \textbf{112}.

\afterpage{
{\renewcommand{\arraystretch}{1.4}
\rowcolors[]{2}{gray!15}{white}
\begin{table}
\fontsize{7.5}{11}\selectfont
\centering
\begin{tabular}{p{.5cm} p{7cm}} 
 \toprule
 \rowcolor{blue!20}\textbf{RQs} &  \textbf{Extracted data}\\ [0.5ex] 
\midrule

RQ1& General categories of main topics and subtopics of the primary studies.\\
RQ2& Details associated with its year of publication, venue, and the topic that it belongs too.\\  

RQ3& Five items for each primary study: aim, experiments, results, \textit{pros}, and \textit{cons}. \\
RQ4& Open problems and opportunities for the categories of the primary studies if available, \textit{i.e.}, if explicitly or implicitly can be extracted.\\ [2ex] 
\bottomrule
\end{tabular}
\caption{This table summarizes the data extracted for each RQ from each primary study.}
\label{table_data_extraction}
\end{table}
}
} 

\afterpage{
{\renewcommand{\arraystretch}{1.4}
\rowcolors[]{2}{gray!15}{white}
\begin{table*}[bp]
\hspace*{-4mm}
\fontsize{7.5}{11}\selectfont
\centering
\begin{tabular}{p{1.3cm} p{2.1cm} p{.8cm} p{4.1cm} p{4cm} p{4.2cm}} 
 \toprule
 \rowcolor{blue!20}\textbf{Survey} & \textbf{Online databases}& \textbf{Type}& Type& RQs & \textbf{Main differences (Us vs. them)} \\ [0.5ex] 
\midrule

He \textit{et al.}~\cite{he2021survey}& IEEE, ACM, Springer, Elsevier, Wiley, and ScienceDirect& SLR 
& `log OR logging OR log parsing OR log compression OR (log + anomaly detection) OR (log + failure prediction) OR (log + failure diagnosis)'
& No clear-cut RQs; the survey discusses logging mechanism, compression, and mining for the goal of reliability analysis.
& Main differences in the search query and RQs. Our work is different on logging cost, log statement automation, evaluation metrics, and taxonomy chart.\\

Chen and Jiang~\cite{chen2021survey} & IEEE, ACM, and DBLP
& SLR
&`logging OR instrumentation OR tracing'
& No clear-cut RQs; the goal of the survey is to techniques used in all three software log instrumentation approaches: conventional logging, rule-based logging, and distributed tracing, and uncovers four categories of challenges. &Main differences in the databases, search query, and RQs. Our work is different on various topics covered in automated log mining.\\

C{\^a}ndido \textit{et al.}~\cite{candido2021log}
& Google Scholar, ACM, IEEE, Scopus, and  Springer
&SM
&`log AND (trace OR event OR software OR system OR code OR detect OR mining OR analysis OR monitoring OR web OR technique OR develop OR pattern OR practice)'
&RQ1: trends in log-based monitoring; RQ2: Different scopes of log-based monitoring&
Main differences in the search query, RQs, and focus of the survey; it focuses on monitoring for studies up to the end of 2019.\\

Gholamian and Ward (This study)
& IEEE, ACM, Springer, ScienceDirect, Scopus, and Google Scholar
&SM/SLR
&'software AND log AND (statement OR file OR record OR event)'
& Four RQs listed in Section~\ref{rq_subsection} on the categorization of subtopics, publication trends, systematic comparison of each topic, and future directions. 
& In sum, we differ in the search query and more comprehensive approach in searching, the proposed RQs, and one-of-a-kind discussion on taxonomy (Figure~\ref{tree}), findings for each RQ, and directions on research opportunities (Table~\ref{summary_future}).\\
\bottomrule
\end{tabular}

\caption{This table summarizes the difference between this survey and other recent survey studies.}
\label{table_survey_comparison}

\end{table*}
}
}

\subsection{Data Extraction and Collection}
After the primary studies (112) are selected, from each study, we extract the data required for analyzing and answering each research question.    
For example for RQ1 and RQ2, from each paper, we extract its primary topic and other secondary topics (\textit{i.e.}, subtopics) which are discussed in the study. 
For RQ3, for each primary study, we provide its \textit{1) aim}, \textit{i.e.}, the problem it is trying to address, \textit{2) experimentation}, \textit{3) results and findings}, \textit{4) advantages}, \textit{i.e., pros}, and \textit{5) disadvantages}, \textit{i.e., cons}. 
We extract \textit{pros} and \textit{cons} based on 1) authors explanation of the main benefit and limitation of their work, 2) findings of the follow-up studies that have performed comparisons, or finally, 3) our gained knowledge during the survey and by reviewing the research in comparison to other related studies. 
The data extraction is initially performed by the first author of this survey, and then reviewed by the second author. 
In cases where there exists a disagreement, both authors discuss the issue until reaching a consensus. 
Table~\ref{table_data_extraction} lists the data extracted from primary studies for each RQ.\looseness=-1

Based on the extracted data, we categorize the selected publications (112) into \textbf{twelve categorizes} after carefully studying them. 
Our methodology for categorizing the publications has been based on a top-down approach. 
Meaning that, we first were able to draw categories that either focus on logging statements or the ones that focus on log files. 
We then further narrowed down each category based on its primary focus.
Next, we also extracted subtopics for each paper, as usually, publications also partially cover some other related topics in their research. For example, Zhao \textit{et al.} work ~\cite{zhao2017log20} primarily focuses on log statement automation, but it also covers topics related to logging cost analysis. The details for each category is available in Table~19 in Section~\ref{paper_full_list}.

\subsection{Summary of Differences}
Our survey differs from other related surveys in various ways as listed in Table~\ref{table_survey_comparison}. 
We summarize how our study differs from the other recent surveys that cover parts of the software logging research domain. 
Not only our approach is different in a variety of ways from the compared surveys (\textit{e.g.}, we extract different types of data from primary studies), our work also differs significantly in the selected primary studies.  
In comparison, out of 112 primary studies that we have reviewed, 57, 70, and 80 of them are not reviewed by He \textit{et al.}~\cite{he2021survey}, C{\^a}ndido \textit{et al.}~\cite{candido2021log}, and Chen and Jiang~\cite{chen2021survey}, respectively. 
Additionally, 39 primary studies that we reviewed are not examined in the three aforementioned surveys \textbf{combined}, \textit{i.e.}, they are exclusive to this survey.

\section{RQ1: How the prior logging research can be categorized to different topics? }\label{rq1_categorize}
In this section, we provide our findings for the first research question by reviewing the available literature and categorizing the logging research into its topics (and subtopics), which enables us to explain the survey scope that follows next.

\subsection{Survey Scope}
Based on our methodology, the scope of our survey is developed as follows. 
To the best of our knowledge, there is no prior work that provides a systematic and comprehensive coverage on log mining and automation techniques in software systems, covering different aspects of logging such as mining source code and log files, automating log printing statements in the source code, their evaluation techniques, as well as a comprehensive review of log mining and log analysis approaches. 
As mentioned before, there are a few existing surveys on the application of execution logs for anomaly and problem detection~\cite{salfner2010survey}, system monitoring~\cite{candido2019contemporary}, reliability engineering~\cite{he2021survey}, and instrumentation~\cite{chen2021survey}. 

Thus, we cover the following sections in this survey:
\begin{itemize}
\item \textbf{Logs and log files.} We explain what are log files, log statements, and log messages, and what sort of applications and analyses they are leveraged for.
\item \textbf{Logging cost.} We point out the quantitative and qualitative costs and benefits associated with logging.
\item \textbf{Logging statement mining and automating.}
Logging research aims to understand current logging practices and use the findings to improve the log printing statements with automatic log insertion and learning to log techniques. 
Thus, we review:
\begin{itemize}
\item \textbf{Logging code practices.} This section includes studies that empirically or automatically investigate how developers insert logging statements into the software's source code and how the logging evolution and improvement can benefit the usage of the logging code.
\item \textbf{Automatic log insertion and learning.} We cover the studies that leverage static code analysis, heuristics, and machine learning techniques to automatically add (or improve) logging statements in the source code to make them more effective in failure diagnosis.
\item \textbf{Evaluation.} As we study automatic ways of addition and learning of log statements, we introduce several metrics to measure the effectiveness of the proposed approaches. 
\end{itemize}
\item \textbf{Mining logs.} 
This part provides insight into methods and tools to analyze log messages and log files. 
This can be further divided into log management, log parsing, and their applications. Thus, we review:
\begin{itemize}
\item \textbf{Log management and maintenance.} Management and collection of logs are important as a pre- or post-step of log analysis. 
\item \textbf{Log parsing.} To enable log message analysis, we first require to parse the log messages and extract their templates. 
\item \textbf{Application of logs.} 
We review a wide range of applications that leverage automated log analysis for various software engineering tasks, such as anomaly detection and failure diagnosis. 
\end{itemize}
\item \textbf{Emerging applications of logs.} We review the recent special interest in applications of logs in other domains, such as mobile devices and big data.
\item \textbf{Opportunities for future work.} Based on the current log-related research, we comment on the future directions and opportunities for each category of logging research.
\end{itemize}

As our aim is to provide a comprehensive and end-to-end survey of the logging in software, we have covered the topics of \textbf{\textit{mining and automation of logging statements}} and \textbf{\textit{mining of log files}} side-by-side in this survey. 
We have seen that these topics go hand-in-hand and the synergy between them has resulted in more effective logs and logging practices. 
In fact, the ultimate goal of a mountain of studies for logging statement automation is to \textbf{\textit{improve various log mining tasks}} (Figure~\ref{method_1}). 
For example, \textit{ErrLog}~\cite{yuan2012conservative} and \textit{LogEnhancer}~\cite{yuan2012improving} automatically introduce new logging statements or add additional variables to the logging statements (\textit{i.e.}, \textbf{automation of the logging statements}) to improve the quality of logs, and, subsequently, \textbf{\textit{improve log mining tasks}} such as error detection and program diagnosability. 
In another example, authors of \textit{Log20}~\cite{zhao2017log20}, an automated log placement tool, explain that the main objective of their log placement tool is to \textit{``disambiguate''} the execution paths, and consequently, improve the effectiveness of log mining methods in \textit{``debugging real-world failures''}. 
Furthermore, we observe that ignoring the cross-cutting concerns of logging has resulted in log-related issues in the past, such as stale and confusing logging statements, and has hindered effective log file mining~\cite{shang2014exploratory,kabinna2018examining}. 

\afterpage{
\begin{landscape}
\begin{figure}[h]
\vspace*{5mm}
\hspace*{-7mm}
\centering 
\includegraphics[scale=.7]{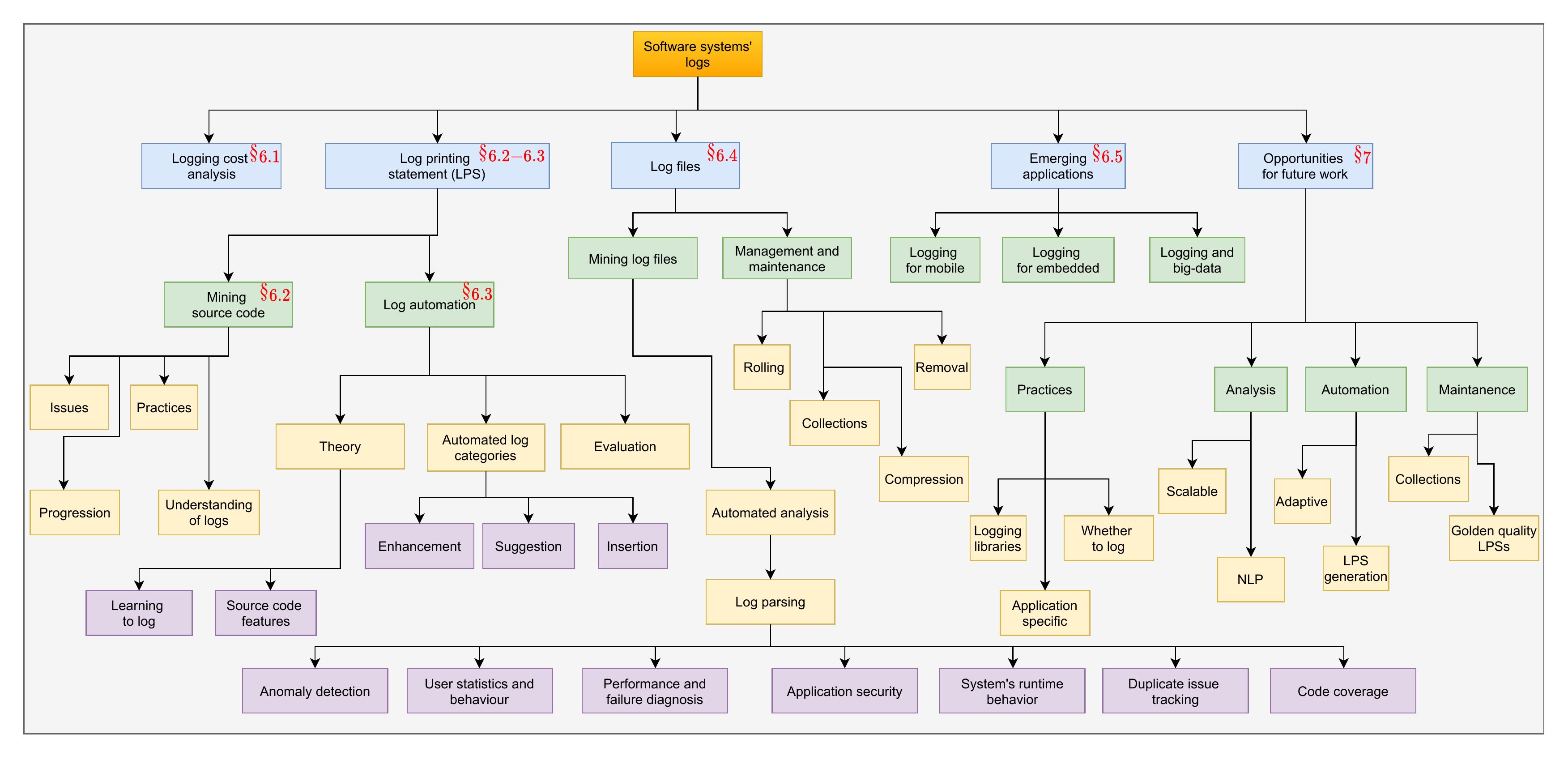}
\caption{This figure provides a taxonomy of the present-day logging research. 
We initially provide a background on logs and log messages in \S\ref{log_file}, our study design in \S\ref{sutdy_design}, and the categorization of research to subpotics in \S\ref{rq1_categorize}. 
We then provide the publication trends in \S\ref{rq2_trends} and explain logging costs and benefits in \S\ref{log_cost}. 
Log printing statement research covers two sections, \S\ref{log_mining}-\ref{automated_logging}, followed by log files' research in \S\ref{mine_log_files}. 
\S\ref{log_apps} and \S\ref{opportunities} provide emerging applications and opportunities for future work, respectively. 
We conclude our survey in \S\ref{conclusions}.}
\label{tree}
\end{figure}
\end{landscape}
}  

It is also important to mention that there exist additional industry products, with an aggregate market cap beyond \$125 billion USD, that perform log analysis for various goals such as performance evaluation, cloud monitoring, and data analytics, to name a few: Datadog~\cite{urldatadog} (\$49.7B)\footnote{Values are collected at the time of this survey from Google Finance. For example for Datadog: \url{https://www.google.com/finance/quote/DDOG:NASDAQ}}, Splunk~\cite{urlsplunk} (\$27.3B), Elasticsearch~\cite{urlelastic} (\$15.5B), Loggly~\cite{urlloggly} (\$5.7B), Dynatrace~\cite{urldynatrace} (\$22.1B), New Relics Inc.~\cite{urlnewrelic} (\$4.9B), and XpoLog~\cite{urlxpolog}. 
However, this paper mainly focuses on surveying academic works, or peer-reviewed publications from the industry. 
We acknowledge that there is a mountain of work in the industry that performs log analysis, however, it is mainly outside the scope of this survey. 
Additionally, logging is also used in other computing systems, such as embedded or hardware devices/sensors, which are generally outside the scope of this survey, as we mainly focus on software systems. 
In addition, we mention some of the mobile devices studies that are closely related and have aimed to replicate the efforts in software logging research in Section~\ref{log_apps}.   
Lastly, Figure~\ref{tree} summarizes a taxonomy of the categorization of modern-day logging research in our survey, and we leverage the classification in this figure to divide the logging research into subtopics and study them in the upcoming sections. 

\begin{tcolorbox}[breakable, enhanced]
\small \textbf{Finding} \textbf{1.} \textit{
Based on the taxonomy and our literature review, the logging research is spread through twelve categories (topics):  
\begin{enumerate*}[label=\protect\circled{\arabic*}] 
\item costs and benefits of logging, 
\item logging practices, 
\item logging progression, 
\item log-related issues, 
\item log printing statement automation, 
\item log maintenance and management,
\item log parsing,
\item log-based anomaly detection,
\item log-based runtime behavior analysis, 
\item log-based performance, fault, and failure diagnosis,
\item log-based user, business, security, and code-coverage analyses, and
\item emerging applications of logs.
\end{enumerate*}
}
\end{tcolorbox}

\section{\textbf{RQ2:} What are the publication trends based on venues, topics, and years? }\label{rq2_trends}
For this RQ, after categorizing the \textbf{112} selected publications, we organize the publications based on their venues and publication years and draw high-level trends. 
Based on our findings from the trends, we summarize some of the log-related challenges, such as automating the logging code and automated log analysis, that the prior research has aimed to address at the end of this section. 

\subsection{Venue Trends}
Table~\ref{pub_venues_breakdown} provides a breakdown of surveyed publications per venue.
Although looked for, we could not find a related work in TOSEM~\footnote{ACM transactions on Software Engineering and Methodology.}. 
The majority of the research in this field is published in EMSE, ICSE, TSE, ASE, and JSS. 
We suggest that the authors of future publications consider the following venues in Table~\ref{pub_venues_breakdown} for submitting their works, and consider the number of related references that their work aligns with in that venue.
This ensures that their works will receive higher visibility and a thorough comparison with the prior work. 

\subsection{Topic Trends}
Figure~\ref{fig_pub_per_topic} shows the percentage of publications per each topic. 
Overall, we have divided the logging research into twelve subcategories (\textit{i.e.}, topics), which we will review throughout this survey.  
Table~\ref{reference_per_topic} (located at the end of the survey) lists the topics and provides the related references for each one. 
The table serves as a quick-access guide to review the research happening in each topic. 
Based on our analysis, the top-5 active and popular research topics in the field of logging based on the number of publications are: \textbf{1) log mining for anomaly detection}, \textbf{2) log printing statement automation}, \textbf{3) log mining for performance and failure diagnosis}, \textbf{4) log maintenance and management}, and \textbf{5) log parsing}.

\subsection{Year Trends}
Figure~\ref{paper_list} illustrates the number of conference (blue), journal (orange), and archived (gray) publications per year until October 31, 2021. 
The upward trend in the plot suggests a continuous and growing interest of the research community to explore various dimensions of logging research. 
The publications are from both academia and industry such as Microsoft~\cite{fu2014developers,zhu2015learning,barik2016bones}, Twitter~\cite{lee2012unified}, Huawei~\cite{he2020loghub}, RIM~\cite{shang2014exploratory}, and others~\cite{pecchia2015industry}. 
Additionally, there exists valuable research from the synergy between academic researchers and industry teams, which further emboldens the efforts by bringing real-world industry experiences~\cite{ding2015log2,he2020loghub,liu2019logzip}. 
As such, we foresee the research in this area will continue to grow and foster in the upcoming years as there are interesting and promising trends for future research, explained in Section~\ref{opportunities}.

\afterpage{\renewcommand{\arraystretch}{1.2}
\newcounter{magicrownumbers}
\newcommand\rownumber{\stepcounter{magicrownumbers}\arabic{magicrownumbers}}
\rowcolors[]{2}{gray!15}{white}
\begin{table*}[bp]
\scriptsize	
\begin{center}
  \begin{tabular}{p{.2cm} p{.3cm} p{3.5cm} |p{.6cm} p{2.5cm}|p{.2cm} p{.3cm} p{4cm} |p{.7cm} p{1.8cm}} 
 \toprule
 \rowcolor{blue!20}  No. & {\fontsize{6}{9}\selectfont Type} & Name (\# of publications) & Abbr. & References &No& {\fontsize{6}{9}\selectfont Type}&Name (\# of publications)& Abbr.& References \\ [0.5ex] 
 \midrule

   \rownumber. &J &  Empirical Software Engineering (13) & EMSE &
   \cite{russo2015mining,yao2020log4perf,chen2017characterizing2,shang2015studying,kabinna2018examining,li2019guiding,hassani2018studying,li2017towards,li2017log,yao2020study,chowdhury2018exploratory,zeng2019studying,chen2021demystifying} 
&   10. &   C&  IEEE International Conference on Data Mining (3)& ICDM&\cite{lim2014identifying,fu2009execution,du2016spell}\\   
   
\rownumber.& C & International Conference on Software Engineering (12) & ICSE&  \cite{syer2013leveraging,li2021deeplv,yang2021semi,yuan2012characterizing,chen2017characterizing,chen2020studying,fu2014developers,pecchia2015industry,zhu2015learning,shang2013assisting,barik2016bones,busany2016behavioral}
& 11. &C& Mining Software Repositories (3) & MSR&\cite{kabinna2016logging,candido2021exploratory,gholamian2021naturalness}\\

  \rownumber. &  J & IEEE Transactions on Software Engineering (6) & TSE & \cite{li2021studying,locke2021logassist,li2020qualitative,
  dai2020logram,liu2019variables,yao2021improving}
   & 12.& C&  International Symposium on Software Reliability Engineering (3) & ISSRE&\cite{bertero2017experience,chen2020logtransfer,pecchia2012detection} 
\\ 

\rownumber. &    C& International Conference on Automated Software Engineering (6) & ASE& \cite{li2020shall,chen2018automated,liu2019logzip,he2018characterizing,bao2019statistical,le2021log}  
 &  13.& C & ACM Symposium on Operating Systems Principles (3) & SOSP&\cite{zhao2017log20,xu2009detecting,zhang2017pensieve} \\

\rownumber. & J & Journal of Systems and Software (5)& JSS &\cite{bao2018execution,farshchi2018metric,mavridis2017performance,fronza2013failure,shang2014exploratory}
& 14. &  C & International Conference on Software Analysis, Evolution and Reengineering (2) & SANER &\cite{kabinna2018examining,jia2018smartlog} \\

\rownumber.& C & International Symposium on Reliable Distributed Systems (5) & SRDS&\cite{fu2012logmaster,zhang2020anomaly,gurumdimma2016crude,chuah2013linking,gholamian2021distributed} 
&15. &  C & International Conference on Architectural Support for Programming Languages and Operating Systems (2) &{\fontsize{6}{9}\selectfont ASPLOS}&\cite{yuan2010sherlog,yu2016cloudseer}  \\
    
\rownumber. &    C & IEEE/IFIP Conference on Dependable Systems and Networks (4) & DSN &\cite{xu2014pod,oliner2011online,oprea2015detection,cinque2010assessing} 
&16.&C&USENIX Annual Technical Conference (2) & ATC&\cite{lou2010mining,ding2015log2}  \\    

\rownumber. & C &  ACM Joint European Software Engineering Conference and Symposium on the Foundations of Software Engineering (4) &  ESEC/ FSE &\cite{he2018identifying,zhang2019robust,amar2018using,wang2021would}  
& 17.& C& ACM Symposium on Applied Computing (2) &ACM SAC&\cite{gholamian2020logging,marty2011cloud}\\

\rownumber. & C &  USENIX Symposium on Operating Systems Design and Implementation (4) & OSDI&\cite{zhao2014lprof,zhao2016non,yuan2014simple,yuan2012conservative} 
& 18   & J &  Software: Practice and Experience (2)& SP\&E&\cite{kim2020automatic,gujral2021exploratory} \\
   
    \bottomrule
\end{tabular}
\end{center}
\vspace*{-4mm}
\caption{\scriptsize List of related venues for our survey sorted from the most to least number of references. 
J stands for journal, and C stands for conference, symposium, or workshop.
The list of additional venues with only one publication include: arXiv~\cite{shin2020effective,he2020loghub}, CSUR~\cite{chen2021survey}, TOCS~\cite{yuan2012improving}, TPDS~\cite{di2018exploring}, IEEE Software~\cite{miranskyy2016operational}, TSMCA~\cite{li2009integrated}, TNSM~\cite{huang2020hitanomaly}, JCST~\cite{zou2016uilog}, HiPC~\cite{chuah2010diagnosing}, CSRD\cite{taerat2011baler}, IJCAI~\cite{meng2019loganomaly}, NSDI~\cite{nagaraj2012structured}, CIKM~\cite{tang2011logsig}, CCS~\cite{du2017deeplog}, SIGKDD~\cite{li2017flap}, IMC~\cite{qiu2010happened}, MASCOTS~\cite{awad2016performance}, WSE~\cite{tang2010approach}, IWQoS~\cite{zhou2020logsayer}, ICWS\cite{he2017drain}, ICPC~\cite{messaoudi2018search}, ICSEM~\cite{anu2019approach}, ICPE~\cite{yao2018log4perf}, ICC~\cite{zeng2015linux}, Middleware~\cite{xu2013detecting}, VLDB~\cite{lee2012unified}, CNSM~\cite{vaarandi2015logcluster}, APSYS~\cite{zhang2011autolog}, HICSS~\cite{salman2017designing}, and COMPSAC~\cite{lal2016logoptplus}. The total number of listed papers is \textbf{112}.
}
\label{pub_venues_breakdown}
\vspace{-5mm}
\end{table*}
}

\afterpage{
\begin{figure}[h]
  \centering
  \includegraphics[width=\linewidth]{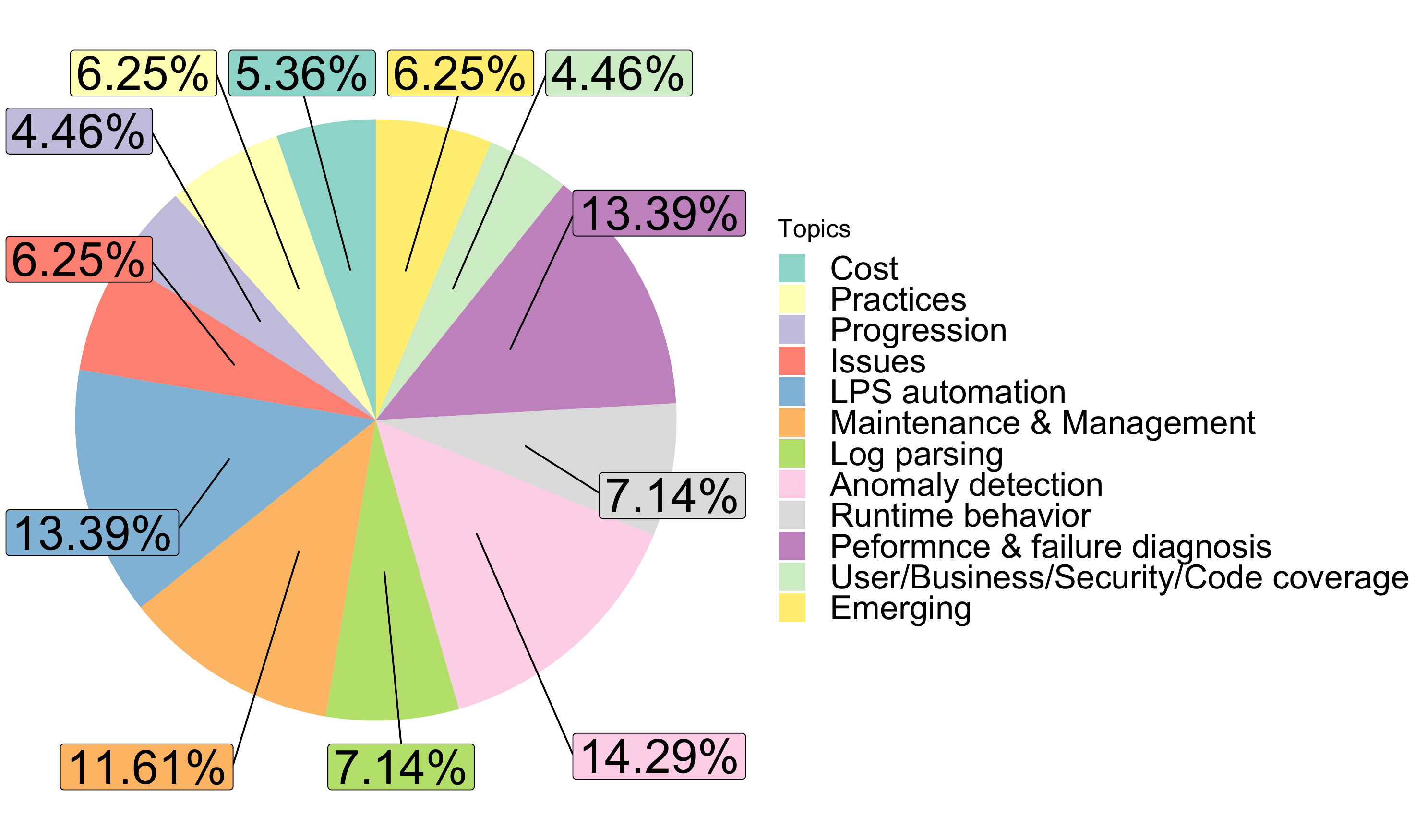}
  \caption{Percentage of publications in each topic.}
  \label{fig_pub_per_topic}
\end{figure}}

\afterpage{
\begin{figure}[h]
\vspace*{-25mm}
\hspace*{-10mm}
\centering 
\includegraphics[scale=.38]{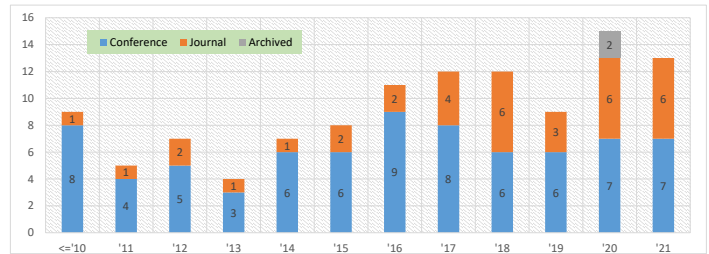}
\vspace*{-25mm}
\caption{Number of publications per year divided into conference and journal categories. Publications up to October 31, 2021, have been observed at the time of this survey.}
\label{paper_list}
\end{figure}
}

\subsection{Logging Challenges}\label{log_prac_chall}
Based on the knowledge gained throughout the survey, we summarize the challenges that the prior literature has aimed to tackle in the following.
Although current research advances have made logs more useful and effective, there are still multiple remaining challenges and avenues for future work and improvement. 
Categories of challenges remain in various aspects of \textbf{logging source code} and \textbf{log file analysis and mining}. 

\subsubsection{Logging Code}
As a result of a lack of well-accepted standards and guidelines for logging practices~\cite{fu2014developers,pecchia2015industry,chen2017characterizing}, currently, developers mostly rely on their personal experience or intuition to perform their logging decisions. 
However, for this manual process, \textit{i.e.}, developers inserting logging statements into the source code, to lead to effective logging practices, we are facing four main challenges:

\begin{enumerate}
\item The first challenge is \textbf{where-to-log}, which is the decision of selecting appropriate logging points. 
Logging statements can be placed in different locations of interest in the source code, such as inside \textit{try-catch} block, \textit{function return value}, \textit{etc}. 
Although log statements provide valuable insight into the running system's state, they are I/O intensive tasks and excessive logging can incur performance and maintenance overhead~\cite{ding2015log2,zhao2017log20,gholamian2021distributed}. 
Consequently, developers are often faced with the challenge of making an informed decision for \textit{where-to-log} in order to avoid introducing unjustified performance degradation and maintenance overhead. 

\item The second challenge, \textbf{what-to-log}, concerns with what information to include in the log message. 
As explained in Figures~\ref{log_example_c} and~\ref{log_example_lib}, the log statement description provides a brief context of the execution and the internal variables provide more insights into the dynamic context of the current execution state. 
Therefore, the logging description and logged variables should satisfy their purposes and be clear and informative about the current state of the program. 
The logging description should also stay up-to-date and in-sync with the feature code updates, as some developers fail to update the logging statements as feature code changes~\cite{kabinna2016logging,shang2014understanding}. 

\item The third challenge is \textbf{how-to-log}, which concerns with how the logging code, as a subsystem, combines with the rest of the software system. 
As the logging code is intertwined across different source code modules, some prior researchers have suggested modularizing the logging code, as an independent subsystem, which becomes compiled into the feature code in the later stages of the system release~\cite{kiczales1997j,urlaspectj}. 
Nevertheless, many industrial and open-source software projects still tend to mix the logging code with the feature code~\cite{pecchia2015industry,chen2017characterizing2,yuan2012characterizing,chen2017characterizing}. 
As a result, maintaining and developing high-quality logging code as the feature code evolves remains challenging and crucial to the overall quality of the software. 
\item The fourth challenge, and also the recent~\cite{mizouchi2019padla} and mostly overlooked in prior research~\cite{zhu2015learning}, discusses the question of \textbf{whether-to-log}, which concerns with dynamically adjusting the degree of logging in response to the runtime requirements. 
For example, if a suspected anomaly is detected, the logging platform can enable more detailed logging, and if the system is acting normal, it minimizes the number of logs to lessen the overhead.   
\end{enumerate}

\begin{figure}[h]
\begin{center}
\includegraphics[scale=.57]{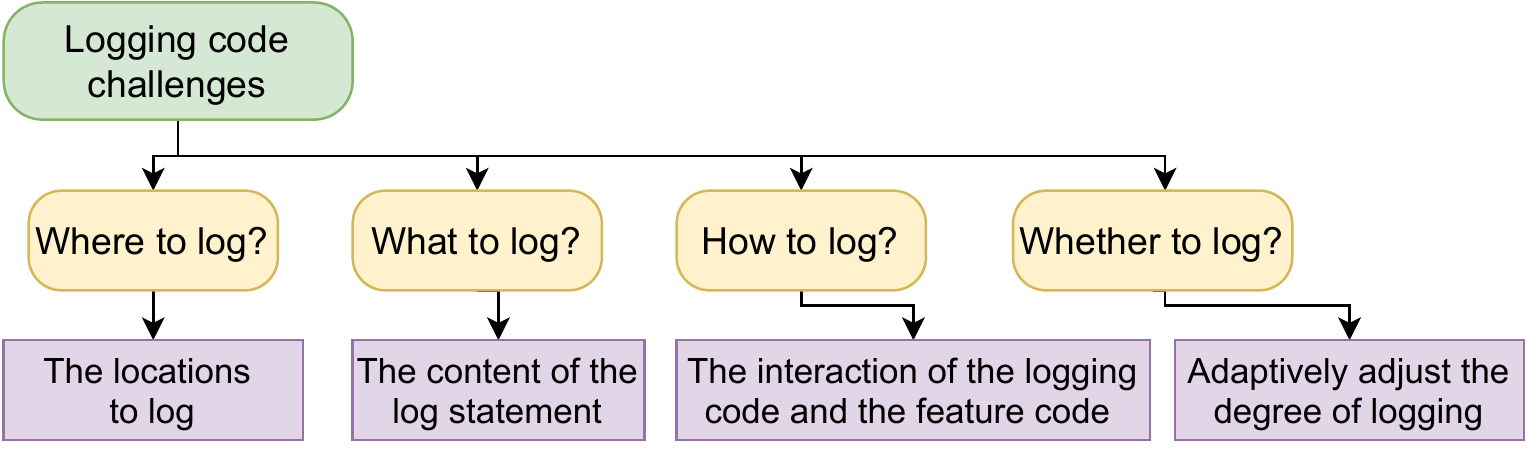}
\caption{Logging code challenges.}
\label{practice_challenges}
\end{center}
\vspace{-5mm}
\end{figure}

Figure~\ref{practice_challenges} summarizes the challenges regarding the logging code. 
In the following, we briefly review the related research concerning each challenge.
Later on, we will revisit these research efforts in more detail in their related section in the survey. 

\textbf{Where to log.} The research in this area is interested in finding appropriate logging points. 
One approach is to analyze the source code and look for specific types of code blocks, \textit{i.e.}, unlogged exception code blocks, and insert logging statements inside them~\cite{yuan2012conservative}. 
Other log placement objectives, such as disambiguating execution paths~\cite{zhao2017log20}, minimizing the I/O and performance overhead~\cite{ding2015log2}, and feature extraction and learning approaches~\cite{fu2014developers,zhu2015learning} also exist in the literature.   

\textbf{What to log.} The research in this area is concerned with what the content of logging statements should be to make the logs more effective for future system observation purposes, such as debugging and failure diagnosis. 
Possible approaches include \textit{automatically adding variables} that can clarify execution paths to the LPSs~\cite{yuan2012improving,liu2019variables}.  
Additionally, concerns on the content of logging statements arise when developers neglect to update the LPSs as the related feature code is updated~\cite{kabinna2016logging}, which is also a common problem in \textit{source code comments}~\cite{tan2007icomment}. 

\textbf{How to log.} This avenue is concerned with the problem of how to develop and maintain a high-quality logging code. 
Additionally, it pays attention to the interaction and integration of the logging code as a subsystem with the rest of the software system.  
Paradigms such as Aspect Oriented Programming (AOP)~\cite{kiczales1997j} aim to look at log statements as a submodule of the software which is separate from the feature code, and unifies with the rest of the software at a later time during the development process. 
However, the current logging practices in the industry and open-source projects commonly fuse the logging code inside the feature code. 
Logging libraries and utilities~\cite{chen2020studying} also take part in organizing the logs and improving their formatting and quality. 
As such, enhancement of logging libraries can positively impact on ensuring \textit{how we log}. 

\textbf{Whether to log.} One approach to tackle the challenge of the number of logs is to enable dynamic filtering of logs during runtime~\cite{ding2015log2,mizouchi2019padla}. 
Enablement of this paradigm would allow to take the pressure off the developers and enable more conservative addition of log statements in the source code, without being concerned about the overhead. 
Thus, depending on the program state, logs are dynamically discarded or collected if they serve any online or postmortem analysis purpose.   

In sum, the challenges associated with the logging code have initiated a variety of research to analyze and mine logging statements, their progression, and their issues. 
In addition, this has also triggered the development of tools and approaches to automatically predict and suggest logging statements and their associated details, such as log verbosity levels and variables, inside the source code. 


\subsubsection{Log File Analysis and Mining} 
The challenges involved in this category stem from the large scale of log files as the computer systems become larger and further distributed. 
In addition, because log files are unstructured (or semi-structured), due to the intermixing of both static (\textit{e.g.}, log description) and dynamic (\textit{e.g.}, variable values) runtime content, this further complicates the automated analysis of logs.

\textbf{Logging cost \& benefit.} As the size of computer systems increases, the voluminousness and heterogeneity of logs, which turns it into a big-data problem, imposes additional challenges. 
Collecting, processing, and storing of logs become more challenging as logging can infer additional computation, storage, and network overhead. This calls for further quantitative~\cite{gholamian2021distributed} and qualitative~\cite{li2020qualitative} cost analysis of logging.  

\textbf{Automated log analysis.} Due to the voluminousness and heterogeneity of generated logs, and in some cases, the need for real-time processing of logs, the development of efficient, scalable, and real-time log analysis tools becomes challenging~\cite{miranskyy2016operational}. 
These tools aim to achieve a variety of goals through logs such as log parsing, anomaly detection, runtime behavior monitoring, and failure detection. Moreover, other less-investigated software territories,  such as mobile systems and big-data applications, face additional or sometimes different log analysis challenges stemming from their different use cases and operating conditions, such as performance and power limitations.  

\textbf{Log maintenance and management.} As the scale of logs increases, their maintenance and management also become more challenging. 
These challenges revolve around efficient collection, organization, compression, and storage of logs. 
In addition, for various machine learning-based analyses  of logs (\textit{e.g.}, anomaly detection), labeled log collections are hard to find. This challenge may be addressed by proposing efficient and scalable automated (or semi-automated) labeling techniques that can infer the labels for large-scale data from a small sample of labels data~\cite{yang2021semi}.

 

\begin{tcolorbox}[breakable, enhanced]
\small \textbf{Finding} \textbf{2.} \textit{ The top five research topics for logging research are: \begin{enumerate*}[label=\protect\circled{\arabic*}] 
\item log mining for anomaly detection,
\item log printing statement automation, 
\item log mining for performance and failure diagnosis,
\item log maintenance and management, and 
\item log parsing.
\end{enumerate*}
}

\small \textbf{Finding} \textbf{3.} \textit{ The top five publications venues for logging research are: \begin{enumerate*}[label=\protect\circled{\arabic*}] 
\item Empirical Software Engineering (EMSE), 
\item International Conference on Software Engineering (ICSE),
\item Transactions on Software Engineering (TSE), 
\item Automated Software Engineering Conference (ASE), and
\item Journal of Systems and Software (JSS).
\end{enumerate*}}

\small \textbf{Finding} \textbf{4.} \textit{ Due to the challenges associated with logging code, \textit{i.e., what, where, whether, and how to log}, and log file analysis and mining challenges, \textit{i.e.}, logging cost, automated log analysis, and log maintenance and management, there exists a continuously growing interest in log-related research and the prior work is published in top-ranked venues in yearly basis (Figure~\ref{paper_list} and Table~\ref{pub_venues_breakdown}).}
\end{tcolorbox}
\vspace*{-3mm}
{\renewcommand{\arraystretch}{1}
\newcommand\rownumber{\stepcounter{magicrownumbers}\arabic{magicrownumbers}}
\rowcolors[]{2}{gray!15}{white}
\begin{table*}[h]
\fontsize{7.5}{11}\selectfont
\centering
\begin{tabular}{ p{1.5cm}|p{15cm}} 
\toprule
\rowcolor{blue!20} \textbf{Overhead} & \textbf{Details} \\
\midrule
Disk I/O bandwidth & Logging causes additional I/O bandwidth (BW) which may interfere with the required I/O BW for the system's core functionality. 
The BW requirement by enabling all logs (\textit{i.e., verbose} level vs. \textit{medium} level) can become significantly higher than the presumed BW. 
For example, in~\cite{ding2015log2}, the extra BW by enabling all logs is 8MB$/$s, which, however, should have been $\leq$200KB/s. \\

Storage & As the logging BW increases, OS might slow down, and other processes that require disk space and BW may crash, and even the logging subsystem could crash. 
Additionally, more logging requires more storage space. 
\\

CPU & As the CPU usage of the logging subsystem is increased, service to other processes is slowed down.  
Once the CPU usage of logging goes up to double digits, the slowdown of the other processes becomes significantly noticeable.
Ding \textit{et al.}~\cite{ding2015log2} recommend an overhead of 3-5\% as the CPU usage upper bound for logging. \\

Memory & Developers noticed unexpected increases in memory usage of the logging subsystem, which was the root cause of one service incident.
Additionally, memory leakage of the logging system caused days of effort in debugging. \\
\bottomrule
\end{tabular}
\vspace{-2mm}
\caption{System's performance overhead associated with logging.}
\label{log_cost_table}
\end{table*}
}

\afterpage{\renewcommand{\arraystretch}{1.2}
\newcommand\rownumber{\stepcounter{magicrownumbers}\arabic{magicrownumbers}}
\rowcolors[]{2}{gray!15}{white}
\begin{table*}
\scriptsize	
\begin{center}
 \begin{tabular}{p{4cm} p{4cm} |P{.7cm} p{3.5cm} p{3.5cm}} 
 \toprule
 
\rowcolor{blue!20} \textbf{Reference - Approach} & \textbf{Results} & \textbf{Type} & \textbf{Pro}& \textbf{Con} \\ [0.5ex] 
 \midrule

Yuan \textit{et al.}~\cite{yuan2012conservative} - Conservatively adds log statements to the source code while aiming to minimize the introduced execution overhead.
&This study categorized seven generic patterns of error sites based on the study on 250 failures, such as exceptions, function return errors, \textit{etc}.
& Quan./ Qual.&  Errlog provides three different levels of configurable logging overhead. & Focuses on Error Logging Statements (ELS).\\

Zeng \textit{et al.}~\cite{zeng2015linux} - Measures the overhead of Linux security auditing through log messages.
&The authors measured up to 5\% performance overhead when the audit logging is enabled.
&Quan.
& Proposes an adaptive approach to reduce the overall system overhead from 5\% to 1.5\%
& Reduced auditing might result in a lower level of security protection for the system.\\

Ding \textit{et al.}~\cite{ding2015log2} - Surveys engineers in Microsoft and applies a constraint solving-based method to calculate an optimal logging placement.
&Maximizes extracted runtime information and, concurrently, minimizes the I/O and performance overhead.
& Quan./ Qual. 
&Two levels of filtering, \textit{i.e.}, local and global filters, to discard less-informative logging messages and simultaneously keep important messages.
& Curtailed to performance analysis of logs, and falls short for logs recording error and failure information.\\

Li \textit{et al.}~\cite{li2020qualitative} - Studies developers' logging considerations when it comes to the costs and benefits associated with logging.
& Main benefits of logging communicated by developers include: \textit{diagnosing runtime failures}, \textit{using logs as a debugger}, \textit{user/customer support} and \textit{system comprehension}.
&Qual.
& Survey of 66 developers and a case study of 223 logging-related issue reports from the issue tracking systems.
&Limits to open-source projects and closed-source projects might evaluate differently on their logging costs and benefits.\\

Yao \textit{et al.}~\cite{yao2020log4perf} - Introduces a statistical approach to map logging statements to the performance of the system, \textit{i.e.}, CPU usage.
& If the performance model's prediction error is noticeably impacted, it implies that the modified log helps to model the CPU usage properly. 
& Quan.
& The approach finds and suggests removing insignificant log statements.
& Logging statements that are not covered by the performance tests cannot be identified by this approach.\\

Gholamian and Ward~\cite{gholamian2021distributed} - Performs experimental analysis on logging cost and information gain for different log verbosity levels in a distributed environment. 
& It presents nine findings for logging cost in different scenarios with and without system failures. The research observes 8.01\% and $\sim$268X overhead in the execution time and storage when the trace log level is enabled versus the info log level.
& Quan.
& The research measures the impact of different types of failures in a distributed environment on the generated logs and performs a case study for anomaly detection from OpenStack logs with entropy values. 
&The evaluation is performed on a small cluster of four nodes. The findings should be confirmed on a large-scale distributed cluster.\\

\bottomrule

\end{tabular}
\caption{Logging cost and benefit research - Topic (A). {`Type'} shows \textit{qualitative}, \textit{quantitative}, or both.}
\label{log_cost_benefit_table}
\end{center}
\end{table*}
}

\section{RQ3: How the research in each topic can be systematically compared with their approaches, pros and cons?}\label{rq3_breakdown}
In this section, we review the available literature in each category of logging research and provide a comparative analysis. 

\subsection{Category A: Logging Cost and Benefit Analysis}\label{log_cost}
Although logs are useful and provide insight into the internal state of the running software, they also impose inherent costs on different subsystems of a computer system. 
We can assess the costs and benefits of logging both \textit{quantitatively} and \textit{qualitatively}, which we review in the following. 

\subsubsection{Quantitative assessment}
Quantitative assessment for benefits of logging measures to what extent logging improves a specific debugging task. 
For example, Yuan et al.~\cite{yuan2012conservative} observed the benefits of improved logs, as they contributed to $\sim$60\% faster diagnosis time when compared with the original logging statements, \textit{i.e.}, prior to the enhancement. 
Log associated overheads can be also evaluated quantitatively~\cite{ding2015log2,zeng2015linux}. 
Table~\ref{log_cost_table} summaries \textbf{logging cost breakdown} on various subsystems of a computing system~\cite{ding2015log2}, including I/O, storage, CPU, and memory. 
For example, one approach~\cite{zhao2017log20} to simplify and \textbf{measure the slowdown} caused by logging statements is to calculate the number of times ($n$) each log statement is being executed and multiply that with the overhead of a single log statement execution ($l$), \textit{i.e.}, $n \times l$. 
Other research efforts in this area have measured the \textbf{overhead of Linux security auditing through log messages} by enabling and disabling audit logging~\cite{zeng2015linux}, and statistically \textbf{mapped logging statements to the performance of the system}, \textit{i.e.}, CPU usage~\cite{yao2018log4perf,yao2020log4perf}. 
Gholamian and Ward~\cite{gholamian2021distributed} evaluated the computation and storage cost associated with logging in a distributed environment. 
In addition, they associated the logging in different verbosity levels with the amount of information gained while synthesizing various failure scenarios.

\subsubsection{Qualitative assessment}
In contrast to quantitative metrics for measuring logging overhead, \textit{e.g.}, system slowdown or I/O cost, qualitative approaches aim to understand the underlying trade-offs from developers' perspectives through surveys or questionnaires. 
A developer survey at Microsoft~\cite{ding2015log2} uncovered main overheads associated with logging, from developers' perspective, as listed as \textit{``Details''} in Table~\ref{log_cost_table}. 
Developers were also inquired about the methods they use for containing the logging overhead for large-scale online service systems.  
They mentioned a variety of methods to limit the logging overhead such as \textbf{adjusting the logging verbosity level} (93\% of developers have applied this approach), \textbf{manual removal of unnecessary logs} (64\%), and \textbf{periodic archiving of log files} to save disk space (43\%). 
Additionally, this study observed the lack of a cost-awareness guideline during log instrumentation. 
Some developers often had little idea about the logging overhead when they planned to add new logging statements to the source code. 
Thus, developers require to be more mindful in adding logging statements in scenarios such as \textit{for-loops}, which iterate a large number of times and could cause high overhead, especially on CPU, I/O, and storage throughput. 
A recent study~\cite{li2020qualitative}, \textbf{qualitatively} examined {logging cost and benefits from developers' perspectives}. 
One qualitative measure of the logging cost, \textit{i.e.}, too much logging, causes \textbf{noisy log files} which hinders program comprehension and results in strenuous log file analysis.   
In contrast to costs associated with logging, main qualitative benefits of logging communicated by developers include \textbf{the capability to diagnose runtime failures with logs}, \textbf{using logs as a debugger}, \textbf{user/customer support}, and \textbf{system comprehension}. 
Table~\ref{log_cost_benefit_table} summarizes the research in Category A. 
\begin{tcolorbox}[breakable, enhanced]
\small \textbf{Finding} \textbf{5.} \textit{
In sum, although logs provide insight into the internal state of the running software, they also impose inherent costs on different subsystems of a computer system. Developers should pay close attention to both quantitative and qualitative costs and benefits of logging while making logging decisions.}
\end{tcolorbox}

\subsection{Mining Log Printing Statements}\label{log_mining}
There has been a significant body of research aiming to mine, understand, and characterize various source code logging practices~\cite{fu2014developers,yuan2012characterizing,chen2017characterizing2,chen2017characterizing}. 
Because, intuitively, understanding previously applied logging practices is the gateway to help developers improve their current logging habits. 
Thus, to derive the in-the-field LPS practices, the first step is to mine the source code's logging statements and extract useful insight and observable patterns. 
Consequently, there are two broad classes of prior studies that have sought after understanding and mining of the logging practices in both industry and open-source projects: 1) \textbf{mining logging code}, and 2) \textbf{mining log files}. 
We review the research for mining logging code in this section and mining of log files in Section~\ref{mine_log_files}.

\begin{figure}[h]
\centering
\includegraphics[scale=.52]{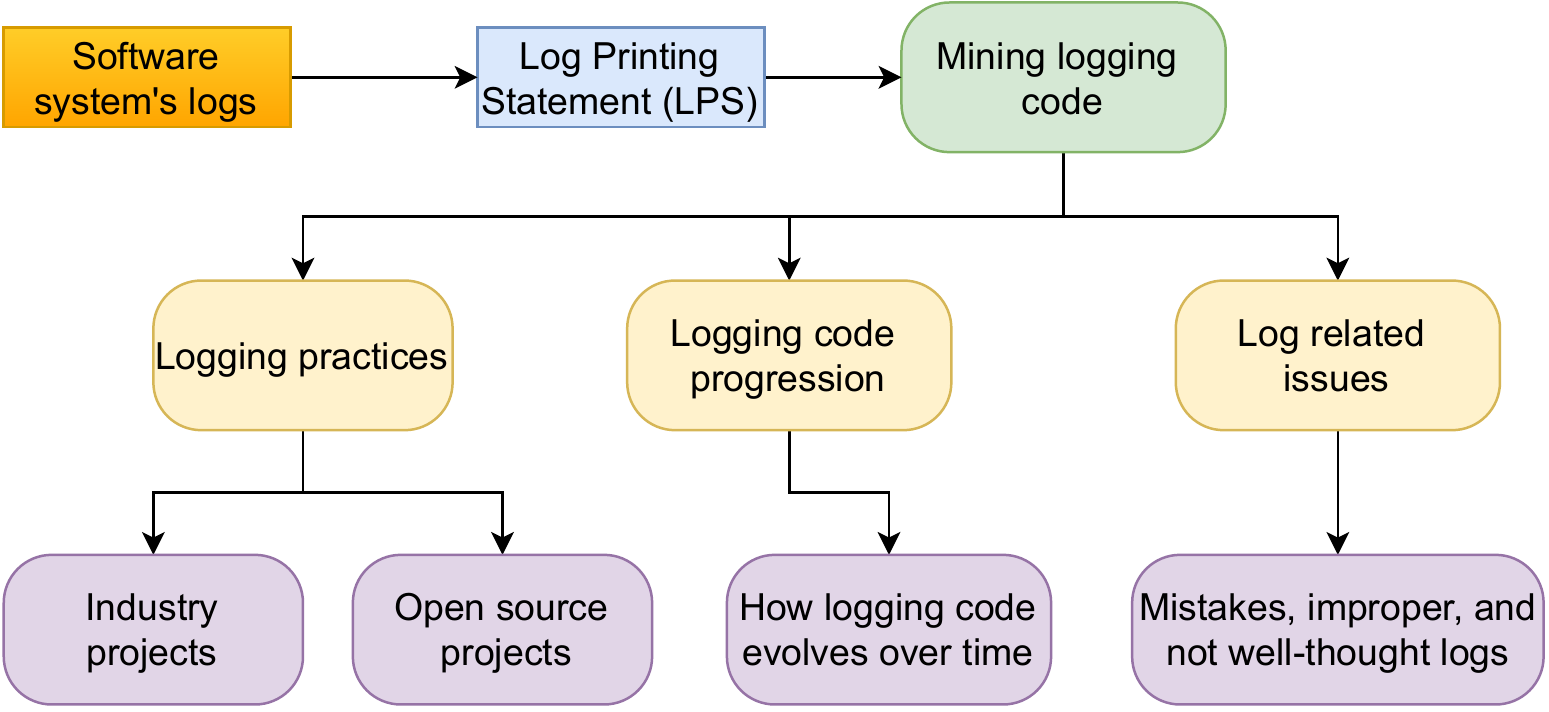}
\caption{Logging code mining research in subcategories.}
\label{code_mining}
\vspace{-2mm}
\end{figure}

Mining log printing statements (LPSs) in the source code principally focuses on understanding how developers insert LPSs into the source code and how logging code evolves over time to gain insight into the common logging practices. 
Figure~\ref{code_mining} shows the categorization of research for logging code mining of software projects which is divided to research on \textbf{logging practices (Category B)}, \textbf{logging code progression (Category C)}, and \textbf{logging-code issues (Category D)}.
In the tables that follow, our convention is that \textit{pro} signifies an \textit{advantage}, or an \textit{improvement}, and \textit{con} signifies a \textit{limitation}, \textit{room for improvement}, or an \textit{avenue for future work}.  
This section includes research in Categories \textit{B, C, and D,} that we review in detail.

\afterpage{\renewcommand{\arraystretch}{1.2}
\newcommand\rownumber{\stepcounter{magicrownumbers}\arabic{magicrownumbers}}
\rowcolors[]{2}{gray!15}{white}
\begin{table}
\scriptsize	
\begin{center}
 \begin{tabular}{p{2cm} P{.5cm} P{.5cm} P{1cm} P{1cm}  P{1cm}} 
 \toprule
 
\rowcolor{blue!20} \textbf{Reference} &\textbf{LSC} &\textbf{LLU} & \textbf{Lang.} & \textbf{Source} & \textbf{\# Proj.}\\ [0.5ex] 
 \midrule

Yuan \textit{et al.}~\cite{yuan2012characterizing}& \checkmark & & C/C++ &OS& 4\\

Fu \textit{et al.}~\cite{fu2014developers}& \checkmark & & C\# & CS&2\\

Pecchia \textit{et al.}~\cite{pecchia2015industry}& \checkmark & & C/C++& CS&3\\

Shang \textit{et al.}~\cite{shang2015studying}& \checkmark
& & Java& OS & 2\\

Chen and Jiang~\cite{chen2017characterizing2}& \checkmark 
&& Java& OS& 21\\

Zhi \textit{et al.}~\cite{zhi2019exploratory}& \checkmark 
&\checkmark & Java & OS/CS & 20\\ 

Chen \textit{et al.}~\cite{chen2020studying}& &\checkmark & 
Java& OS& $\sim$11,000\\

\bottomrule
\end{tabular}
\end{center}
\vspace{-2mm}
\caption{\footnotesize Comparison of logging practices research - Topic (B). LCS: Logging Source Code; LLU: Logging Libraries and Utilities; Lang: Programming Language of the project; OS: Open Source; CS: Closed Source; \# Proj.: Number of Projects reviewed.}
\label{logging_practice_table_OW}
\end{table}
}

\subsubsection{Category B: Logging Practices}
Mining logging practices aims to gain insight into the current logging habits of developers both in open-source and industrial proprietary software projects. 

\textbf{Open-source projects.} 
Prior work in this category includes empirical studies to characterize current logging practices in open-source projects such as Apache Software Foundation (ASF)~\cite{urlapache} projects~\cite{yuan2012characterizing,chen2017characterizing2}. 
Other works~\cite{chen2017characterizing2,shang2015studying} aimed to find recurrent mistakes in the logging code and its relationship to overall source code quality. 
Another tread of research~\cite{chen2020studying,zhi2019exploratory} has examined logging configurations, libraries, and utilities.

\textbf{Industry projects.} Similar to the open-source software, logging is a widely adopted practice in industry software projects. 
Fu et al.~\cite{fu2014developers} conducted a study on logging practices of two software systems at Microsoft, and Pecchia et al.~\cite{pecchia2015industry} examined application-critical software logging practices at Selex ES.

Table~\ref{logging_practice_table_OW} provides a high-level comparison of primary studies concerning logging practices. 
In sum, prior research has considered logging practices in logging source code (LSC) and logging libraries and utilities (LLU), for different programming languages (Lang.) and for different number of projects (\# Proj.) in both open-source (OS) and closed-source (CS) software. 
Table~\ref{logging_practice_table} provides additional details for each study and compares the research on mining of the logging source code for open-source and industrial projects.

\afterpage{\renewcommand{\arraystretch}{1.2}
\newcommand\rownumber{\stepcounter{magicrownumbers}\arabic{magicrownumbers}}
\rowcolors[]{2}{gray!15}{white}
\begin{table*}
\scriptsize	
\begin{center}
 \begin{tabular}{p{3cm} p{3cm} p{4cm} p{3cm} p{3cm}} 
 \toprule
 
\rowcolor{blue!20} \textbf{Reference - Aim} &  \textbf{Experiments}& \textbf{Results} & \textbf{Pro}& \textbf{Con} \\ [0.5ex] 
 \midrule

Yuan \textit{et al.}~\cite{yuan2012characterizing} - Study and characterize logging practices in four open-source C/C++ based projects.
&Four software projects: \textit{Apache httpd}, \textit{OpenSSH}, \textit{PostgreSQL}, and \textit{Squid}. 
& Observes ten findings and their implications that software logging is pervasive and developers spend significant time maintaining logging code.
& Provide a simple checker to detect verbosity level inconsistencies.
& A follow-up study observed contradictory findings in some cases~\cite{chen2017characterizing2}.\\

Fu \textit{et al.}~\cite{fu2014developers} - Conducts source code analysis on two software systems at Microsoft, to categorize logged and unlogged snippets.
&A questionnaire and a decision-tree classifier to detect whether a code snippet requires a logging statement.&The research uncovers five categories of logged code snippets, including \textit{return-value-check} and \textit{exception-catch} snippets.
& Extracts contextual features and proposes a decision-tree classifier, which can detect whether a code snippet requires a logging statement.
&Logging categories can be broken down further into subcategories.\\

Pecchia \textit{et al.}~\cite{pecchia2015industry} - Studies the logging practices on a critical industrial software at Selex ES.
& Experimented with the software at Selex ES in three product lines, \textit{i.e}., middleware (MW), business logic (BL), and human-machine interface (HMI).
&The study uncovers three main reasons for logging in the industrial domain: state dump, execution tracing, and event reporting.
&Observed logging is highly developer-dependent, and company-wide \textbf{log policies} and \textbf{guidelines} are needed.
&The study is limited to a very particular closed-source software system, and the findings might not generalize to software in other application domains.\\

Shang \textit{et al.}~\cite{shang2015studying} - Explores the relationship between logging characteristics and the code quality.
& A case study on four releases of Hadoop and JBoss projects.
&Logging characteristics provide a strong indicator of post-release defects, \textit{i.e.}, files with more logging statements have a higher rate of post-release defect compared to the files without logging.
& Developers' code improvement efforts should focus more on the source code files with high logging density or high rate of log churn.
& The study cannot establish a causal relationship, \textit{i.e.}, there might be a large portion of defects not captured due to not being logged extensively.\\

Chen and Jiang~\cite{chen2017characterizing2} - A replication work
of Yuan \textit{et al.}'s work~\cite{yuan2012characterizing} on 21 Java projects.
&21 open-source Java projects in three different domains: \textit{server}, \textit{client}, and \textit{supporting components}.
& Similar findings as~\cite{chen2017characterizing2} regarding logging pervasiveness and that developers' significant amount of time
spent on maintaining the logging statements.
& A high portion of code updates are for improving the quality of logs and contrary to~\cite{chen2017characterizing2}, this research finds developers spend more time fixing reported failures when log messages are present.
&Contradictory findings compared to the prior work~\cite{chen2017characterizing2} raises the concern of how useful the findings are, and if logging practices are project, programming language, and domain dependent.\\

Zhi \textit{et al.}~\cite{zhi2019exploratory} - Conducts an exploratory study on the logging configuration practices and how they evolve over time. 
& 10 open-source and 10 industrial java projects in various domains and sizes.
& The research's main findings show that current practices of logging configurations concerns with logging management, logging storage, logging formatting, and logging-configuration quality.
& Provides a simpler checker to statically analyze and detect log configuration issues. The authors found some issues on open-source projects by applying the checker.
& Further research to improve the quality of logging configurations is required to detect and resolve logging configuration smells.\\

Chen \textit{et al.}~\cite{chen2020studying} - Studies logging utilities (LUs) usage
in Java project.
& Over 11,000 projects and 3,850 Java LUs (e.g., SLF4J~\cite{urlslf4j}) from GitHub.
& With a heuristic-based technique, the study observed a positive correlation between the size of the project and the complexity of LUs.
&Some projects still use multiple LUs to bring in more flexibility, and, additionally, support and enable logging behavior of the imported packages.
& Currently, configuring different LUs is a manual and error-prone task. Thus, error-free and automatic checkers and techniques to configure LUs are required.\\
\bottomrule
\end{tabular}
\end{center}
\vspace{-2mm}
\caption{Logging practices research - Topic (B).}
\label{logging_practice_table}
\end{table*}
}

\begin{tcolorbox}[breakable, enhanced]
\small \textbf{Finding} \textbf{6.} \textit{
In sum, logging is a \textbf{pervasive convention} in various software domains (\textit{e.g.}, server, client, and support applications) and developers utilize \textbf{various logging practices} and spend a \textbf{significant amount of time} updating logging statements.}
\end{tcolorbox}

\afterpage{\renewcommand{\arraystretch}{1.2}
\newcommand\rownumber{\stepcounter{magicrownumbers}\arabic{magicrownumbers}}
\rowcolors[]{2}{gray!15}{white}
\begin{table}[h]
\scriptsize	
\begin{center}
 \begin{tabular}{p{1.5cm} p{2cm} P{1cm} P{1cm}  P{1cm}} 
 \toprule
 
\rowcolor{blue!20} \textbf{Reference} &\textbf{Category} & \textbf{Lang.} & \textbf{Source} & \textbf{\# Proj.}\\ [0.5ex] 
 \midrule

Shang \textit{et al.}~\cite{shang2014exploratory,shang2011exploratory}& log addition, deletion, modification & C/Java &OS/CS& 3
\\

Kabinna \textit{et al.}~\cite{kabinna2016logging}& logging library migration& Java &OS& 223
\\

Kabinna \textit{et al.}~\cite{kabinna2018examining,kabinna2016examining}& log context and owner & Java &OS& 4
\\

Li \textit{et al.}~\cite{li2018logtracker,li2019guiding}& log revisions & C/C++ &OS& 12\\

Rong \textit{et al.}~\cite{rong2020can}&
logging intentions and concerns & Java &CS& 3\\

\bottomrule
\end{tabular}
\end{center}
\vspace{-2mm}
\caption{\footnotesize Comparison of logging code progression research - Topic (C). Category: the category of logging progression observed in the research.}
\label{logging_code_progression_table_OW}
\end{table}
}

\subsubsection{Category C: Logging Code Progression}
Thus far, we have discussed the research investigating logging practices in both open-source and industry projects. 
Prior research has also studied the progression (\textit{i.e.}, evolution) of the logging code in software projects. 
Progression means that how logging code changes over time. 
Prior studies have concluded that logging code \textbf{evolves significantly} (\textit{i.e.}, high churn rate), even at a higher rate than the \textit{feature code} over the lifespan of the software development~\cite{yuan2012characterizing,shang2011exploratory,shang2014exploratory,zhao2017log20,kabinna2016examining,kabinna2018examining}. 
Additionally, many projects go through logging library migrations throughout their lifetime~\cite{kabinna2016logging}, and research has proposed tools to predict likely logging code revisions, \textit{e.g.}, LogTracker~\cite{li2018logtracker,li2019guiding}. 
Table~\ref{logging_code_progression_table_OW} provides a high-level summary of logging code progression research, and Table~\ref{logging_prograssion_table} provides additional details and compares the research on the progression of the logging code. 

\begin{tcolorbox}[breakable, enhanced]
\small \textbf{Finding} \textbf{7.} \textit{
In sum, prior studies have investigated the evolution of the logging code and libraries from different angles as the feature code evolves. These angles include \begin{enumerate*}[label=\protect\circled{\arabic*}] 
\item addition, deletion, and modification of logs, 
\item logging library migrations,
\item impact of log context and owner on log changes,
\item log revision analysis, and
\item intentions and concerns in log evolution.
\end{enumerate*}}
\end{tcolorbox}

\afterpage{\renewcommand{\arraystretch}{1.2}
\newcommand\rownumber{\stepcounter{magicrownumbers}\arabic{magicrownumbers}}
\rowcolors[]{2}{gray!15}{white}
\begin{table*}
\scriptsize	
\begin{center}
 \begin{tabular}{p{3cm} p{3cm} p{4cm} p{3cm} p{3cm}} 
 \toprule
 
\rowcolor{blue!20} \textbf{Reference - Aim} &  \textbf{Experiments}& \textbf{Results} & \textbf{Pro}& \textbf{Con} \\ [0.5ex] 
 \midrule
 
Shang \textit{et al.}~\cite{shang2014exploratory,shang2011exploratory} - Explores the progression of logging code in execution (\textit{i.e.}, log files) and source code levels. 
& Two open-source \textit{(Hadoop and PostgreSQL)} and one industrial \textit{(EA)} software projects.
&The logging code changes at a high rate across versions, which might break the functionality of log processing applications (LPA).
& Developers could avoid the majority of the logging code modifications through better logging designs.
& The broad range of the avoidable logging code changes raises the concern of if the observed values are software system dependent.\\

Kabinna \textit{et al.}~\cite{kabinna2016logging} - Studies the logging library migrations in Apache Software Foundation (ASF) projects.
& Studies 223 ASF projects with their issue tracking systems in JIRA.
& As more flexible logging libraries with additional features emerge, many ASF projects have undergone logging library migrations or upgrades.
&Although adding more flexibility and performance improvement are cited as the primary drivers for logging library migrations, performance after library migration is rarely improved.
&A questionnaire survey from developers involved in logging migration efforts can bring additional value and more insight into their rationale behind the logging updates and best practices.\\

Kabinna \textit{et al.}~\cite{kabinna2018examining,kabinna2016examining} - Investigates the stability of logging statements over time, \textit{i.e.}, whether a logging statement will go under revisions in the future. 
& Four open-source projects: \textit{Liferay}, \textit{ActiveMQ}, \textit{Camel} and \textit{CloudStack}.
& A significant portion of logging statements change throughout their lifetime, and factors such as file ownership can affect the stability of logging statements.
& Developers of LPAs should rely on more stable logging statements for designing their tools.
&The research considers only the first change of logging statements. However, the already changed logging statements might become more stable after going through modifications and prior fixes.\\

Li \textit{et al.}~\cite{li2018logtracker,li2019guiding} - Studies the co-evolution process of logging statements as bug fixes and feature code updates are committed.
&12 open-source projects in C/C++ language from various domains, including \textit{Httpd}, \textit{Rsync}, \textit{Collectd}, \textit{Postfix}, and \textit{Git}.
&Proposes LogTracker, a tool that proactively predicts log revisions by correlating the rules learned from historical log revisions, \textit{e.g.}, the logging context, and the feature code.
&Utilizes code clones to learn log revision rules with the insight that semantically similar codes will likely require similar logging revisions.
&The tool can only guide log revisions for codes that share similar logging context, and the percentage of these revisions is not substantial.\\

Rong \textit{et al.}~\cite{rong2020can} - Investigates the status
of developers' \textit{intention} and \textit{concerns} (I\&C) on logging practices. 
& Developers' interviews and code analysis on three industrial software projects. 
& Major gaps and inconsistencies exist between the developers'
I\&C and real log statements in the source code. 
&For reasons such as \textit{lack of supporting facilities} and \textit{the version evolution of source code}, the developers' I\&C are poorly reflected in the log statements.
&Only missing log statements are considered as inconsistencies, and unnecessary log statements are not evaluated.\\
\bottomrule
\end{tabular}
\end{center}
\vspace{-4mm}
\caption{Logging code progression research - Topic (C).}
\label{logging_prograssion_table}
\vspace{1mm}
\end{table*}
}

\afterpage{\renewcommand{\arraystretch}{1.2}
\newcommand\rownumber{\stepcounter{magicrownumbers}\arabic{magicrownumbers}}
\rowcolors[]{2}{gray!15}{white}
\begin{table}[h]
\scriptsize	
\begin{center}
 \begin{tabular}{p{1.5cm} p{2cm} P{1cm} P{.7cm}  P{1cm}} 
 \toprule
 
\rowcolor{blue!20} \textbf{Reference} &\textbf{Category} & \textbf{Lang.} & \textbf{Source} & \textbf{\# Proj.}\\ [0.5ex] 
 \midrule

Yuan \textit{et al.}~\cite{yuan2014simple}&
 noisy logs, \textit{Aspirator} checker & C/Java &OS& 5\\

Shang \textit{et al.}~\cite{shang2014understanding}&
 JIRA tickets, log intentions & Java &OS& 3\\
 
Chen and Jiang~\cite{chen2017characterizing}&
 anti-patterns, \textit{LCAnalyzer} & Java &OS& 3\\

Hassani \textit{et al.}~\cite{hassani2018studying}&
 inappropriate and missing log statements & Java &OS& 2\\

Li \textit{et al.}~\cite{li2021studying,li2019dlfinder}&
 repetitive logging statements, \textit{DLFinder} & Java &OS& 5\\
 
Gujral \textit{et al.}~\cite{gujral2021exploratory}& logging questions semantics &  Q\&A websites &OS& 6~\cite{url_gujral2021exploratory}\\

Chen \textit{et al.}~\cite{chen2021demystifying}& challenges and benefits of analyzing user logs &  bug reports &OS& 10~\cite{url_chen2021demystifying}\\

\bottomrule
\end{tabular}
\end{center}
\vspace{-2mm}
\caption{\footnotesize Comparison of log-related issues research - Topic (D). We provide a link under `\# Proj.' if the data or implementation is available.}
\label{logging_issues_table_OW}
\end{table}
}

\subsubsection{Category D: Log-related Issues}\label{log_issue}
The extensive usage of logs comes with mistakes, improper, and not well-thought logging practices, which results in logging issues and low-quality logging statements. 
Some of the research in this thread overlaps with \textit{logging practices} and \textit{logging code progression}, as some of the logging issues are uncovered during the examination of logging practices and their evolution. 
Yaun \textit{et al}.~\cite{yuan2014simple} presented a characteristic study on real-world failures in distributed systems, and observed that the majority of failures print explicit failure-related log messages which can be used to replay (\textit{i.e.}, recreate) the failures. 
However, recorded log messages are noisy, which makes the analysis of logs tedious.
Several efforts have aimed to identify and reduce log-related issues, such as finding recurrent \textbf{logging mistakes}, \textit{i.e.}, anti-patters~\cite{chen2017characterizing}, \textbf{adjusting verbosity levels} (sometimes back-and-forth), \textbf{adding missing variables}, \textbf{modifying static text} to fix inconsistencies~\cite{yuan2012characterizing}, and finding \textbf{logging code smells}~\cite{li2019dlfinder,li2021studying}, \textit{i.e.}, duplicates. 
Hassani \textit{et al}.~\cite{hassani2018studying} empirically categorized log-related issues in open-source projects, among them \textbf{inappropriate log messages} and \textbf{missing logging statement itself} (also~\cite{zhao2017log20}) in locations that have to be logged.
The detection of logging code issues would be helpful, as developers can add revisions to logging statements, and hence improve the quality of log statements. 
As such, in addition to tools that automatically detect log-related issues such as DLFinder~\cite{li2019dlfinder} and LCAnalyzer~\cite{chen2017characterizing}, future research will benefit from developing tools that can automatically \textit{fix} log-related issues. 

Moreover, due to the lack of proper communication with developers of large-scale software systems, practitioners, who review logs for software maintenance tasks, might encounter challenges in understanding the logging messages. 
Such challenges may hamper the effectiveness and correctness of leveraging logs. 
Therefore, utilizing development knowledge~\cite{shang2014understanding}, in particular issue reports for log statements, \textit{e.g.}, JIRA tickets~\cite{urljira}, can help practitioners to better understand log messages. 
Shang \textit{et al.}~\cite{shang2014understanding} identified five categories of information that practitioners often look for to understand in log messages: \textbf{meaning}, \textbf{cause}, \textbf{context}, \textbf{impact} of the log message, and the \textbf{solution} for the log message. 
The key takeaway is that leveraging development knowledge, such as issue reports and code commit information, helps in clarifying the log messages.
Tables~\ref{logging_issues_table_OW} and~\ref{logging_issues_table} summarize and compare the research in Category D. 
From Table~\ref{logging_issues_table_OW} becomes obvious that future research that can examine log-related issues in proprietary software is of high value as it is missing at the moment.  

\begin{tcolorbox}[breakable, enhanced]
\small \textbf{Finding} \textbf{8.} \textit{
In sum, prior studies have investigated the log-related issues form different perspectives with providing some automated tools to detect log issues: \begin{enumerate*}[label=\protect\circled{\arabic*}] 
\item noisy logs interfere with failure diagnosis,
\item development knowledge (JIRA tickets) can help with log intention discovery, and
\item anti-patterns, duplicates, and missing LPSs are among log-related issues.
\end{enumerate*}}
\end{tcolorbox}

\afterpage{\renewcommand{\arraystretch}{1.2}
\newcommand\rownumber{\stepcounter{magicrownumbers}\arabic{magicrownumbers}}
\rowcolors[]{2}{gray!15}{white}
\begin{table*}
\scriptsize	
\begin{center}
 \begin{tabular}{p{3cm} p{3cm} p{4cm} p{3cm} p{3cm}} 
 \toprule
 
\rowcolor{blue!20} \textbf{Reference - Aim} &  \textbf{Experiments}& \textbf{Results} & \textbf{Pro}& \textbf{Con} \\ [0.5ex] 
 \midrule
Yuan \textit{et al.}~\cite{yuan2014simple} - Presents a characteristic study on real-world failures in distributed systems to understand how faults evolve to user-visible failures.
&198 user-reported failures that occurred on \textit{Cassandra}, \textit{HBase}, \textit{HDFS}, \textit{Hadoop MapReduce}, and \textit{Redis}.
&The majority of failures print explicit log messages which can be used to replay (\textit{i.e.}, recreate) the failures. However, the recorded log messages are noisy, which makes the analysis of logs tedious. 
& Provides a simple rule-based static checker, \textit{Aspirator}, to detect the location of the code bug patterns, including log-related issues. 
&The study limited to a set of data-intensive systems in their production quality, \textit{i.e.}, not during the development phase.\\

Shang \textit{et al.}~\cite{shang2014understanding} - Utilizes development knowledge [118], \textit{e.g.}, JIRA tickets~\cite{urljira} to understand the intention of log statements. 
&300 randomly sampled logging statements, and manually examining the email threads in the mailing list for three open-source systems: \textit{Hadoop}, \textit{Cassandra}, and \textit{Zookeeper}.
& Identifies five categories of information that practitioners often look for to understand in log messages: \textbf{meaning}, \textbf{cause}, \textbf{context}, \textbf{impact} of the log message, and the \textbf{solution} for the log message.
& The approach can be used to identify the experts for a particular log line and seek their help.
&Development knowledge is considered for log lines at the method level. The higher the level, the more development knowledge that can be attached, but the more overwhelming such attached knowledge might become.\\

Chen and Jiang~\cite{chen2017characterizing} - Characterizes anti-patterns (AP) (\textit{i.e.}, recurrent mistakes) in the logging source code.
& 352 log changes from three systems: \textit{ActiveMQ}, \textit{Hadoop}, and \textit{Maven}. 
& Finds six different anti-patterns in the logging code, such as wrong log levels and logging nullable objects, and proposes a tool, \textit{LCAnalyzer}, to detect anti-patterns.
&The approach learns anti-patterns from how developers fix the defects in their logging code.
&The work detects APs based on the independent historical changes to the logging code and falls short in detecting APs in cases that there has not been an update to the logging code.\\

Hassani \textit{et al.}~\cite{hassani2018studying} - Studies log-related issues for open-source software projects.
&563 log-related JIRA issues from \textit{Hadoop} and \textit{Camel} projects.
&As per authors findings, among the most common logging code issues are: \textit{1) inappropriate log messages}, \textit{2) missing logging statements}, \textit{3) inappropriate log verbosity levels}, and \textit{4) log library configuration issues}. 
&Developed a tool to detect incorrect log verbosity levels based on the words that appear in the logging statement's description.
&Log-issue checkers are threshold-dependent and in some cases result in a low number of detected issues.\\

Li \textit{et al.}~\cite{li2021studying,li2019dlfinder} - Studies issues with duplicate logging statements, which are logging statements that have the same static text messages.
&4K duplicate logging statements in five open-source projects: \textit{Hadoop}, \textit{CloudStack}, \textit{Elasticsearch}, \textit{Cassandra}, and \textit{Flink}.
& Repetitive logging statement descriptions can be potential logging code smells~\cite{zhang2011code}, \textit{i.e.}, a problematic duplicate logging code, which can have a detrimental or misleading effect in the understanding of the dynamic state of the system.
&Uncovers five categories of duplicate logging code smells and proposes a static analysis tool, DLFinder, to automatically detect duplicate logging code smells.
&The research eliminate the top 50 most frequent words when detecting log message mismatch (LM), which might cause false negatives.\\

Gujral \textit{et al.}~\cite{gujral2021exploratory} - Aims to identify a variety of logging issues faced by different software practitioners.
& Six Q\&A websites: \textit{Stack Overflow}, \textit{Serverfault}, \textit{Superuser}, \textit{Database Administrators}, \textit{Software Engineering}, and \textit{Android Enthusiasts}.
& Logging issues are prevalent across various domains (\textit{e.g.}, database, networks, and mobile), and at the same time, practitioners from different domains encounter different logging issues. 
&It performs semantic analysis of logging questions with topic modeling, which reveals several topics, such as \textit{logging conversion pattern}, android device logging, \textit{database logging}, \textit{logging level}, \textit{etc.}
&It solely considers Q\&A websites and its link with other software artifacts, such as \textit{issue tracker} and \textit{version control} systems requires further exploration.\\

Chen \textit{et al.}~\cite{chen2021demystifying} - Conducts an empirical study on the benefits and also challenges that developers face while reviewing the user-provided logs.
& 10 open-source systems including \textit{ActiveMQ}, \textit{AspectJ}, \textit{Hadoop Common}, \textit{HDFS}, and MapReduce, with the total of 1,561  and 7,287 logged and unlogged bug reports, respectively.
& In contrast to prior studies, this research finds that bug reports with logs take longer more time to resolve compared to bug reports without logs, as developers often ask for additional logs in those cases. In addition, the authors find that there exists a high degree of matching (73\%) between the classes that generate the logs and the actual buggy classed that causes the bug report. 
& It also includes a manual study of bug reports and finds that a noticeable portion of user-provided logs only contain the failure and do not provide the required context (\textit{e.g.}, the execution details) to locate the root cause. 
&Some of the findings, \textit{e.g.}, bugs with user logs take longer to be resolved and that it is common for log statements to be removed from the source code, are in contrast with prior work~\cite{yuan2012characterizing}, and future research should investigate the reasons behind these discrepancies.\\

\bottomrule
\end{tabular}
\end{center}
\vspace{-3mm}
\caption{Log-related issues research - Topic (D).}
\label{logging_issues_table}
\end{table*}
}
\subsection{Category E: Log Statement Automation}\label{automated_logging}
As mentioned earlier, execution logs, which are the output of logging statements in the source code, are a valuable source of information for system analysis and software debugging. 
Thus, high-quality logging statements are the precursor of effective log file mining and analysis. 
Conversely, low-quality LPSs result in log-related issues (Section~\ref{log_issue}), and they hinder the understanding of software problems whenever they happen. 
Currently, due to the \textit{ad-hoc} nature of logging, lack of general guidelines, and because developers mostly insert logging statements based on their personal experiences, the quality of log statements can hardly be guaranteed~\cite{zhang2011autolog}. 
Therefore, automated logging which aims to add or enhance log statements inside the source code either proactively or interactively is a well-motivated effort and can improve the quality of logging statements and, ultimately, result in more effective log mining tasks.\looseness=-1 

\begin{figure}[h]
\centering
\includegraphics[scale=.46]{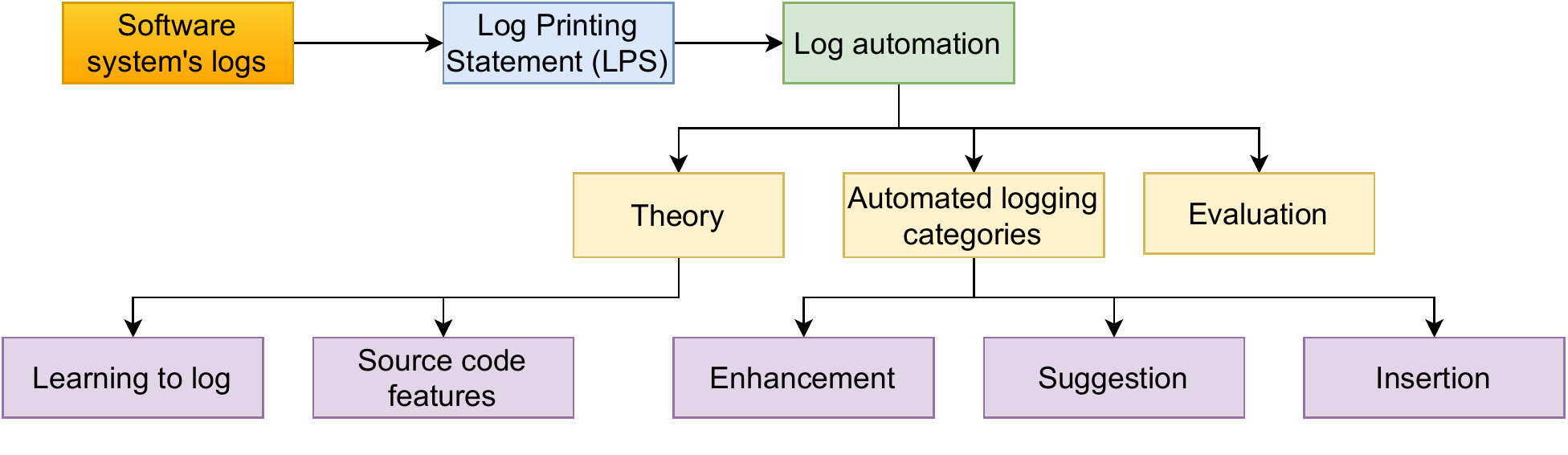}
\caption{Logging research with the emphasis on logging code automation research, Category E.}
\label{tree2_log_automation}
\end{figure}

Figure~\ref{tree2_log_automation} presents the log statements' research with emphasis on automated logging.  
As per this figure, we review the theory behind automated logging, automated logging approaches, and how they are evaluated. 
Prior studies have suggested creating and utilizing statistical models from common logging practices, and learning logging heuristics from experience, and using them to provide \textbf{new logging suggestions} or \textbf{enhance the already existing LPSs}.

\subsubsection{Log Automation - Motivation and Theory}\label{log_auto_theory}
One of the common approaches for log automation is the application of machine learning methods to predict whether a code snippet needs a logging statement by training a model on a set of logged code snippets, and testing it on a new unlogged code set, \textit{i.e.}, supervised learning. 
In this section, we first review the background and theory for machine learning methods and continue with automated logging approaches. 

\textbf{Motivation.} With the ever-increasing size of software systems, it is most likely that a single developer is in charge of developing only a small subsystem of the whole software system. 
Under this situation, making wise logging decisions becomes quite challenging as developers do not have full knowledge of the whole system~\cite{zhu2015learning}. 
As logs are quite pervasive and useful for system maintenance~\cite{shilin2016expr}, if the logging decisions can be learned automatically, a log suggestion tool can be constructed to help developers make better decisions. 
Ultimately, such a tool can increase the quality of logs and save developers time.     

\begin{figure}[h]
\vspace{-2mm}
\centering
\includegraphics[scale=.43]{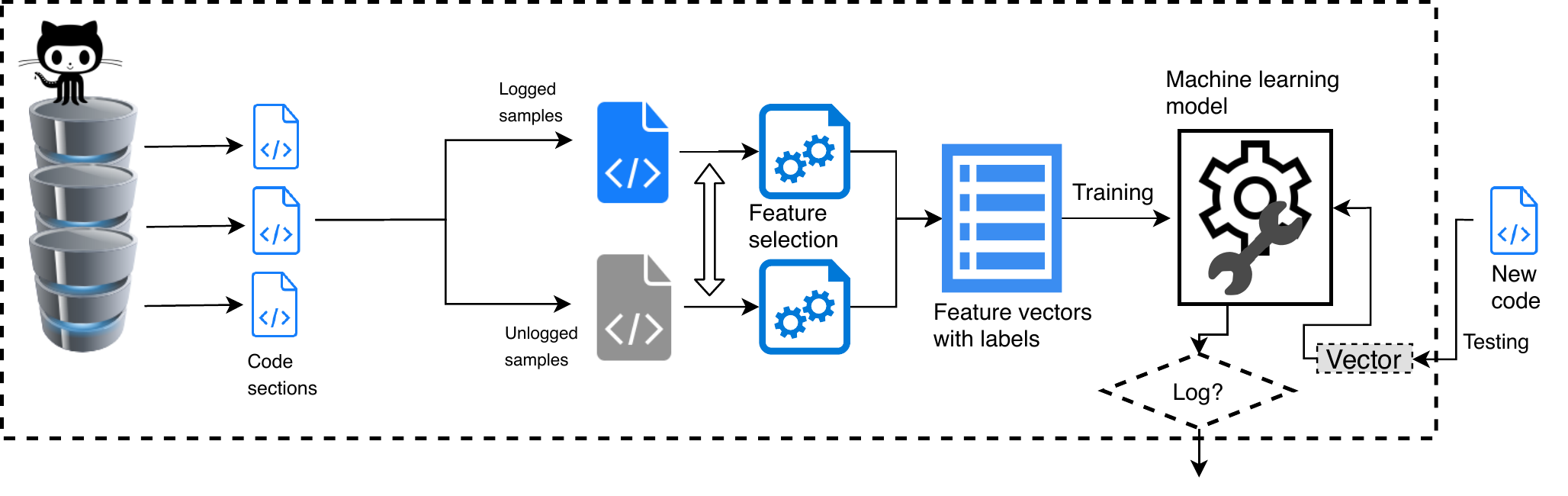}
\vspace*{-7mm}
\caption{\textit{Learning to log} platform.}
\label{learning_to_log}
\vspace{-5mm}
\end{figure}

\textbf{Learning to Log.} The idea of \textit{learning to log} is to construct a machine learning (ML), or deep learning (DL), model that can learn common logging practices and provide logging suggestions to the developers or directly make logging decisions and insert logging statements into a newly-developed source code snippet. 
A typical learning to log tool~\cite{zhu2015learning} is outlined in Figure~\ref{learning_to_log}. 
The log learning steps are: \textbf{1)} \textit{code collection from repositories}, \textbf{2)} \textit{labeling the collected source code}, \textbf{3)} \textit{feature extraction and selection}, \textbf{4)} \textit{feature vectors and model training}, and finally \textbf{5)} \textit{logging enhancement, suggestion or automatic insertion}. 
Based on this platform, once the \textit{training phase} is completed, during the \textit{testing phase}, the learning model decides whether a new code snippet requires a logging statement by extracting its features and feeding it to the ML model, and observing the model's output.  
Learning algorithms apply a wide range of techniques such as: pattern or rule-based~\cite{yuan2012conservative,yuan2012improving,cinque2012event,zhao2017log20}, machine learning such as Naive Bayes, Bayes Net, Logistic Regression, SVM, and Decision Trees~\cite{zhu2015learning}, Random Forrest~\cite{li2017towards,candido2021exploratory}, Ordinal Regression~\cite{li2017log}, and most recently, Deep Learning~\cite{gholamian2020logging,li2020shall,liu2019variables,li2021deeplv}.

\subsubsection{Source Code Feature Formulation}\label{app1}
\begin{figure}[h]
\vspace{-2mm}
\centering
\includegraphics[scale=.33]{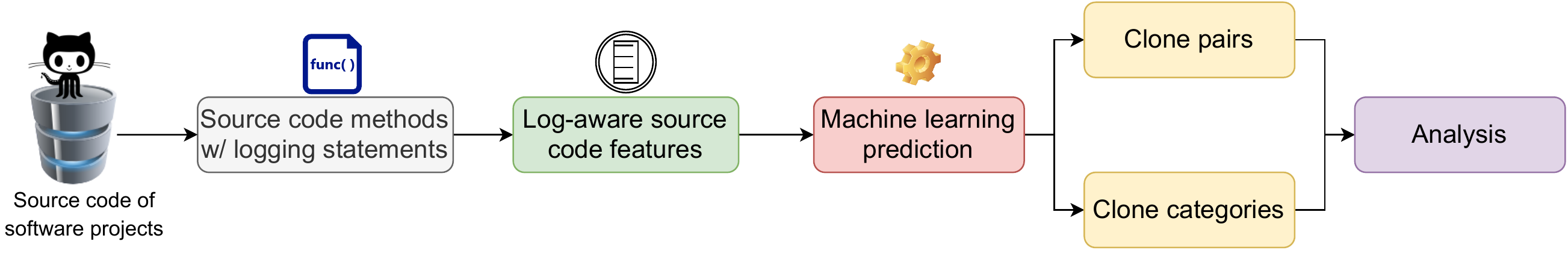}
\caption{Log prediction with source code features and code clones.}
\label{code_clone}
\vspace*{-1mm}
\end{figure}

In order to be able to learn and predict log statements, prior research~\cite{zhu2015learning,gholamian2020logging} have proposed to define related source code features and utilize them for predicting whether a code block requires a log statement. 
Source code features can be \textbf{structural} (type of the code blocks, \textit{e.g.}, \textit{catch clause, if-else}), \textbf{functional} (e.g., metrics such as \textit{code complexity, dependencies, fan-in, and fan-out}), \textbf{contextual} (\textit{e.g.}, variables and keywords in the code snippet), and source code \textbf{semantic} features~\cite{li2020shall,jia2018smartlog}, \textit{i.e., what the code snippet is trying to do}. 
What category of features to select and how well they can distinguish the logged and unlogged code snippet is an active research topic~\cite{li2020shall,gholamian2020logging,zhu2015learning,lal2016logopt,jia2018smartlog}. 
Additionally, the logging automation research has benefited from leveraging the findings in adjacent software tasks such as source code clone detection~\cite{saini2018oreo}, and code commenting~\cite{huang2019learning} for feature selection as the idea is that similar code snippets should follow similar logging patterns. 
Figure~\ref{code_clone} shows a log prediction platform based on similar code snippets (\textit{i.e.}, clone pairs), which are then later utilized for log prediction. 
Source code features are extracted from \textit{method definitions} with logging statements. 
Then, once the machine learning model is trained and clone pairs are extracted, they are leveraged for log location prediction. 

\afterpage{\renewcommand{\arraystretch}{1.2}
\newcommand\rownumber{\stepcounter{magicrownumbers}\arabic{magicrownumbers}}
\rowcolors[]{2}{gray!15}{white}
\begin{table}[h]
\scriptsize	
\begin{center}
 \begin{tabular}{p{1.3cm} p{2.6cm} P{.8cm} P{.8cm}  P{.8cm}} 
 \toprule
 
\rowcolor{blue!20} \textbf{Reference} &\textbf{Category} & \textbf{Lang.} & \textbf{Source} & \textbf{\# Proj.}\\ [0.5ex] 
 \midrule

Zhang \textit{et al.}~\cite{zhang2011autolog}& LPS insertion, \textit{AutoLog} & Java &OS&1\\

Yuan \textit{et al.}~\cite{yuan2012improving}& add variables to LPSs, \textit{LogEnhancer} & C &OS&8\\

Zhu \textit{et al.}~\cite{zhu2015learning}& LPS suggestions, exception and RVC& C\# &CS/OS&2+2\\

Lal \textit{et al.}~\cite{lal2016logoptplus}& LPS prediction for \textit{try-catch} and \textit{if} blocks& Java & OS&2\\

Zhao \textit{et al.}\cite{zhao2017log20} & optimal LPS placement, \textit{Log20}& Java & OS&4\\ 

Li \textit{et al.}~\cite{li2017towards} & RF classifier for log change suggestion &Java& OS&4\\ 

Li \textit{et al.}~\cite{li2017log} & hierarchical clustering for LVL &Java& OS&4\\ 

Jia \textit{et al.}~\cite{jia2018smartlog} & intention-aware ESL prediction&C/C++& OS&6\\ 

Anu \textit{et al.}~\cite{anu2019approach} & RF with context features for LVL prediction-aware ESL prediction&Java& OS&4\\ 

Liu \textit{et al.}~\cite{liu2019variables} & DL recurrent neural network(RNN) for log variable prediction&Java& OS&9\\

Gholamian and Ward~\cite{gholamian2020logging} & source-code clones for log statement location prediction&Java& OS&3\\

Kim \textit{et al.}\cite{kim2020automatic} & LVL prediction with semantic and syntactic features &Java& OS&22\\ 

Li \textit{et al.}~\cite{li2020shall} & DL code-block level log location suggestion&Java& OS&7\\ 

C\^{a}ndido \textit{et al.}~\cite{candido2021exploratory}  & transfer learning for log location precitionDL code-block level log location suggestion&Java& CS/OS&1+29\\ 

Li \textit{et al.}~\cite{li2021deeplv} & DL approach for LVL prediction with RNN &Java& OS&9\\

\bottomrule
\end{tabular}
\end{center}
\vspace{-2mm}
\caption{\footnotesize Comparison of log printing statement automation research - Topic (E). Lang: programming Language; Source: if the studies projects are OS or CS; OS: Open Source; CS: Closed Source.}
\label{log_statement_automation_table_OW}
\end{table}
}

\subsubsection{Automated Logging Categories}
\begin{figure}[h]
\vspace{-3mm}
\includegraphics[scale=.55]{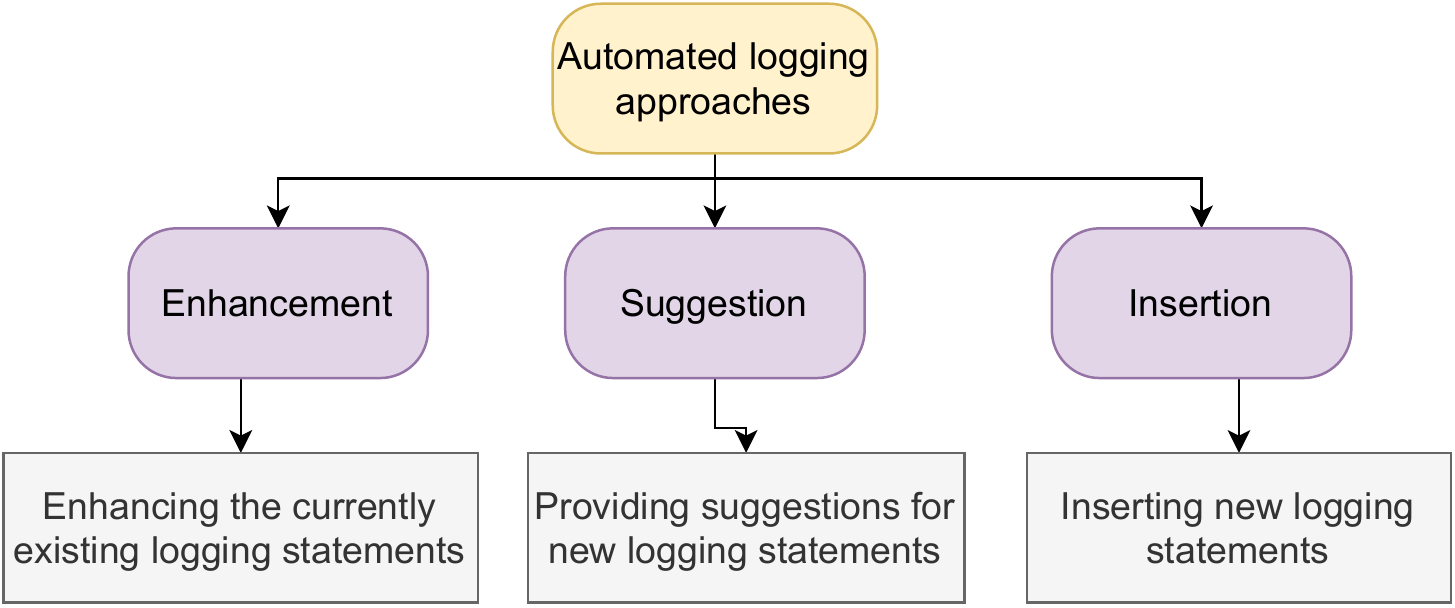}
\caption{Auto logging of the software systems' source code.}
\label{auto_logging}
\vspace{-2mm}
\end{figure}

Figure~\ref{auto_logging} highlights the research in this area categorized into three subtopics: \textit{log enhancement, log suggestion,} and \textit{log insertion}. 
These approaches are primarily concerned with log \textbf{location} prediction, \textit{i.e., where to log}, and secondarily the \textbf{content} to include in the logging statements, \textit{i.e., what to log}. 
Log enhancement aims to improve the quality of existing logging statements, such as adding more runtime context~\cite{yuan2012improving}. 
Log suggestion aims to provide suggestions for logging locations that might have been missed, and log insertion aims to proactively insert logging statements into the source code~\cite{zhao2017log20}. 
The approaches can also be categorized based on the targeted LPSs. Some approaches focus on error logging statements (ELS), such as logging when an error happens inside a \textit{catch clause}~\cite{yuan2012conservative,jia2018smartlog}, and others focus on normal logging statements (NLS), \textit{e.g.}, such as \textit{method-level logging}~\cite{gholamian2020logging}.
Although the goals of these approaches are similar, the intended way of their implementation can be different. 
For example, one practical scenario of implementing log suggestion approaches is as IDE plugins that provide just-in-time suggestions~\cite{gholamian2020logging}. 
However, the log insertion techniques are implemented as post-processing tools that scan the source code and insert logs for various criteria of interest, such as disambiguating execution paths~\cite{zhao2017log20} or logging catch clauses~\cite{yuan2012conservative}. 
This categorization is not mutually exclusive and some of the prior work overlap in their approaches.  

\textbf{Log Verbosity Level (LVL) and Description Predictions (LSD).} In addition to predicting log location, prior research has also investigated approaches for prediction of the appropriate logging \textbf{verbosity level} for newly composed logging statements~\cite{li2021deeplv,li2017log,hassani2018studying,anu2019approach,mizouchi2019padla}. The approaches either apply some type of learning to predict the log verbosity level~\cite{anu2019approach}, or perform dynamic adjustment of the log level during the runtime~\cite{mizouchi2019padla}. 
Other research also aims to predict the description~\cite{he2018characterizing}, or variables included in the logging statements~\cite{liu2019variables,yuan2012improving}. 
In Table~\ref{log_statement_automation_table_OW}, we provide a high-level comparison of automated logging research, and Table~\ref{log_statement_automation_table} provides a more detailed comparison of each research. 


{\renewcommand{\arraystretch}{1}
\begin{table}[h]
\small
\centering
\begin{tabular}{l|l|c|c|}
\multicolumn{2}{c}{}&\multicolumn{2}{c}{\cellcolor{blue!20}Model prediction}\\
\cline{3-4}
\multicolumn{2}{c|}{}&Positive&Negative\\
\cline{2-4}
\multirow{2}{*}{Actual (ground truth)}& Positive & $TP$ & $FN$ \\
\cline{2-4}
& Negative & $FP$ & $TN$ \\
\cline{2-4}
\end{tabular}
\caption{Confusion matrix for log prediction.\\}
\label{conf_matrix}
\end{table}
}

{\renewcommand{\arraystretch}{1.1}
\newcommand\rownumber{\stepcounter{magicrownumbers}\arabic{magicrownumbers}}
\rowcolors[]{2}{gray!15}{white}
\begin{table}[h]
\scriptsize
\centering
\begin{tabular}{  m{2cm}  P{2cm} m{3.5cm} } 
\toprule
\rowcolor{blue!20}\textbf{Metric} & \textbf{Formula}& \textbf{Description} \\ 
\midrule
Precision &  $\frac{TP}{TP+FP}$ & The ratio of correctly identified positive instances to the number of all positive predictions.  \\ 

Recall (a.k.a. \textit{sensitivity}, \textit{hit rate}, and \textit{true positive rate}) & $\frac{TP}{TP+FN}$ & The ratio of the correctly predicted instances to the number of existing positive instances.  \\ 

False Negative Rate & $\frac{FN}{FN+TP}$ & The ratio of false negatives to the total number of existing positive instances.\\

True Positive Rate &   $\frac{TP}{TP+FN}$ & The ratio of true positives to the total number of existing positive instances.\\ 

True Negative Rate (a.k.a. specificity, and selectivity)&   $\frac{TN}{TN+FP}$ & The ratio of true negative to the total number of existing negative instances.\\

False Positive Rate (a.k.a. fall-out) &   $\frac{FP}{FP+TN}$ & The ratio of false positives to the total number of existing negative instances.\\ 

F-Measure (a.k.a, F-Score, F1-Score) & \hspace*{-2mm} $\frac{2\times Precision\times Recall}{Precision+Recall}$ & Harmonic mean of Precision and Recall.   \\ 

Accuracy &   $\frac{TP+TN}{TP+TN+FP+FN}$ & Accuracy is the proportion of correctly identified logged instances to the total number of cases.\\

Balanced Accuracy (BA) &   $\frac{1}{2} \times (\frac{TP}{TP + FN} + \frac{TN}{TN + FP} )$ & Balanced accuracy (BA) is the average of the proportion of logged instances and the proportion of unlogged instances that are correctly classified.\\ 

Receiver Operating Characteristic curve (ROC)  &  $\frac{TPR}{FPR}$ & The plot of the true positive rate against the false positive rate.\\

Area Under Curve  &  $area_{under}(ROC)$ & \textit{Area under the curve} is the area under the Receiver Operating Characteristic curve.\\

BLEU (BiLingual Evaluation Understudy)  & \hspace*{-3mm}$
\frac{count\_tokens{(C\cap R)}}{count\_tokens(C)}$* & Similar to Precision but for auto-generated text. The ratio of the candidate tokens (C) that exist in reference tokens (R) over the total candidate tokens.\\

ROUGE (Recall-Oriented Understudy for Gisting Evaluation)  & \hspace*{-3mm}  $ \frac{count\_tokens{(C\cap R)}}{count\_tokens(R)}$* & Similar to Recall but for auto-generated text. The ratio of the reference tokens that exist in the candidate tokens over the total reference tokens. \\

Edit Distance  & \tiny \hspace*{-5mm}  $D(i,j)=min \big(D(i,j-1)+1, D(i-1,j)+1, D(i-1,j-1)+{0|2}\big)$ & Minimum edits require to convert sequence i to sequence j. \\

\bottomrule
\end{tabular}
\vspace*{-1mm}
\begin{flushleft}
{{\normalsize *} \scriptsize{Simplified formulas are presented. There is also weight corresponding to the size of n-grams that comes in the general formula for BLEU~\cite{papineni2002bleu}. Similarly, for ROUGE, refer to~\cite{lin2004rouge}.}}
\end{flushleft}
\caption{Evaluation metrics for automated log prediction.}
\label{metrics}
\end{table}
}

\subsubsection{Evaluation Metrics}\label{metrics_section}
After training the learning model, its performance should be evaluated during the testing phase by applying new code instances as input and finding out the prediction outcome that whether or not this new code snippet requires a logging statement, \textit{e.g}., Figure~\ref{learning_to_log}. 
This is an example of a \textit{binary classification} problem~\cite{freeman2008comparison}. 
Differently, log verbosity level prediction is evaluated as a \textit{multi-class classification} problem~\cite{aly2005survey}, as generally several verbosity levels are available for the log statements, \textit{e.g.}, \textit{WARN, INFO, DEBUG}, \textit{etc}. Furthermore, ordinal multi-class classification~\cite{li2007ordinal} considers an order between the possible prediction labels. 
For example, for verbosity level prediction, $WARN < INFO < DEBUG$, such that $WARN<INFO$ means $INFO$ is more verbose than $WARN$~\cite{li2021deeplv}.     
Thus, different evaluation metrics are applied to assess the quality of learning models and their prediction accuracy. 
In general, the performance of a logging prediction method is evaluated by first extracting the confusion matrix.

{
\renewcommand{\arraystretch}{1.2}
\newcommand\rownumber{\stepcounter{magicrownumbers}\arabic{magicrownumbers}}
\rowcolors[]{2}{gray!15}{white}
\begin{table*}
\scriptsize	
\begin{center}
 \begin{tabular}{p{3cm} p{3cm} p{4cm} p{3cm} p{3cm}} 
 \toprule
 
\rowcolor{blue!20} \textbf{Reference - Aim} &  \textbf{Experiments}& \textbf{Results} & \textbf{Pro}& \textbf{Con} \\ [0.5ex] 
 \midrule
Zhang \textit{et al.}~\cite{zhang2011autolog} - Proposes \textit{AutoLog}, which generates additional informative logs to help developers discover the root cause of a software failure.
& Performs a proof-of-concept case study on Apache Hadoop Common.
&AutoLog embeds a two-stage process of \textit{log slicing} and \textit{log refinement} of the program to narrow down the execution paths that could have led to the system's failure.
&The approach narrows down the execution paths that could have led to a system's failure. AutoLog is targeted for interactive in-house development.
&The program needs to be re-executed every time new log statements are added, which is time-consuming.\\

Yuan \textit{et al.}~\cite{yuan2012improving} - Proposes a tool, \textit{LogEnhancer}, to find and add useful variables to log statements.
& Evaluated on a total of 9,125 log messages from eight applications in different domains, including \textit{apache httpd}, \textit{postgresql}, and \textit{cvs}.
&LogEnhancer is effective in automatically adding a high percentage of log variables, on average, 95.1\%, that programmers manually included.
&The tool performs static analysis on the source code starting from the log statement and navigates backward to find variables that are causally along the path that results in the execution of the log statement.
&As LogEnhancer's improvement is limited to the existing log statements, its effectiveness diminishes if the logging statements are missing.\\

Zhu \textit{et al.}~\cite{zhu2015learning} - Proposes \textit{LogAdvisor}, which aims to provide logging suggestions for \textit{exception} and \textit{return-value-check} code blocks.
&Two industrial software systems from Microsoft and two open-source software systems from GitHub (\textit{SharpDevelop} and \textit{MonoDevelop}).
&LogAdvisor achieves a high \textit{balanced accuracy}, ranging from 84.6\% to 93.4\%, to match developers' logging decisions, and the decision tree model achieves the highest scores.
& Trains a machine learning model (\textit{e.g.}, SVM and decision trees) to predict whether a focused code snippet requires a logging statement. 
&It is focused and limited to two categories of code snippets: 1) \textit{exception snippets} and 2) \textit{return-value-check}.\\

Lal \textit{et al.}~\cite{lal2016logoptplus} - Introduces \textit{LogOptPlus} tool for automated \textit{catch} and \textit{if} code block logging prediction. 
&Two open-source projects: \textit{Apache Tomcat} and \textit{CloudStack}.
&The prediction model with random forest achieves the highest F1-score
80.70\% (Tomcat) and 92.25\% (CloudStack) for \textit{if-block} logging prediction.
& Applies five different learning techniques, \textit{e.g.,} AdaBoost, Gaussian Naive Bayesian, and Random Forests achieve the highest Precision and Recall.
& It is limited to specific code blocks, \textit{i.e.}, \textit{if-block} and \textit{catch clause}.\\
 
Zhao \textit{et al.}\cite{zhao2017log20} - Introduces \textit{Log20}, a tool that finds a placement of logging statements to minimize execution path ambiguity.
&Evaluated on four open-source Java projects: \textit{HDFS}, \textit{HBase}, \textit{Cassandra}, and \textit{ZooKeeper}.
&Log20 achieves a lower logging overhead with the same level of informativeness (\textit{i.e.}, entropy) compared to existing logging statements by developers.
& It applies Shannon's information theory equation to measure the entropy of the program by approximately considering all of the possible execution paths. 
& The approach does not consider developers' concerns and practices, does not explain the static content of LPSs, and change of workload can cause extra logging overhead.\\

Li \textit{et al.}~\cite{li2017towards} - Analyzes log changes in open-source projects and proposes commit-time logging suggestions.
& Four open source projects: \textit{Hadoop}, \textit{Directory Server}, \textit{Commons HttpClient}, and \textit{Qpid}.
&Performs a manual analysis on four software systems and categorizes the changes to logging statements into four major groups: 1) \textit{block change}, 2) \textit{log improvement}, 3) \textit{dependence-driven
change}, and 4) \textit{logging issues}.
& Proposes a random forest (RF) classifier for each code commit to suggest whether a log change is required. 
The classifier's balanced accuracy for within-project suggestions is 0.76 to 0.82.
& As the model is trained on prior log changes, it might miss scenarios that there are no prior logging changes to learn from.\\   

Li \textit{et al.}~\cite{li2017log} - Determines the appropriate log verbosity level for the newly-developed logging statement.
& Analyzes four open source projects: \textit{Hadoop}, \textit{Directory Server}, \textit{Hama}, and \textit{Qpid}.
& Collected five categories of quantitative metrics that play important roles in determining the appropriate log level: \textit{logging statements metrics, containing block metrics, file metrics, change metrics, and historical metrics}. Achieves AUC in the range of 0.75 to 0.81 for log level prediction. 
&Metrics from the block which contains the logging statement, \textit{i.e.}, the surrounding block of a logging statement, play the most important role in the ordinal regression models for log levels. 
& The results show that the ordinal regression models for log level prediction are project-dependent.\\

Jia \textit{et al.}~\cite{jia2018smartlog} - Proposes, \textit{SmartLog}, which is an intention-aware error logging statement (ELS) suggestion tool with two intention models: IDM and GIDM.
&Experiments on six open-source projects in C/C++:  \textit{Httpd}, \textit{Subversion}, \textit{MySQL}, \textit{PostgreSQL}, \textit{GIMP}, and \textit{Wireshark}.
& It improves on the Recall values (average of 0.61) and achieves higher scores compared to LogAdvisor~\cite{zhu2015learning} (average of 0.45) and Errlog~\cite{yuan2012conservative} (average of 0.18). 
&This work improves on prior work by going beyond code patterns and syntax features, and considers source code intentions, \textit{i.e.}, semantics.
& This work is limited to ELS prediction, \textit{i.e.}, \textit{exception} and \textit{function return-value} logging.\\
 
Anu \textit{et al.}~\cite{anu2019approach} - Proposes a method to make the logging level decisions by understanding the logging intentions. 
&Four open-source software projects: \textit{Hadoop}, \textit{Tomcat}, \textit{Qpid}, and \textit{ApacheDS}.
& It reaches AUC values higher than 0.9 in log level prediction. The approach extracts the contextual features from logging code snippets and leverages a machine learning model (\textit{i.e.}, a random forest model) to automatically predict the verbosity level of logging statements. 
&As a proof of concept, the authors also implement a prototype tool, \textit{VerbosityLevelDirector}, to provide guidance on log verbosity level selection in focused code blocks.
&The work is limited to focused code blocks: \textit{exception handling blocks} and \textit{condition check blocks}.\\

Liu \textit{et al.}~\cite{liu2019variables} - Presents an approach to recommend the variables to include in logging statements.
& Evaluates on nine open-source Java projects: \textit{ActiveMQ}, \textit{Camel}, \textit{Cassandra}, \textit{CloudStack}, \textit{DirectoryServer}, \textit{Hadoop}, \textit{HBase}, \textit{Hive} and \textit{Zookeeper}.
&The approach first learns ``rules'' from existing logged code snippets by extracting contextual features with deep learning recurrent neural networks (RNN). The approach outperforms five baselines, including random guessing and IR methods in log variable prediction.
& The tool provides a ranked list of variables that probably are required logging to the developer.  
&The method only considers the code preceding the logging statement. 
As such, extending this approach to include the code succeeding the logging statement can improve on logging variable recommendation.\\

\bottomrule
\end{tabular}
\end{center}
\vspace{-3mm}
\caption{Log printing statement automation research - Topic (E).}
\label{log_statement_automation_table}
\end{table*}
}

\afterpage{\renewcommand{\arraystretch}{1.2}
\newcommand\rownumber{\stepcounter{magicrownumbers}\arabic{magicrownumbers}}
\rowcolors[]{2}{gray!15}{white}
\begin{table*}
\ContinuedFloat
\scriptsize	
\begin{center}
 \begin{tabular}{p{3cm} p{3cm} p{4cm} p{3cm} p{3cm}} 
 \toprule
 
\rowcolor{blue!20} \textbf{Reference - Aim} &  \textbf{Experiments}& \textbf{Results} & \textbf{Pro}& \textbf{Con} \\ [0.5ex] 
 \midrule

Kim \textit{et al.}\cite{kim2020automatic} - Proposes an approach to verify the appropriateness of the log verbosity levels. 
& 22 open-source projects from three different domains: \textit{message queuing}, \textit{big data}, and \textit{web application server}.
& Applies semantic and syntactic features and recommends a new log level in case the current level is deemed inappropriate. It reaches 77\%  precision and 75\% recall in log level validation.
& Creates \textit{domain word model} from all of the log messages in application domains, which enables knowledge sharing between different projects.   
&In some cases, the appropriateness of log levels is dependent on developers' opinions and is quite arguable.\\

Gholamian and Ward~\cite{gholamian2020logging} - Proposes a log-aware code clone detection (LACC) approach for log suggestions.
& Three open-source Java projects: \textit{Tomcat}, \textit{Hadoop}, and \textit{Hive}.
& Performs an experimental study of logging characteristics of source code clones and observes that code clones match in their logging behavior. 
Achieves 90\% accuracy in log location prediction. 
& It applies source code features and machine learning methods to detect log-aware code clones for log statement prediction.
& The approach can only suggest logs for code snippets that can find their clone pairs in the software code base.\\

Li \textit{et al.}~\cite{li2020shall} - Discusses the locations that need to be logged, and proposed a learning approach to provide code block level logging suggestions.
&Seven open-source systems: \textit{Cassandra}, \textit{Elasticsearch}, \textit{Flink}, \textit{HBase}, \textit{Kafka}, \textit{Wicket}, and \textit{ZooKeeper}. 
& The authors discover six categories of logging locations in different types of code blocks from developers' logging practices. It achieves balanced accuracy of 80.1\%) using syntactic source code features.
& Utilizes a pipeline of word embedding, RNN layer, and a dropout layer in its deep learning model for log location prediction.
& The achieves acceptable prediction by leveraging syntactic information only. Additional studies are required to combine syntactic and semantic features of the source code blocks.\\

C\^{a}ndido \textit{et al.}~\cite{candido2021exploratory} - Proposes a log suggestion approach based on machine learning methods.
&An enterprise software, Adyen, and 29 Apache projects. 
&The authors extract source code metrics from methods and evaluate the performance of five different learning approaches on log suggestions. The best performing model achieves 72\% of balanced accuracy on Adyen's log statements set.
& Performs a study on 29 Java projects and leveraged learning transfer to generalize to an industry project.
& The applied transfer-learning approach shows a lower performance when trained on open-source projects and tested on Ayden enterprise project.\\

Li \textit{et al.}~\cite{li2021deeplv} - Proposes a deep learning approach for log level prediction with an ordinal-based output layer.
&Nine large-scale open-source projects: \textit{Cassandra}, \textit{ElasticSearch}, \textit{Flink}, \textit{HBase}, \textit{JMeter}, \textit{Kafka}, \textit{Karaf}, \textit{Wicket}, \textit{Zookeeper}.
& The authors initially perform a manual study and categorize five different logging locations. The model trained with syntactic features achieves an average AUC of 80.8\%.
&Their findings infer that the log levels that fall far apart on the verbosity scale manifest different characteristics. 
&Log levels that are closer in order, \textit{e.g.}, warn and error are more difficult to distinguish with this approach.\\
\bottomrule
\end{tabular}
\end{center}
\vspace{-3mm}
\caption{Log printing statement automation research - Topic (E) (continued).}
\label{log_statement_automation_table}
\end{table*}
}

In Table~\ref{conf_matrix}, \textit{``Model prediction''} values are from the learning model and the \textit{``Actual''} values are the \textit{ground truth}. 
Prior research often considers the developers' inserted logging statements as \textit{ground truth}. 
To create a set of training and testing data for the machine learning process and have a proper \textit{ground truth} to compare with, one approach is to collect all of the code snippets with logging statements, and some samples of unlogged code, to include both positive and negative cases. 
Then, after deciding the train-test split and training the ML model, prior work removes the log statements from the test data. 
During the testing phase, the model's performance is evaluated on the test code snippets with their logging statements being removed. 
This way we measure how well the model can \textbf{remember} which code snippets should have and which ones should not have logging statements, compared to the developers' originally-inserted LPSs. 
Multiple iterations of the training-testing can be applied, \textit{e.g.}, cross-validation~\cite{browne2000cross}, to confirm the results. 
From Table~\ref{conf_matrix}, \textit{TP} means that the model correctly predicted a code snippet that requires a logging statement, and \textit{FN} denotes that the model incorrectly predicted that a code snippet does not require a logging statement. 

Based on the confusion matrix, we can define some of the common metrics for evaluating the performance of log learning models in Table~\ref{metrics}.  
The definitions for \textit{Precision} and \textit{Recall} are straightforward. 
In order to ensure a prediction model benefits from equally good or comparable Precision and Recall values, \textit{F-Measure} is defined as the harmonic mean of {Precision and Recall}. 
Qualitatively, good performance of \textit{F-Measure} implies good performance on both {Precision and Recall}. 
\textit{Accuracy} represents correctly identified logged instances to the total number of cases. 
\textit{Balanced Accuracy (BA)} is the average of the proportion of logged instances and the proportion of unlogged instances that are correctly classified. 
In case there is an imbalance in the data, \textit{e.g.}, in Table~\ref{conf_matrix}, if \textit{TN} is much larger than \textit{TP}, \textit{Balanced Accuracy (BA)} is widely used to evaluate the modeling results~\cite{zhang2005ensembles,zhu2015learning,li2017towards}, because it avoids the over-optimism that \textit{accuracy} might experience. 
Receiver Operating Characteristic (ROC) plots \textit{true positive rate} against \textit{false positive rate}. 
AUC (area under the curve) is the area under the ROC curve. 
Intuitively, the AUC evaluates how well a learning method can distinguish logged code snippets and unclogged ones. 
The AUC ranges between 0 and 1. 
A high value for the AUC indicates a high discriminative ability of the learning model; an AUC of 0.5 indicates a performance that is no better than random guessing~\cite{li2017towards}. BLEU~\cite{papineni2002bleu} and ROUGE~\cite{lin2004rouge} scores are equivalent to Precision and Recall and are leveraged to evaluate the auto-generated text compared to the original text developed by developers. 
These scores have applications in evaluating the auto-generated \textbf{l}og \textbf{s}tatement \textbf{d}escriptions (LSDs), which are sequences of tokens, \textit{i.e.}, words. For example, for a candidate LSD, $C$ and the reference LSD, $R$, BLEU measures the ratio of tokens of $C$ that also appear in $R$ (analogous to Precision), and ROUGE measures the ratio of tokens of $R$ that have appeared in $C$ (analogous to Recall). 
The range of values for BLEU and ROUGE is \textit{[0,1]}, with 1 being the perfect score. 
These two measures combined explain the quality of the auto-generated LSDs. 
\textit{Edit distance} can be also used for checking the distance between auto-generated text and the developer inserted LSD.

\textbf{Examples of Metrics Used.} Li \textit{et al}. ~\cite{li2019dlfinder} utilized \textit{Precision} and \textit{Recall} to calculate the performance of \textit{DLFinder} in detecting logging code smells. 
Zhu \textit{et al}.~\cite{zhu2015learning} used \textit{BA} to evaluate the accuracy of \textit{LogAdvisor}, which advises the developer if logging statements are required for a focused code snippet. 
Li \textit{et al}.~\cite{li2017towards,li2017log,li2018studying} used \textit{ROC} and \textit{AUC} to evaluate their methods in \textit{log verbosity level prediction} and \textit{logging commit change suggestion}.
Kim \textit{et al.}~\cite{kim2020automatic} used \textit{F-Measure} to evaluate their log verbosity level recommendation approach, and Gholamian and Ward~\cite{gholamian2020logging} utilized \textit{Accuracy} to evaluate the performance of their log-aware clone detection approach. 
He \textit{et al.}~\cite{he2018characterizing} leveraged BLEU and ROUGE scores to evaluate the effectiveness of the candidate log statement descriptions when compared to the developer-inserted log descriptions. 
Edit distance~\cite{levenshtein1966binary} also has applications in finding similar code snippets for enabling logging suggestions~\cite{he2018characterizing}.

\begin{tcolorbox}[breakable, enhanced]
\small \textbf{Finding} \textbf{9.} \textit{
In sum, prior studies have investigated different learning paradigms, such as clustering, random forest, deep learning, and transfer learning. 
The approaches aim to predict the \textbf{location} (LSL), \textbf{verbosity level} (LVL), \textbf{variables} (VAR), and \textbf{description} (LSD) of the logging statements.}
\end{tcolorbox}
\subsection{Mining Log Files}\label{mine_log_files}
Priorly, we mentioned the purpose of logging statements added by developers is to expose valuable runtime information. 
The output of logging statements is written to log files, which are used by a plethora of log processing tools to assist developers and practitioners in different tasks such as software debugging and testing~\cite{andrews1998testing,jiang2008automatic}, performance monitoring~\cite{yao2018log4perf}, and postmortem failure detection and diagnosis~\cite{yuan2010sherlog,nagaraj2012structured,syer2013leveraging,xu2014pod}. 
We review log mining techniques and approaches in the following.

\afterpage{\renewcommand{\arraystretch}{1.2}
\newcommand\rownumber{\stepcounter{magicrownumbers}\arabic{magicrownumbers}}
\rowcolors[]{2}{gray!15}{white}
\begin{table}[h]
\scriptsize	
\begin{center}
 \begin{tabular}{p{1.2cm} p{2.6cm} p{1.3cm} P{.7cm}P{.7cm}} 
 \toprule
 
\rowcolor{blue!20} \textbf{Reference} &\textbf{Category} & \textbf{Log source} & \textbf{Type} & \textbf{Org.}\\ [0.5ex] 
 \midrule

Li \textit{et al.}~\cite{li2009integrated}& framework for knowledge acquisition from historical log data & Windows&IND& IBM\\

Marty~\cite{marty2011cloud}& cloud logging management challenges  & cloud logs from AWS &IND& Loggly\\

Li \textit{et al.}~\cite{li2017flap}& analysis of logs, log review and correlation & \textit{Network X} &IND& Huawei\\

Amar \textit{et al.}~\cite{amar2018using}& log differencing  & user study and FSA logs&ACA& N/A \\

Bao \textit{et al.}~\cite{bao2019statistical}& statistical log differencing  & user study and FSA logs&ACA&  N/A\\ 

Liu et al.~\cite{liu2019logzip}& effective log compression for log management & five system logs &ACA& \cite{urllogzipgithup}\\ 

Yao \textit{et al.}~\cite{yao2020study} & compression of logs vs. natural text& 9 systems + 2 (Wiki, Gutenberg) &ACA& \cite{kindigithup}\\

Shin \textit{et al.}~\cite{shin2020effective}& \textit{LogCleaner} to remove redundant log lines& 10+2 & ACA& N/A\\

He \textit{et al.}~\cite{he2020loghub}& a log hub for various system logs& 17 & ACA& \cite{loghubgithup}\\ 

Chen and Jiang~\cite{chen2021survey}& survey of log instrumentation techniques.& N/A & ACA& \cite{urlsurveychen}\\

Locke \textit{et al.}~\cite{locke2021logassist}& organizing and summarizing logs& HDFS, Zookeeper, ES & ACA/ IND& N/A\\

Yao \textit{et al.}~\cite{yao2021improving}& pre-compression processing of logs& 16 different systems logs& ACA& N/A\\

Wang \textit{et al.}~\cite{wang2021would}& \textit{DPLOG} for big data monitoring& Spark benchmarks & ACA /IND& \cite{urldplog}\\

\bottomrule
\end{tabular}
\end{center}
\vspace{-2mm}
\caption{\footnotesize Comparison of log maintenance and management research - Topic (F). Log source: source of logs for the study; Type: IND: Industrial; ACA: Academic; Lang: programming Language; Org.: Organization or link if available.}
\label{log_management_table_OW}
\end{table}
}

\afterpage{\renewcommand{\arraystretch}{1.2}
\newcommand\rownumber{\stepcounter{magicrownumbers}\arabic{magicrownumbers}}
\rowcolors[]{2}{gray!15}{white}
\begin{table*}
\scriptsize	
\begin{center}
 \begin{tabular}{p{3cm} p{3cm} p{4cm} p{3cm} p{3cm}} 
 \toprule
 
\rowcolor{blue!20} \textbf{Reference - Aim} &  \textbf{Experiments}& \textbf{Results} & \textbf{Pro}& \textbf{Con} \\ [0.5ex] 
 \midrule
Li \textit{et al.}~\cite{li2009integrated} - Proposes a data-driven management framework by knowledge acquisition from historical log data.
&Log files collected from several Windows machines in a university network.
& Performs experiments on categorizing dependent and independent log events, and applies text mining techniques to categorize log messages, mines temporal data, and performs event summarization. 
& Provides a graphical representation of temporal relationship among events as an event relationship network (ERN)~\cite{perng2003data}.
&Common categories of log messages are manually determined, which can be automated from historical data.\\

Marty~\cite{marty2011cloud} - Proposes a proactive logging guideline to support forensic analysis in cloud environments. 
& N/A (the research does not provide experimentation).
& Discusses the challenges of logging in the cloud environments such as \textit{decentralization} and \textit{volatility} of logs.
& Outlines the guidelines for when the logging is required: \textit{business relevant}, \textit{operational}, \textit{security}, \textit{compliance}. 
&The guideline can be expanded to include forensic timeline analysis of logs, log review, and log correlation.\\

Li \textit{et al.}~\cite{li2017flap} - Introduces FLAP, a web-based integrated system to utilize data mining techniques for log analysis and knowledge discovery. 
& \textit{Network X} event logs at Huawei Technologies.
& Performs a case study and the results show the approach's applicability for different tasks, such as event summarization (graph) and root cause analysis.
& It provides learning-based log event extraction and provides event summarization and visualization. 
&Some of the tasks, \textit{e.g.}, root cause mining, rely on domain knowledge to manually diagnose possible problems.\\

Amar \textit{et al.}~\cite{amar2018using} - Investigate the usage of finite-state models for log differencing. 
& Mutated logs from FSA models and a user study of 60 participants.
& The proposed approaches (\textit{i.e.}, \textit{2KDiff\& nKDiff}) can expedite the process of identifying behavior differences between logs.
&The proposed models present \textit{sound}, \textit{complete}, and \textit{concise} comparisons for log differencing, and presents two algorithm: 2KDiff that compares two and nKDiff that compares multiple logs at once.
&The presented work is limited to identifying k-differences between a set of comparing logs and does not consider temporal invariants.\\

Bao \textit{et al.}~\cite{bao2019statistical} - Proposes a statistical log differencing approach, which calculates the frequencies of behaviors found in the logs.
& Controlled user study with 20 participants and log traces generated from 13 publicly available FSA models.
& The proposed approaches achieve the required guarantees of the defined statistical test with an acceptable overhead. 
& It is a follow-up work to~\cite{amar2018using} and improves by enabling developers to control the sensitivity of log differencing by setting the statistical significance value (s2KDiff and snKDiff). 
&The approach is limited to highly structured logs, which might not be the case in practice. 
Thus, helper tools are required for pre-processing and structurizing of logs. \\

Liu et al.~\cite{liu2019logzip} - Proposes a new log compression method, logzip, to allow for more effective log compression.
& Five log datasets: \textit{HDFS}, \textit{Spark}, \textit{Windows}, \textit{Android}, and \textit{Thunderbird}.
&Achieves higher log compression ratios compared to general-purpose compressors, \textit{e.g.}, \textit{bzip2}, and can generate compressed files around half of the size of general-purpose compressors.
&Performs iterative clustering with template extraction and parameter mapping and can compress in three incremental levels: L1: \textit{field extraction}, L2: \textit{template extraction}, and L3: \textit{parameter mapping}.
&The performance of the decompression should be also evaluated and compared with other compressors.\\

Yao \textit{et al.}~\cite{yao2020study} - Studies the performance of general compressors on compressing log data relative to their performance on compressing natural language data. 
&Nine system logs, such as \textit{HDFS} and \textit{LinuxSyslogs}, and two natural language (NL) data, \textit{Wiki} and \textit{Gutenberg}. 
& Reviews twelve widely used general compressors to compress nine log files collected from various software systems. The observation is that log data is more repetitive than natural language, and log data can be compressed and decompressed faster than NL with higher compression ratios.

&One of the findings is that general compressors perform better on small log sizes, and their default compression level is not optimal for log data. 
&The findings and implications of this research have not been utilized to propose a log-aware compressor.\\

Shin \textit{et al.}~\cite{shin2020effective} - Introduces \textit{LogCleaner}, which performs periodicity and dependency analyses for removing repetitive logs.
& Two proprietary and eleven publicly available log datasets including:  \textit{CVS}, \textit{RapidMiner}. 
& The approach can accurately detect and remove 98\% of the operational messages and preserve 81\% of the transactional log messages, and reduces the execution time of the \textit{model inference} task from logs.
& Segregates and only keeps transactional messages, which record the functional behavior of the system from operational messages of the system. 
&The performance of LogCleaner is heavily dependent on the quality of upstream log parser and template extraction, and requires manual analysis and domain knowledge.\\

He \textit{et al.}~\cite{he2020loghub} - Provides a repository, \textit{Loghub}, of logs from various software systems.
& Provides 17 log datasets from various application domains, including   \textit{distributed systems}, \textit{supercomputers}, and \textit{operating systems}. 
& Provides a framework for AI-powered log analysis and applies a practical usage scenario of Loghub for anomaly detection for supervised and unsupervised approaches. 
&Loghub datasets have been widely utilized for research both in academia and industry.
&There is still a shortage of labeled datasets to facilitate the evaluation of supervised log analysis tasks.\\

Chen and Jiang~\cite{chen2021survey} - Performs a survey on log instrumentation techniques.
& N/A (the research does not provide experimentation).
& Focuses on the three log instrumentation steps: \textit{logging approaches}, \textit{logging utility integration}, and \textit{logging code composition}.
& Defines four categorizes of challenges for instrumentation: \textit{usability}, \textit{diagnosability}, \textit{logging code quality}, and \textit{security compliance}.
&The research can be improved by providing a connection between \textit{logging source code} and its corresponding \textit{log messages} in the \textit{log files}.\\

Locke \textit{et al.}~\cite{locke2021logassist} - Proposes \textit{LogAssist} to assist practitioners with organizing and summarizing logs.
& Logs from one enterprise (ES) and two open source systems: \textit{HDFS} and \textit{ZooKeeper}.
& Groups logs into event sequences to extract workflows and illustrate the system's runtime execution paths. LogAssist shrinks the log events by 75.2\% to 93.9\% and the unique workflow types by 70.2\% to 89.8 in HDFS and Zookeeper logs.
& \textit{LogAssist} is able to reduce the number of log events of interest to practitioners, thus it saves time and improves the practitioners' experience in log analysis.
&In some cases, the searched keywords (\textit{e.g.}, \textit{“error”} or \textit{“exception”}) for finding problematic log lines result in a large quantity of logs for practitioners to manually review.\\

\bottomrule
\end{tabular}
\end{center}
\vspace{-3mm}
\caption{Log maintenance and management research - Topic (F).}
\label{log_management_table}
\end{table*}
}

\afterpage{\renewcommand{\arraystretch}{1.2}
\newcommand\rownumber{\stepcounter{magicrownumbers}\arabic{magicrownumbers}}
\rowcolors[]{2}{gray!15}{white}
\begin{table*}
\ContinuedFloat
\scriptsize	
\begin{center}
 \begin{tabular}{p{3cm} p{3cm} p{4cm} p{3cm} p{3cm}} 
\toprule
\rowcolor{blue!20} \textbf{Reference - Aim} &  \textbf{Experiments}& \textbf{Results} & \textbf{Pro}& \textbf{Con} \\ [0.5ex] 
 \midrule

Yao \textit{et al.}~\cite{yao2021improving} - Proposes \textit{LogBlock} a pre-compression approach to preprocess small log blocks to achieve a higher log compression ratio. 
& Evaluation on logs of 16 systems in different domains, such as \textit{Proxifier}, \textit{Android}, \textit{BGL}, and \textit{HDFS}.
& \textit{LogBlock}'s preprocessing approach improves the relative compression ratios by a median of 5\% to 21\%, while it also achieves a faster compression time.
& \textit{LogBlock} results in higher compression ratios for logs by reducing the log repetitiveness through log header preprocessing and rearranging the log content.  
&Although \textit{LogBlock} performs well for small log blocks, other compression approaches, \textit{e.g.}, \textit{Logzip} and \textit{gzip}, outperform LogBlock for larger block sizes.\\

Wang \textit{et al.}~\cite{wang2021would} - Proposes \textit{DPLOG,} an approach to assist monitoring of big data applications. 
& 1000 randomly sampled spark-related questions on StackOverflow, six Spark
benchmarks, and a user study of 20 developers. 
& \textit{DPLOG} introduces less than 10\% overhead while reducing the big data application debugging time by 63\%. 
&The approach provides the intermediate information of big data benchmarks, which allows the developers to better identify an issue's root cause for the chained methods and lazy evaluations.
&DPLOG can be improved by including visualization and customizability of the recorded information.\\

\bottomrule
\end{tabular}
\end{center}
\vspace{-3mm}
\caption{Log maintenance and management research - Topic (F) (continued).}
\label{log_management_table}
\end{table*}
}

\subsubsection{Category F: Log Maintenance and Management}
As the size of logs increases, the job of methods and tools which manage and maintain logs becomes more crucial, cumbersome, and of value. 
For example, FLAP~\cite{li2017flap} provides an end-to-end platform for log collection, maintenance, and analysis.  
In the following, we review \textit{log collections}, \textit{log compression}, \textit{log rolling}, and \textit{log removal}, and Table~\ref{log_management_table} summarizes the research in this category. 

\textbf{Log Collections.} The aim of maintaining a log collection is for auditing~\cite{lee2013loggc} or enabling benchmarking for different types of log analysis~\cite{he2018identifying,farshchi2018metric,mavridis2017performance}. 
For example, Loghub~\cite{he2020loghub} provides a repository of logs from various software systems, and Cotroneo \textit{et al.}~\cite{urlopenstackdata} have released an OpenStack failure dataset containing injected faults. 
The logs are used in various prior works for evaluating tasks such as compression techniques~\cite{yao2020study,skibinski2007fast}, failure analysis~\cite{xu2014pod,yuan2010sherlog,he2018identifying} and bug detection~\cite{cotroneo2019bad}. 
We observe that although execution logs of different systems are conveniently available, it is difficult to find large-scale collections of labeled datasets, which are especially crucial for supervised learning of log mining tasks~\cite{xu2009detecting}. 
This is because manually labeling large datasets of logs is quite cumbersome. 
Thus, we see significant value in curating a database of labeled logs (\textit{e.g.}, normal vs. anomaly/failure log records), and the development of automated log labeling techniques~\cite{yang2021semi}.   

\textbf{Log Compression.} With the continued growth of large-scale software systems, they tend to generate larger volumes of log data every single day, which makes the analysis of logs challenging. 
As such, to cope and contain this challenge, developers and practitioners apply tools for compression and continuous archiving of logs~\cite{yao2020study,liu2019logzip}. 
Hassan et al.~\cite{hassan2008industrial} applied log compression to extract common usage scenarios. 
Yao \textit{et al}.~\cite{yao2020study} studied the performance of general compressors on compressing log data relative to their performance on compressing natural language data. 
Their work reviews 12 widely used general compressors to compress nine log files collected from various software systems. 
Because log files generally benefit from higher repetition than natural text~\cite{yao2020study}, there is an avenue of outstanding work to develop log-aware compression techniques, that consider log characteristics in their algorithms and their parameter selections and achieve a higher compression/decompression performance.

\textbf{Log Rolling.} As log data generally grows rapidly during the system's execution~\cite{lemoudden2015managing,yao2020study}, logging libraries such as Logback~\cite{urllogback}, Log4j/2~\cite{urllog4x}, and SELF4J~\cite{urlslf4j} often support the continuous rolling \textit{i.e.}, archiving of log files as new logs become available. 
For example, as the size of the generated log goes beyond a user-defined value on the storage or a specific time interval has passed (\textit{e.g.}, daily, weekly), a Log4j Appender~\cite{urlslf4japp} can zip, rename, and store the log with a timestamp, \textit{e.g.}, ``\texttt{logs/app-\%d{MM-dd-yyyy}.log.gz}''. 
Log archiving helps with the long-term maintenance and organization of logs. 
There is certainly room for further research on improving and automating log archiving policies and techniques. 

\textbf{Log Removal.} Although logs are useful, but due to the large volume of them, they can become noisy, hard to analyze, and cause inaccuracy in log analysis. 
As such, prior research has aimed to detect and remove duplicate log messages~\cite{shin2020effective,conforti2016filtering,sun2019filtering}. 
For example, Shin \textit{et al.}~\cite{shin2020effective} introduced \textit{LogCleaner}, which performs \textit{periodicity} and \textit{dependency} analyses to filter out and remove periodic and dependent log messages. 
In sum, the approaches that are applied for log maintenance and management facilitate automated analysis of logs, and furthermore, will yield more accurate log analysis. 
Table~\ref{log_management_table_OW} provides an overview of the log maintenance and management research, and Table~\ref{log_management_table} provides additional details.

\begin{tcolorbox}[breakable, enhanced]
\small \textbf{Finding} \textbf{10.} \textit{
In sum, log collections, log compression, log rolling, and log removal are the pillars of log maintenance and management.  
Prior studies have investigated the approaches to compress, clean, and summarize logs to help developers and practitioners achieve more efficient interactions with logs.}
\end{tcolorbox}

\subsubsection{Automated Log File Analysis - Challenges and Motivation}
Logs record system runtime information and are widely used and examined for assessing the systems' health and availability~\cite{shilin2016expr}. 
Traditionally, developers or operators often inspect the logs manually with keyword searches (\textit{e.g.}, ``fail'', ``exception'') and rule matching (\textit{e.g., ``grep $<$RegEx$>$''}) to find any potential problems in case of a system failure.
However, manual or keyword inspection of log files becomes impractical with the ever-increasing complication of software systems because of the following reasons~\cite{shilin2016expr}: 
\begin{enumerate}[label=\protect\circled{\arabic*}]
\item As current computer systems generate a massive volume of logs, \textit{e.g.}, at the rate of 50 gigabytes per hour (around 120$\sim$200 million log lines) on Alibaba Cloud Computing Mailing System (Aliyun Mail)~\cite{mi2013toward}, this makes it close to impossible to manually extract useful information from the log files and track down any system issues.   

\item The complex and concurrent nature of software systems makes it unmanageable for a single developer to comprehend the entire functionality of the system, as a single developer might be only responsible for the development of a small sub-module of the entire project.   
For example, hundreds of developers take part in the development of parallel computing platforms, such as Apache Spark~\cite{urlspark1}; thus makes it quite challenging, if not impossible, for a single developer to pin down an issue from concurrent and massive log files.

\item Parallel and distributed software systems generally apply various methods of fault-tolerant and performance optimization techniques in order to recover from a hardware failure or perform load-balancing and scheduling. 
For example, a resource manager daemon, \textit{e.g.}, on a Hadoop or Spark cluster, may intentionally terminate a running application and move it to another node in the system in order to expedite the execution of that task. 
As a result, the traditional and manual way of searching in the log files for keywords such as \textit{killed, terminated, failure} might not be useful and can lead to multiple false-positive cases~\cite{lin2016log} and further muddle the manual inspection. 
\end{enumerate}

Moreover, although automatic log analysis helps developers and practitioners significantly to speed up the process (\textit{e.g.},~\cite{zhao2016non,zhang2019robust,xu2014pod}), the automatic log analysis itself is still very challenging because log messages are usually unstructured free-form text strings, and application behaviors are often very heterogeneous and complicated~\cite{fu2009execution}.
As a result, effective automated log analysis methods are well sought after. 
To enable automatic log analysis, the very first step is \textbf{log parsing, Category G}, followed by applications of automated log analysis, \textit{i.e.}, \textbf{anomaly detection, Category H}, \textbf{runtime behavior, Category I}, and \textbf{performance, failure, and fault diagnosis, Category J}, that we review in the following.

\begin{figure}
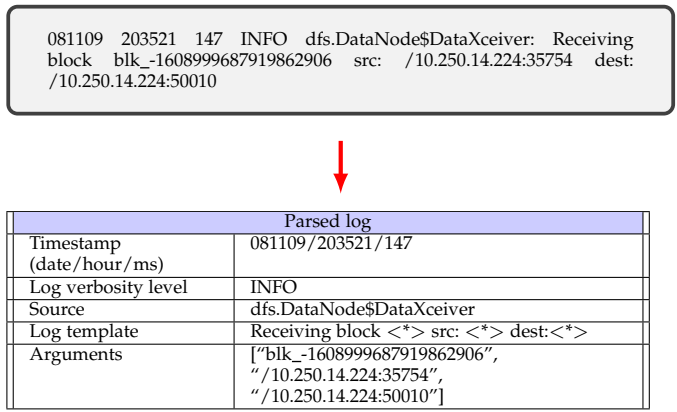

\small
\begin{tcolorbox}\scriptsize
081109 203521 147 INFO dfs.DataNode\$DataXceiver: Receiving block blk\_-1608999687919862906 src: /10.250.14.224:35754 dest: /10.250.14.224:50010
\end{tcolorbox}
\begin{center}
\textbf{$\DownArrow[20pt][>=latex,red, ultra thick]$}\\
\end{center}
\scriptsize
\begin{tabular}{ ||p{2.5cm}|p{5cm}|| }
\hline
\multicolumn{2}{||c||}{\cellcolor{blue!20}Parsed log} \\
\hline
Timestamp (date/hour/ms)    & 081109/203521/147 \\
\hline
Log verbosity level & INFO \\
\hline
Source & dfs.DataNode\$DataXceiver \\
\hline
Log template & Receiving block $<$*$>$ src: $<$*$>$ dest:$<$*$>$  \\
\hline
Arguments& [``blk\_-1608999687919862906'', ``/10.250.14.224:35754'', ``/10.250.14.224:50010'']  \\
\hline
\end{tabular}
\caption{Log parsing for a raw log message to a parsed log from HDFS logs~\cite{xu2009detecting}.}
\label{logs_parsing}
\vspace{-3mm}
\end{figure}

\afterpage{\renewcommand{\arraystretch}{1.2}
\newcommand\rownumber{\stepcounter{magicrownumbers}\arabic{magicrownumbers}}
\rowcolors[]{2}{gray!15}{white}
\begin{table}[h]
\scriptsize	
\begin{center}
  \begin{tabular}{p{1.2cm} p{.8cm} p{2cm} P{1.4cm}P{.7cm}}
 \toprule
 
\rowcolor{blue!20} \textbf{Reference} &\textbf{Tool} & \textbf{Approach} & \textbf{Systems} & \textbf{Org.}\\ [0.5ex] 
 \midrule

Qiu~\cite{qiu2010happened} \textit{et al.} & \textit{Syslog Digest}& word frequency &Syslogs& AT\&T\\

Taerat \textit{et al.}~\cite{taerat2011baler} & \textit{Baler}& token attributes &4 (BGL, Liberty)& N/A\\

Tang \textit{et al.}~\cite{tang2011logsig} & \textit{LogSig}& term pair and \textit{k} partition &5 (FileZilla, ThunderBird)& N/A\\

Vaarandi \textit{et al.}~\cite{vaarandi2015logcluster} & \textit{LogCluster}& associate rule mining and clustering &6 OS +1 CS& \cite{urllogcluster}\\

Du \textit{et al.}~\cite{du2016spell}  & \textit{Spell}& longest common subsequence&2 (HPC, BGL)& \cite{urlspell}\\

He \textit{et al.}~\cite{he2017drain}  & \textit{Drain}&  fixed-depth parse tree&5 (BGL, HDFS)& \cite{urldrain}\\

Messaoudi \textit{et al.}~\cite{messaoudi2018search}  & \textit{MoLFI}&  
search-based evolutionary approach&6 (HDFS, Proxifier)& \cite{urlmolfi}\\

Dia~\cite{dai2020logram}  & \textit{Logram}&  n-grams dictionaries&16 (Andriod, Apache)& N/A\\

\bottomrule
\end{tabular}
\end{center}
\vspace{-2mm}
\caption{\footnotesize Comparison of log parsing research - Topic (G). Systems: number of systems evaluates and some examples; Org.: Organization or link if available.}
\label{log_parsing_table_OW}
\end{table}
}

\afterpage{\renewcommand{\arraystretch}{1.2}
\newcommand\rownumber{\stepcounter{magicrownumbers}\arabic{magicrownumbers}}
\rowcolors[]{2}{gray!15}{white}
\begin{table*}
\scriptsize	
\begin{center}
 \begin{tabular}{p{3cm} p{3cm} p{4cm} p{3cm} p{3cm}} 
 \toprule
 
\rowcolor{blue!20} \textbf{Reference - Aim} &  \textbf{Experiments}& \textbf{Results} & \textbf{Pro}& \textbf{Con} \\ [0.5ex] 
 \midrule
Qiu~\cite{qiu2010happened} \textit{et al.} - Designs \textit{SyslogDigest} that extracts log events from the router's syslogs.
& Syslog data from two large operational networks: a tier-1 ISP backbone network and a nationwide commercial IPTV backbone network.
& It combines an \textit{offline} and \textit{online} domain knowledge learning components automatically extracts relevant domain knowledge from raw syslog data. The authors showcase the applications in network \textit{troubleshooting}, and \textit{health monitoring and visualization}.  
& Applies associated rule mining, then transforms and compresses low-level raw syslog messages into their prioritized high-level events.
&The work is limited only to event template extraction from a specific system, \textit{i.e.}, syslog messages.\\

Taerat \textit{et al.}~\cite{taerat2011baler} - Introduces \textit{Baler}, a token-based log parsing tool.
& Logs of four supercomputers: \textit{BG/L}, \textit{Liberty}, \textit{Spirit}, and \textit{Tbird}.
& For clustering, Baler relies on token attributes rather than frequency or entropy of token positions that are applied in other log parsers. Baler handles large datasets better than compared tools and more efficiently, \textit{i.e.}, faster execution time. 
& Requires only one pass to cluster log messages based on their event templates. 
&Baler relies on the user to provide a dictionary of words.\\

Tang \textit{et al.}~\cite{tang2011logsig} - Proposes \textit{LogSig}, message signature based algorithm to generate events from textual log messages.
& Logs of five real-world systems, including \textit{FileZilla}, \textit{PVFS2}, and \textit{Hadoop}. 
& It searches for the most frequent message signatures and then categorizes them into a set of event types. LogSig performs better in \textit{the quality of the generated log events} (F-Measure) and \textit{scalability} when compared to prior work. 
& LogSig converts each log line into a set of ordered token pairs and then partitions log messages into \textit{k} different groups based on the extracted term pairs.
&LogSig has a prolonged execution time on large log datasets and reaches low accuracy on the BGL data~\cite{he2016evaluation}.\\

Vaarandi \textit{et al.}~\cite{vaarandi2015logcluster} - Presents \textit{LogCluster}, a data clustering and pattern mining algorithm for textual log lines.
& A set of six system logs from a large national institution, including database systems, mail servers, and firewall logs.  
& LogCluster improves SLCT~\cite{vaarandi2003data} such that each Cluster $C_i$ is uniquely identified by its pattern $P_i$, and each pattern consists of words and wildcards, which makes it insensitive to word shifts. LogCluter performs more accurate clustering and finds fewer groups compared to SLCT. 
&The \textit{support} of a cluster is calculated as the number of elements in that cluster: $supp(p_i)= supp(C_i)= |C_i|$. 
Finally, clusters with support of equal or higher than a threshold value, \textit{s}, are selected. 
& The algorithm requires a two-pass process to categorize the list of frequent patterns.\\

Du \textit{et al.}~\cite{du2016spell} - Proposes \textit{Spell}, an online log parsing method based on longest common subsequence (LCS).
& Two supercomputer logs: \textit{Los Alamos HPC log} and \textit{BlueGene/L log}.
& Parses unstructured log messages into structured events types and parameters in an unsupervised streaming fashion with linear time complexity. Spell with pre-filtering has a faster computation time and achieves a higher accuracy compared to prior work.
& The LCS approach achieves a faster template searching process by enabling subsequent matching and prefix trees.
& The prefix tree depth can grow arbitrarily without limitation, which can lead to lengthy computation time on large datasets.\\

He \textit{et al.}~\cite{he2017drain} - Proposes \textit{Drain}, a fixed-depth tree-based online log parsing method. 
& Five real-world data sets: \textit{BGL}, \textit{HPC}, \textit{HDFS}, \textit{Zookeeper}, and \textit{Proxifier}.
& Constructs a tree data structure and groups the logs that belong to similar log events (\textit{i.e.}, templates) into the same \textit{leaf node} of the tree. The approach achieves higher or equal accuracy and obtains 51.85\% to 81.47\% faster runtime compared to other online log parsers.
& Drain is specifically useful for web services as it enables log parsing in a streaming manner, and evaluation shows that Drain outperforms prior offline and online log parsing approaches.
&It appears that Drain does not fully handle positional shifts in the log templates, and log messages that belong to the same log event but have different lengths will be grouped separately.\\

Messaoudi \textit{et al.}~\cite{messaoudi2018search} - Introduces \textit{MoLFI} (Multi-objective Log message Format Identification), which leverages an evolutionary algorithm for log message format identification.
& Six real-world datasets: \textit{HDFS}, \textit{BGL}, \textit{HPC}, \textit{Zookeeper}, \textit{Proxifier}, and one industrial software logs.
& MoLFI achieves a higher performance than the compared alternative algorithms in detecting the correct log message templates.
& Formulates the log template identification task as a
\textit{multi-objective optimization problem} and propose a search-based solution based on the NSGA-II algorithm~\cite{deb2002fast}, \textit{i.e.}, a sorting genetic algorithm.
& MolFI suffers from low efficiency (i.e., high execution time) on large datasets as its iterative genetic algorithm, NSGA-II, is computationally intensive~\cite{zhu2019tools,el2020systematic}.\\

Dia~\cite{dai2020logram} - Introduces \textit{Logram}, which uses \textit{n-grams dictionaries} to perform log parsing. Logram initially calculates the number of appearance of each n-gram (\textit{i.e.}, token) in the log file. 
& Eventuated on 16 publicly available logs including \textit{Android, BGL, HDFS, Spark, and Zookeeper} logs. 
& Achieves a higher parsing accuracy than the prior work, and it is 1.8X to 5.1X faster than in its calculation when compared to prior work.
& Based on the threshold of an n-gram occurrence, Logram decides dynamic and static parts of the log messages and extracts the log templates. 
&One caveat for this log parser is the threshold selection for n-gram appearance, and if a dynamic n-gram occurs frequently, it might be mistakenly picked as a part of the log template.\\
\bottomrule
\end{tabular}
\end{center}
\vspace{-3mm}
\caption{Log parsing research - Topic (G).}
\label{log_parsing_table}
\end{table*}
}

\subsubsection{Category G: Log Parsing}\label{log_parsing}
Log parsing is the process of converting the free-form text format of log files to structured events. 
Figure~\ref{logs_parsing} provides an example of a raw log message from a log file that is parsed to its individual elements. 
Each log message is printed by a logging statement in the source code, which records a specific system event. 
Then, a log parser applies techniques to convert the free-form text format of the log messages to a structured format, as presented in Figure~\ref{logs_parsing}. 
More specifically, the log parser can extract useful information, such as timestamp, log verbosity level, variable arguments, and log template.
The goal of the log parser is to convert each log message to a specific log template (\textit{e.g.}, {Received block $\langle*\rangle$ src: $\langle*\rangle$ dest: $\langle*\rangle$} in Figure~\ref{logs_parsing}). 
Ideally, there is a corresponding logging statement in the source code for each extracted log template, \textit{e.g., log.info(``Received block $\%s$ src: $\%s$ dest: $\%s$'', obj.blk\_id, obj.src, obj.dest)}. 
The better the log parser can match the log templates with actual log printing statements in the source code, the merrier the quality of log parsing, and consequently, the more accurate log analysis tasks that follow.

\begin{figure*}
\vspace{-1mm}
\centering
\includegraphics[scale=.6]{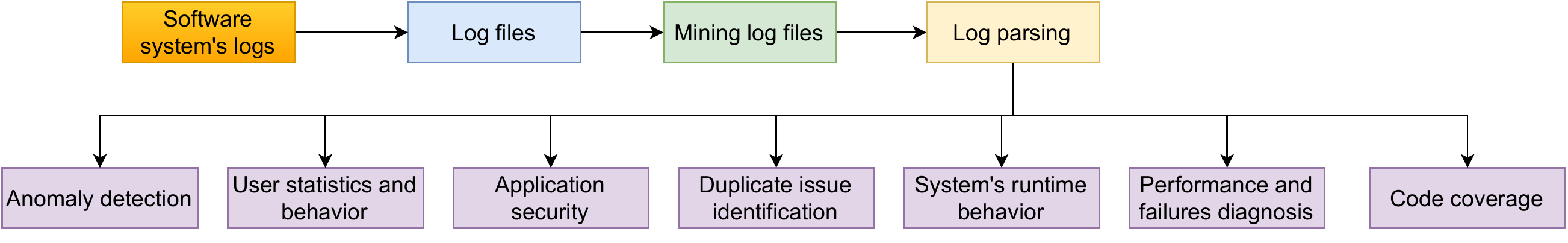}
\caption{Mining of log files for different applications.}
\label{app_logs}
\vspace{-3mm}
\end{figure*}

Log parsing, traditionally, started with defining manual regular expressions to extract log templates and arguments. 
However, this approach alone is no longer efficient due to the huge number of log templates as well as their continuous evolution~\cite{zhu2019tools}.
For example, Xu \textit{et al.}~\cite{xu2009detecting} explained that in a Google's system, on average, thousands of new logging statements are added every month. 
Therefore, automating the log parsing process is very well obliged.   
Some studies have also proposed, as an alternative, the use of static methods to curate the software's source code and extract log patterns directly from the logging statements within the source code~\cite{xu2009detecting,nagappan2009efficiently}. 
These approaches are only useful if the source code of the system is available. 
To extend the application of log parsing to the scenarios that the source code is not available, \textit{e.g.}, proprietary software, other studies have proposed various data mining approaches to extract log templates from the log files instead, such as frequent pattern mining in SLCT~\cite{vaarandi2003data} and its extension, LogCluster~\cite{vaarandi2015logcluster}, iterative partitioning in IPLoM~\cite{makanju2009clustering}, clustering in LKE~\cite{fu2009execution}, longest common subsequence in Spell~\cite{du2016spell}, search-space multi-objective optimization approach in MoLFI~\cite{messaoudi2018search}, parsing trees in Drain~\cite{he2017drain}, and n-gram dictionary-based in Logram~\cite{dai2020logram}.
Contrary to the regular-expression-based and static analysis methods, these techniques are capable of extracting log templates from log files without needing access to the source code.  
After the logs are parsed, they are used for various applications, such as anomaly detection and failure diagnosis. 
Zhu et al.~\cite{zhu2019tools} presented a log parsing benchmark available here~\cite{urllogparserlogpai}, and  El-Masri \textit{et al.}~\cite{el2020systematic} performed a survey of log abstraction techniques.  
Table~\ref{log_parsing_table_OW} provides an overview of log parsing research, and Table~\ref{log_parsing_table} compares and summarizes the related research. 
Figure~\ref{app_logs} categorizes various trends and goals for log file analysis after log parsing is applied.

\begin{tcolorbox}[breakable, enhanced]
\small \textbf{Finding} \textbf{11.} \textit{
In sum, prior studies have applied various heuristics and approaches for automated parsing of numerous systems logs. These approaches include: \begin{enumerate*}[label=\protect\circled{\arabic*}] 
\item associate rule mining, 
\item token frequency and k partitioning,
\item parse trees,
\item search-based evolutionary, and
\item n-gram dictionaries,
\end{enumerate*}. 
The higher performance of log parsing, as the precursor step, can enable more effective downstream log mining tasks. }
\end{tcolorbox}

\afterpage
{\renewcommand{\arraystretch}{1.2}
\newcommand\rownumber{\stepcounter{magicrownumbers}\arabic{magicrownumbers}}
\rowcolors[]{2}{gray!15}{white}
\begin{table}
\scriptsize	
\begin{center}
 \begin{tabular}{p{1.3cm} p{2.5cm} p{2cm} P{.8cm} } 
 \toprule
 
\rowcolor{blue!20} \textbf{Reference} &\textbf{Category} & \textbf{Systems} & \textbf{Org.} \\ [0.5ex] 
 \midrule

Xu \textit{et al.}~\cite{xu2009detecting}& Principal Component Analysis & HDFS \& Darkstar &ACA \cite{urlpca}\\

Fu \textit{et al.}~\cite{fu2009execution}&  finite state automaton (FSA)&  Hadoop\& SILK &IND/ ACA \\

Lou \textit{et al.}~\cite{lou2010mining}&  invariant mining&  Hadoop, Microsoft CloudDB &IND/ ACA \cite{urlim}\\

Chuah \textit{et al.}~\cite{chuah2013linking}& resource usage and log correlation&    ratlogs, syslogs &IND/ ACA \\

Du \textit{et al.}~\cite{du2017deeplog}& online LSTM-based approach,  \cite{urldeeplog} & HDSF, OpenStack &ACA\\

Bertero \textit{et al.}~\cite{bertero2017experience}& NLP and word2vec& Syslogs &ACA\\

Bao \textit{et al.}~\cite{bao2018execution}& probabilistic suffix tree & HDFS, CloudStack &ACA\\

Farshchi \textit{et al.}~\cite{farshchi2015experience,farshchi2018metric}& regression-based statistical approach & AWS &ACA\\

Meng \textit{et al.}~\cite{meng2019loganomaly}& NLP and template2Vec & BGL, HDFS &ACA\\

Zhang \textit{et al.}~\cite{zhang2019robust}& semantic vectors of logs & HDFS, Service X&ACA/ IND\\

Zhang \textit{et al.} \cite{zhang2020anomaly}& numerical workflow relations& BGL, HDFS &ACA\\

Huang \textit{et al.}~\cite{huang2020hitanomaly}& transformer-base DL model& BGL, HDFS, OpenStack &ACA\\

Zhou \textit{et al.}~\cite{zhou2020logsayer} & pattern extraction& HDFS, OpenStack &ACA\\

Chen \textit{et al.}~\cite{chen2020logtransfer} & transfer learning and word embedding (GloVe)& HDFS, Hadoop, proprietary log &IND/ ACA \\

Yang \textit{et al.} 2021 \cite{yang2021semi} & semi-supervised with probabilistic labeling& HDFS, BGL &ACA\\

Le \textit{et al.} 2021 \cite{le2021log} & transformer-based anomaly detection model & HDFS, BGL, Thunderbird, Spirit   &ACA \cite{urlneurallog}\\

\bottomrule
\end{tabular}
\end{center}
\vspace{-2mm}
\caption{\footnotesize Comparison of log anomaly detection research - Topic (H). ACA: academic project, IND: industrial project; Systems: studies systems' logs. Systems: number of systems evaluates and some examples; Org.: Organization or link if available.}
\label{log_anomaly_detection_table_OW}
\end{table}
\renewcommand{\arraystretch}{1.2}
\rowcolors[]{2}{gray!15}{white}
\begin{table}
\scriptsize	
\begin{center}
 \begin{tabular}{p{1.3cm} p{2.5cm} p{2cm} P{.8cm} } 
 \toprule
 
\rowcolor{blue!20} \textbf{Reference} &\textbf{Category} & \textbf{Systems} & \textbf{Org.} \\ [0.5ex] 
 \midrule

Tang \textit{et al.}~\cite{tang2010approach}& service composition patterns & 74 service-oriented applications & ACA\\

Oliner and Aiken~\cite{oliner2011online}& component interaction in large scale systems& eight system logs & ACA\\

Fu \textit{et al.}~\cite{fu2012logmaster}& LogMaster, event correlation from logs  & Hadoop, HPC, BlueGene& ACA \\

Shang \textit{et al.}~\cite{shang2013assisting}& log analytics for cloud environment Analytics  & three Hadoop-based applications& ACA \\

Busany and Maoz\cite{busany2016behavioral}& statistical log analysis & four FSA from industrial logs& ACA\\

Awad and Menasc\'e~\cite{awad2016performance}& performance model extraction through logs & Apache Tomcat access logs& ACA\\

Di \textit{et al.}~\cite{di2017logaider,di2018exploring} & potential correlation detection among system events  & BlueGene/Q Mira supercomputer logs& ACA\\

He et al. \textit{et al.}~\cite{he2018identifying} & clustering-based approach to detect potential system issues  & an online large-scale Service X from Microsoft&ACA \cite{urlog3c} \\

\bottomrule
\end{tabular}
\end{center}
\vspace{-2mm}
\caption{\footnotesize Comparison of system's runtime behavior research - Topic (I).}
\label{log_runtime_behavior_table_OW}
\end{table}
}

\afterpage{\renewcommand{\arraystretch}{1.2}
\newcommand\rownumber{\stepcounter{magicrownumbers}\arabic{magicrownumbers}}
\rowcolors[]{2}{gray!15}{white}
\begin{table}[h]
\scriptsize	
\begin{center}

\begin{tabular}{p{1.3cm} p{2.5cm} p{2cm} P{.8cm} } 
\toprule
 
\rowcolor{blue!20} \textbf{Reference} &\textbf{Category} & \textbf{Systems} & \textbf{Org.} \\ [0.5ex] 
 \midrule 
Cinque \textit{et al.}~\cite{cinque2010assessing}& fault-injection approach for log-based recovery & Apache Web Serve, MySQL, TAO Database & ACA\\

Chuah \textit{et al.}~\cite{chuah2010diagnosing} & failure root cause discovery  with log events & supercomputer logs, \textit{e.g.}, BGL RAS, Ranger& ACA\\

Yuan \textit{et al.}~\cite{yuan2010sherlog}  & SherLog, finding probable failure's execution path & eight real-world software failures & ACA\\

Pecchia \textit{et al.}~\cite{pecchia2012detection}  & impactful factors from logs for failure detection & Apache Web Server, TAO Open DDS, MySQL DBMS & ACA\\

Nagaraj \textit{et al.}~\cite{nagaraj2012structured}  & DISTALYZER, log data for diagnosing performance problems &  TritonSort, HBase, BitTorrent & ACA\\

Fronza \textit{et al.}~\cite{fronza2013failure}  & failure classification through logs & logs of a European enterprise & IND\\

Syer \textit{et al.}~\cite{syer2013leveraging}  & logs and performance counters for memory diagnosis & Hadoop and an enterprise system & ACA/ IND\\

Zhao \textit{et al.}~\cite{zhao2014lprof}  & reconstructing execution flow for performance diagnosis  &  distributed systems, \textit{e.g.}, \textit{HDFS} and \textit{Yarn} & ACA \\

Xu \textit{et al.}~\cite{xu2013detecting,xu2014pod}  & logs for cloud rolling updates faults discovery   &  Amazon Web Services (AWS) & ACA\\

Russo \textit{et al.}~\cite{russo2015mining}  & error predictors from system logs with SVM   &  software system for telemetry and performance of cars & IND \\

Yu \textit{et al.}~\cite{yu2016cloudseer}  & CloudSeer, an  approach for monitoring of interleaved logs   &  OpenStack &  ACA\\

Gurumdimma \textit{et al.}~\cite{gurumdimma2016crude}  & CRUDE, combines console logs with resource usage for error detection  &  supercomputer ratlogs &  ACA\\

Zhao \textit{et al.}~\cite{zhao2016non}  & Stitch, a distributed performance profiler with flow reconstruction  & \textit{Hive}, \textit{Spark}, and \textit{OpenStack}  &  ACA\\

Zou \textit{et al.}~\cite{zou2016uilog}  & UiLog, log-based fault analysis with various components logs & \textit{StrongCloud} logs& ACA/ IND  \\

Zhang \textit{et al.}~\cite{zhang2017pensieve}  & Pensieve, flow reconstruction tool for performance failure reproduction & \textit{HDFS}, \textit{HBase}, \textit{ZooKeeper}, and \textit{Cassandra} & ACA  \\
\bottomrule
\end{tabular}
\end{center}
\vspace{-2mm}
\caption{\footnotesize Comparison of performance, fault, and failure diagnosis research - Topic (J). ACA: academic project, IND: industrial project; Systems: studies systems' logs. Systems: number of systems evaluates and some examples; Org.: Organization or link if available.}
\label{log_performance_failure_table_OW}
\end{table}
}

\afterpage{\renewcommand{\arraystretch}{1.2}
\newcommand\rownumber{\stepcounter{magicrownumbers}\arabic{magicrownumbers}}
\rowcolors[]{2}{gray!15}{white}
\begin{table}[h]
\scriptsize	
\begin{center}
  \begin{tabular}{p{1.3cm} p{2.5cm} p{2cm} P{.8cm} } 
 \toprule
\rowcolor{blue!20} \textbf{Reference} &\textbf{Category} & \textbf{Systems} & \textbf{Org.} \\ [0.5ex] 
 \midrule 

Lee \textit{et al.}~\cite{lee2012unified}& user behavior and analytics from session logs  & Twitter sessions & IND\\

Lim \textit{et al.}~\cite{lim2014identifying}& identification of recurrent and unknown performance issues & (TPC-W) Benchmark~\cite{urltpc} and System X & IND\\

Oprea \textit{et al.}~\cite{oprea2015detection}& log analysis for enterprise network& confidential DNS logs & IND\\
 
Barik \textit{et al.}~\cite{barik2016bones}& log utilization for business decision making & Microsoft logging platform & IND\\

Chen \textit{et al.}~\cite{chen2018automated}& LogCoCo, code coverage analysis with logs  & commercial software from Baidu and an OS project & IND/ ACA\\

\bottomrule
\end{tabular}
\end{center}
\vspace{-2mm}
\caption{\footnotesize Comparison of user, business, security, and code coverage research - Topic (K).}
\label{log_user_business_table_OW}
\end{table}
}

\afterpage{\renewcommand{\arraystretch}{1.2}
\newcommand\rownumber{\stepcounter{magicrownumbers}\arabic{magicrownumbers}}
\rowcolors[]{2}{gray!15}{white}
\begin{table*}
\scriptsize	
\begin{center}
 \begin{tabular}{p{3cm} p{3cm} p{4cm} p{3cm} p{3cm}} 
 \toprule

\rowcolor{blue!20} \textbf{Reference - Aim} &  \textbf{Experiments}& \textbf{Results} & \textbf{Pro}& \textbf{Con} \\ [0.5ex] 
 \midrule

Xu \textit{et al.}~\cite{xu2009detecting} - Applies the Principal Component Analysis (PCA) method to find unusual patterns in logs and identifies log segments that are likely to indicate runtime anomalies and system problems. 
& Logs of the \textit{Darkstar online game server} and \textit{HDFS}.
& PCA extracts \textit{k} principal components by finding the axes with the highest variance among high-dimensional data. The approach can detect anomalous logs with high accuracy and few false positives while being efficient in its computation time.
&For anomaly detection with PCA, two subspaces, \textit{i.e.}, normal, $S_n$, and abnormal, $S_a$, are created. 
$S_n$ is created with the first k principal components, and $S_a$ is constructed with the remaining components. 
&Relies heavily on log parsing step to extract log structure from the logs and detect event sequences, and will fail if log messages do not follow the preferred structure.\\

Fu \textit{et al.}~\cite{fu2009execution} - Introduces a method to detect anomalies by converting unstructured log files to log keys.
& Two distributed systems: \textit{Hadoop} and \textit{SILK}.
&The research learns a finite state automaton (FSA) from the training set log keys to model the normal behavior of the system. The results show that the approach can detect system issues, such as workflow errors. 
& With the FSA and a performance model, the authors can identify anomalies in newly generated log files. The work detects three types of anomalies: 1) \textit{work flow error}, 2) \textit{transition time low performance}, and 3) \textit{loop low performance}.  
& The approach does not work properly for \textit{loop low performance} detection, results in false positives, and it is workload dependent.\\
 
Lou \textit{et al.}~\cite{lou2010mining} - Proposes an approach to detect anomalies by mining program invariants (IM) that have a clear physical manifestation.
& Experiments on \textit{Hadoop} and \textit{Microsoft CloudDB}. 
& It detects anomalies if the new logs break certain invariants, \textit{e.g.}, if an ``open file'' log message appears without observing a ``close file'', this invariant is violated and an anomaly is detected. Generally, produces fewer false-positive cases compared to the PCA-based approach. 
& Improved upon the PCA-based method~\cite{xu2009detecting}, this approach provides the operators with intuitive insight (\textit{i.e.}, what invariant is breached) on anomalies, and, hence, facilitate faster anomaly track down. 
& This approach is not able to detect anomalies that no invariant is broken, \textit{e.g.}, many files are opened and closed continuously.\\

Chuah \textit{et al.}~\cite{chuah2013linking} - Presents \textit{ANCOR} that connects resource usage anomalies with system problems with logs.
& \textit{Ranger} supercomputer logs in two formats: \textit{syslogs} and \textit{ratlogs} (rationalized logs)~\cite{hammond2010end}.
& Evaluates the effectiveness of three different algorithms, \textit{PCA}, \textit{ICA}, and \textit{Mahalanobis distance}. The results reveal a list of events with a strong correlation with system problems, such as \textit{soft lockup}. 
& Performs anomaly and correlation analyses to detect the cluster nodes and jobs that are associated with the extra system resource usage that lead to system failures.
& The approach cannot detect system problems that are not manifested as extra resource usage.\\

Du \textit{et al.}~\cite{du2017deeplog} - Presents \textit{DeepLog}, an online LSTM-based approach, to model system log files as natural language sequences.
& \textit{HDSF} and \textit{OpenStack} log datasets.
& DeepLog decodes the log message, including timestamp, log key, and parameter values, and applies both \textit{deep learning} and \textit{density clustering} approaches. The approach outperforms PCA and IM methods and produces a lower number of false-positive and false-negative cases. 
& Learns log patterns from the normal execution and constructs workflows, and detects anomalies when running log patterns deviate from the normal execution.
& DeepLog is evaluated on systems with highly regulated logs and with a limited log key space, \textit{i.e.}, Hadoop (29 keys) and OpenStack (40 key)~\cite{shen2018tiresias}.\\

Bertero \textit{et al.}~\cite{bertero2017experience} - Leverages natural language processing techniques for anomaly detection. 
&660 syslog log files, half of them (330 files) for normal system executions, and the other half are abnormal runs. 
& Explores the performance of three learning classifiers, \textit{i.e.}, \textit{Naive Bayes}, \textit{Random Forest} (RF), and \textit{Neural Networks}, and evaluates their performance on predicting normal versus stressed (\textit{i.e.}, abnormal) log files. RF has the best performance. 
& Applies a word embedding technique, \textit{i.e.}, \textit{word2vec}, to map log message words to metric space, and then utilizes machine learning classifiers to summarize log files to single points. 
&The approach is limited to supervised learning and requires the pre-labeling of log files.\\

Bao \textit{et al.}~\cite{bao2018execution} - Utilizes both the source code analysis and the log file mining for anomaly detection. 
&  \textit{CloudStack} and \textit{HDFS} logs.
& It presents a probabilistic suffix tree-based statistical approach to detect anomalies from console logs. The results show the proposed approach can detect the largest number of anomalies compared to prior work. 
& The source code analysis employs control flow and log statement analysis to extract \textit{``schema''} for the subsequent log parsing stage. 
& For feature extraction, the approach only takes into consideration the number of occurrences of an event and does not consider the sequential relationship of the traces. 
\\

Farshchi \textit{et al.}~\cite{farshchi2015experience,farshchi2018metric} - Proposes a regression-based statistical approach to correlate operation behavior with cloud metrics.
& Experiments on \textit{Amazon Web Services} (AWS) logs.
& For anomaly evaluation, the work injects faults in 22 iterations of a rolling upgrade task and utilizes the learned model for fault prediction. 
Two-minute-time-window (2 mTW) metric observation prior to the anomaly achieves the highest F-Measure. 
&The authors utilize a regression-based method to detect the most statistically significant metrics for anomaly detection and observe cloud metrics changes and signal anomalies in case of divergence.
& The evaluation is performed with synthetically injected faults. Further evaluation on actual system faults can help to better validate the approach.\\

Meng \textit{et al.}~\cite{meng2019loganomaly} - Proposes \textit{LogAnomaly}, an approach to model log messages as natural language sequences for anomaly detection.
& Two datasets: \textit{BGL} and \textit{HDFS}. 
& LogAnomaly achieves a better F1-Score (0.96 on BGL and 0.95 on HDFS) compared to DeepLog (0.90 on BGL and 0.88 on HDFS).
& It leverages a new word embedding approach, \textit{template2Vec}, to model the sequential and quantitative patterns of logs and extract the semantic information of log templates.
& The approach does not take into account the runtime parameter values.\\

Zhang \textit{et al.}~\cite{zhang2019robust} - Proposes a log-based anomaly detection approach, \textit{LogRobust}, which can handle unstable log lines. 
& Real and synthetic \textit{HDFS} log data, and Service X logs from Microsoft.
& It extracts semantic information of log events as semantic vectors. LogRobust achieves the highest F-Measure in detecting anomalous log lines in unstable datasets, and the performance decreases as the unstable logs change further (\textit{i.e.}, become more unstable).
& It can identify new but semantically similar log events that emerge from logging evolution and processing noise. 
&If there are drastic and significant changes to the entire code base or logging mechanism, the LogRobust would perform poorly in anomaly detection.\\

\bottomrule
\end{tabular}
\end{center}
\vspace{-3mm}
\caption{Log anomaly detection research - Topic (H).}
\label{log_anomaly_detection_table}
\end{table*}
}

\afterpage{\renewcommand{\arraystretch}{1.2}
\newcommand\rownumber{\stepcounter{magicrownumbers}\arabic{magicrownumbers}}
\rowcolors[]{2}{gray!15}{white}
\begin{table*}
\ContinuedFloat
\scriptsize	
\begin{center}
 \begin{tabular}{p{3cm} p{3cm} p{4cm} p{3cm} p{3cm}} 
 \toprule
 
\rowcolor{blue!20} \textbf{Reference - Aim} &  \textbf{Experiments}& \textbf{Results} & \textbf{Pro}& \textbf{Con} \\ [0.5ex] 
 \midrule

Zhang \textit{et al.} \cite{zhang2020anomaly}  - Proposes Anomaly Detection by workflow Relations (ADR). 
& \textit{BGL} and \textit{HDFS} logs. 
& The approach mines numerical relations from logs and uses the relations for anomaly detection. ADR detects a higher number of relations in less time compared to the invariant mining (IM) approach.
& For faster online anomaly detection, ADR leverages an optimization approach, Particle Swarm Optimization (PSO)~\cite{kennedy1995particle}, to find the proper window size to split the log entries.
& The experimentation is performed on highly regulated logs with a low number of log keys (\textit{i.e.}, Hadoop and BGL), and for BGL data, simpler approaches, \textit{e.g.}, SVM, outperform ADR.\\

Huang \textit{et al.}~\cite{huang2020hitanomaly} - Proposes \textit{HitAnomaly}, that is a transformer-based~\cite{vaswani2017attention} log anomaly detection method.   
& Three system logs: \textit{HDFS}, \textit{BGL}, and \textit{OpenStack}.
&The approach achieves F1-Scores higher than prior work for stable logs, and for unstable logs performs best under 10\% instability injection into the log lines. 
& The approach allows to capture the semantic information in both log template sequence and parameter values and provides an attention-based classifier for log anomaly detection.
&HitAnomaly's performance drops lower than LogRubost~\cite{zhang2019robust} for higher rates of instability in log sequences.\\

Zhou \textit{et al.}~\cite{zhou2020logsayer} - Proposes \textit{LogSayer}, a log-based anomaly detection approach with pattern extraction for cloud environment.
&One \textit{HDFS} log set and two \textit{OpenStack} log data sets. 
& The key observation is that different components of cloud systems show different levels of system resource usage during anomalous behavior. The approach performs the best in detecting transient anomalies with an accuracy of 93\% and outperforms DeepLog~\cite{du2017deeplog} and CloudSeer~\cite{yu2016cloudseer}.  
& Applies a back-propagation (BP) LSTM-based approach to learn and correlate the historical logs with current logs, and deviations are signaled as potential anomalies.  
&LogSayer's performance is dependent on the \textit{time window} size, and its performance towards unstable logs~\cite{zhang2019robust} is not evaluated.\\

Chen \textit{et al.}~\cite{chen2020logtransfer} - Proposes \textit{LogTransfer}, to transfer anomalous log knowledge from the source system to the target system.
& Proprietary switch logs over a two-year period, and \textit{Hadoop} application and \textit{HDFS} logs.
&LogTransfer still requires anomalous instances of the target system for optimal performance. It achieves 0.84 \textit{F1-score}, better results than unsupervised and supervised approaches, such as DeepLog and LogAnomaly.
& It applies \textit{GloVe}~\cite{pennington2014glove}, an unsupervised word representation technique, to convert words in log templates to fixed-dimension vectors. 
&In the comparison section, unsupervised methods, \textit{e.g.}, DeepLog is supposed to be trained on normal logs and not on a mix of normal and abnormal logs~\cite{du2017deeplog}.\\

Yang \textit{et al.} 2021 \cite{yang2021semi} - Introduces \textit{PLELog}, a semi-supervised anomaly detection through execution logs. 
& Experimented on \textit{BGL} and \textit{HDFS} logs.  
& With probabilistic label estimation (PLE), it can automatically assign labels to unlabeled datasets. PLELog outperforms compared semi-supervised and unsupervised anomaly detection approaches in terms of F-Measure. 
& It leverages an attention-based GRU neural network to detect anomalies.
& The effectiveness of PLELog falls short compared to some prior anomaly detection approaches, \textit{e.g.}, LogRobust~\cite{zhang2019robust}.\\

Le \textit{et al.} 2021 \cite{le2021log} - Introduces \textit{NeuralLog}, a BERT-based deep learning anomaly detection approach.
& Experimented on \textit{BGL}, \textit{HDFS}, \textit{Thunderbird}, and \textit{Spirit} logs.
& The experiments show that NeuralLog achieves a higher F1-score in anomaly detection compared to LogRobust. However, LogRobust achieves a faster training and prediction time. 
& NeuralLog removes the need for log parsing by directly using the log messages, which can potentially remove the log content loss that happens during the log parsing stage.
&NeuralLog ignores numbers and parameter values such as node ID, task ID, and IP address, which might contain important information.\\

\bottomrule
\end{tabular}
\end{center}
\vspace{-3mm}
\caption{Log anomaly detection research - Topic (H) (continued).}
\label{log_anomaly_detection_table}
\end{table*}
}

\afterpage{\renewcommand{\arraystretch}{1.2}
\newcommand\rownumber{\stepcounter{magicrownumbers}\arabic{magicrownumbers}}
\rowcolors[]{2}{gray!15}{white}
\begin{table}[h]
\scriptsize	
\begin{center}
  \begin{tabular}{p{1.3cm} p{2.5cm} p{2cm} P{.8cm} } 
 \toprule
 
\rowcolor{blue!20} \textbf{Reference} &\textbf{Category} & \textbf{Systems} & \textbf{Impl.} \\ [0.5ex] 
 \midrule

Miranskyy \textit{et al.}~\cite{miranskyy2016operational}& challenges of big data log analysis  & N/A & ACA\\

Salman \textit{et al.}~\cite{salman2017designing}& PhelkStat, log analytics for Apache Spark& supercomputers and network data, \textit{e.g.}, Spirit & ACA\\

Mavridis \textit{et al.}~\cite{mavridis2017performance}& various log file analysis tasks in cloud& Hadoop and Spark & ACA\\

Chowdhury \textit{et al.}~\cite{chowdhury2018exploratory}& energy consumption of logging for mobile & 24 Andriod apps & ACA \cite{urlchowdhury2018exploratory}\\

He \textit{et al.}~\cite{he2018characterizing}& natural language attributes of log statement descriptions & C++ and Java projects & ACA \cite{urlhe2018characterizing}\\

Zeng \textit{et al.}~\cite{zeng2019studying}& logging practices for mobile devices  & 1,444 Andiod apps & ACA\\

Gholamian and Ward~\cite{gholamian2021naturalness}& naturalness and locallness of software logs & system logs and natural corpora, \textit{e.g.}, Spark and Wikipedia & ACA \cite{urlgholamian2021naturalnessdata}\\

\bottomrule
\end{tabular}
\end{center}
\vspace{-2mm}
\caption{\footnotesize Comparison of emerging log research - Topic (L).}
\label{log_emerging_research_table_OW}
\end{table}
}

\afterpage{\renewcommand{\arraystretch}{1.2}
\newcommand\rownumber{\stepcounter{magicrownumbers}\arabic{magicrownumbers}}
\rowcolors[]{2}{gray!15}{white}
\begin{table*}
\scriptsize	
\begin{center}
 \begin{tabular}{p{3cm} p{3cm} p{4cm} p{3cm} p{3cm}} 
 \toprule
 
\rowcolor{blue!20} \textbf{Reference - Aim} &  \textbf{Experiments}& \textbf{Results} & \textbf{Pro}& \textbf{Con} \\ [0.5ex] 
 \midrule

Tang \textit{et al.}~\cite{tang2010approach} - Proposes a log-based approach to identify \textit{service composition patterns} by finding associated services using \textit{Apriori algorithm}.
& A case study on 74 service-oriented applications.
& The approach can detect service composition patterns from control flow with a high accuracy.
&The approach first starts with collecting and preprocessing of execution logs, and continues with identification of frequent web services, and then extrapolates the control flows.
& The approach fails to extract service patterns in cases that control flow has alternative branches.\\

Oliner and Aiken~\cite{oliner2011online} - Proposes an approach to infer the interactions among the components of large-scale systems by analyzing logs.
& Log of eight systems: four supercomputers (\textit{Blue Gene/L}, \textit{Thunderbird}, \textit{Spirit}, and \textit{Liberty}), two data clusters (\textit{Mail Cluster} and \textit{Junior}), and two autonomous vehicles (\textit{Stanley} and \textit{SQL Cluster}).
& Log data signal compression allows for the scalability of \textit{`lag'} correlation, and with minimal loss, this approach identifies system's behavioral model.
&Performs a two-stage analysis: 1) PCA compression to summarize the anomaly signals, and 2) lag correlation to identify if the signals relate to each other with a time lag.
&The extracted signals show correlation and not a causal relationship, and in addition, manual analysis of a system administrator is required to make sense of the data.\\

Fu \textit{et al.}~\cite{fu2012logmaster} - Proposes \textit{LogMaster}, a tool to mine correlations of events in log files of large-scale cloud and high-performance computing (HPC) systems.
& Experimented on three system logs, namely: \textit{Hadoop}, \textit{HPC} cluster and \textit{BlueGene/L}.
& Results show the approach is successful in correlating events related to failures with acceptable \textit{precision} scores but with lower \textit{recall} rates.  
& LogMaster parses the log lines into event sequences where each event creates an informative nine-tuple, and then uses an algorithm, named Apriori-LIS, to mine event rules from logs, and measures the events correlations.
&Experiments on cloud and HPC systems shows LogMaster can predict failures with high \textit{Precision}, however, the \textit{Recall} scores are low and require improvement.\\

Shang \textit{et al.}~\cite{shang2013assisting} - Suggests using execution logs from the cloud environment to assist developers of Big Data Analytics (BDA) applications.  
& A case study on three Hadoop-based apps: \textit{WordCount}, \textit{PageRank}, and \textit{JACK} (industrial).
& The approach reduces the verification effort and reaches comparable precision with traditional keyword search methods in verifying cloud deployment procedures. 
& This approach exposes the differences between pseudo and large-scale cloud deployments and it points the developers' to examine the inconsistencies, and therefore, facilitates the deployment verification effort.
& The approach suffers from a high number of false positives in flagging presumptive problematic log sequences, and results in low precision. \\

Busany and Maoz~\cite{busany2016behavioral} - Proposes an approach for behavioral analysis of logs.  
& Logs generated from four finite-state automaton models.
& For logs of 2000 traces, reduces the analysis requirement to between 60 to 800 traces and unobserved log ratio of 0.01\% and 2.45\%. 
& The approach leverages sampling and statistical inference to provide scalable behavioral analysis of large logs. 
& The approach assumes that the log is sampled independently and adequately reflects the behavior of the system under review. \\

Awad and Menasc\'e~\cite{awad2016performance} - Proposes an approach to use system logs and configuration files to automatically extract performance models of the system.
& Experiments on \textit{Apache Tomcat access logs} from a multi-tier server.
&The results show the method is effective in extracting the workloads and system model by parsing the system configuration files and log files. 
& The approach extracts the interaction patterns between servers and devices and the probability associated with each interaction. 
&The work assumes the log templates are known for the systems under analysis.\\

Di \textit{et al.}~\cite{di2017logaider,di2018exploring} - Proposes \textit{LogAider}, an analysis tool that mines potential correlations between various system events for the diagnosis purpose.
& Logs of \textit{BlueGene/Q Mira} supercomputer~\cite{mirasystem}.
&The approach shows effective correlation between fatal system events and job events, with both high precision and recall values (99.9-100\%).
& LogAider can reveal three types of potential correlations between log events: \textit{acrossfield}, \textit{spatial}, and \textit{temporal}. 
&It uses a threshold to find correlation candidates, and the evaluation scores, such as \textit{Precision} and \textit{Recall}, appear to be threshold dependent and vary significantly with the threshold. \\
 
He et al. \textit{et al.}~\cite{he2018identifying} - Proposes \textit{Log3C}, a clustering-based approach to detect system problems. 
& Three datasets of an online large-scale service system X from Microsoft (confidential).
&Log3C achieves F1-measures values 0.91, 0.86, and 0.868 for three datasets, and outperforms PCA and Invariant Mining approaches. 
& It applies cascading clustering to cluster and match the log sequences efficiently, and then correlates the log sequence clusters with KPIs to identify the impactul problems. 
&The research only considers a single KPI, \textit{i.e.}, \textit{failure rate}, to correlate with the log sequences. The research can be enriched with the inclusion of additional KPIs.\\
\bottomrule
\end{tabular}
\end{center}
\vspace{-3mm}
\caption{System's runtime behavior research - Topic (I).}
\label{log_runtime_behavior_table}
\end{table*}
}

\subsubsection{Category H: Anomaly Detection}\label{anomoly_sub_application}
Execution logs are extensively leveraged to monitor the health of software systems, identify abnormal situations, and detect anomalies that can lead to system failures. 
As such the goal of anomaly detection from logs is to find cues in the log records that are tied to the identification of abnormal system behavior. 
If anomaly detection is used in online fashion, early detection of abnormal log lines could result in stopping anomalies before they turn into a partial or complete failure, \textit{e.g.}, slower system or shutdown, respectively. 
role in incident management of large-scale systems. 
Table~\ref{log_anomaly_detection_table_OW} provides an overview of anomaly detection from logs, and Table~\ref{log_anomaly_detection_table} provides a detailed comparison and \textit{pros} and \textit{cons} of various research for anomaly detention with logs. 
As a reference for further reading, He \textit{et al}. \cite{shilin2016expr} performed a quantitative comparison of various log-based anomaly detection approaches. 

\begin{tcolorbox}[breakable, enhanced]
\small \textbf{Finding} \textbf{12.} \textit{
In sum, anomaly detection methods include various approaches, such as: 
\begin{enumerate*}[label=\protect\circled{\arabic*}]
\item creating a state machine of normal execution and comparing the failure runs with normal models~\cite{jiang2008automatic,fu2009execution}, \item PCA-based approach which projects event logs to normal and abnormal subspaces~\cite{xu2009detecting}, \item deep learning approaches which learn an LSTM model from normal execution workflows~\cite{du2017deeplog}, and unstable logs~\cite{zhang2019robust}, \item semi-supervised deep learning approaches with probabilistic label estimation~\cite{yang2021semi}, \item a statistical approach using probabilistic suffix trees~\cite{bao2018execution}, \item cloud deployment by correlating logs and resource metrics~\cite{farshchi2018metric}, and transformer-based approaches~\cite{le2021log}.\end{enumerate*}}
\end{tcolorbox}

\subsubsection{Category I: System's Runtime Behavior}\label{srbehavior}
Researchers have also utilized logs for monitoring the system's runtime behavior. 
This category of research aims to gain insight into how the system behaves while it is running, what the operational profiles are, and how the system analytics can be leveraged for managing cloud provisioning tasks. 
Examples of log data that are used to extract runtime information generally falls into categories such as supercomputers, \textit{e.g.} BlueGene, large-scale distributed systems, \textit{e.g.}, Hadoop, and online cloud-base services, \textit{e.g.}, Service X at \textit{Microsoft}.  
Some of the research overlaps with approaches for \textit{`anomaly detection'} and \textit{`performance and failure diagnosis'}. 
Table~\ref{log_runtime_behavior_table_OW} provides an overview of the approaches, and we provide further comparison for this research category in Table~\ref{log_runtime_behavior_table}.

\begin{tcolorbox}[breakable, enhanced]
\small \textbf{Finding} \textbf{13.} \textit{
In sum, the approaches for monitoring system's runtime behavior include:
\begin{enumerate*}[label=\protect\circled{\arabic*}] \item using logs to customize operational profiles for industry software~\cite{hassan2008industrial}, \item web service composition~\cite{tang2010approach}, \item detecting inter-component interaction~\cite{oliner2011online},\item mining system events correlation~\cite{fu2012logmaster,di2018exploring}, \item assisting developers in cloud deployments~\cite{shang2013assisting}, \item performance model derivation~\cite{awad2016performance}, \item statistical approach~\cite{busany2016behavioral}, \item big-data analytics for cloud deployment~\cite{shang2013assisting}, and \item detecting impactful system problems~\cite{he2018identifying}.\end{enumerate*}}
\end{tcolorbox}

\afterpage{\renewcommand{\arraystretch}{1.2}
\newcommand\rownumber{\stepcounter{magicrownumbers}\arabic{magicrownumbers}}
\rowcolors[]{2}{gray!15}{white}
\begin{table*}
\scriptsize	
\begin{center}
 \begin{tabular}{p{3cm} p{3cm} p{3.5cm} p{3.5cm} p{3cm}} 
 \toprule
 
\rowcolor{blue!20} \textbf{Reference - Aim} &  \textbf{Experiments}& \textbf{Results} & \textbf{Pro}& \textbf{Con} \\ [0.5ex] 
 \midrule

Cinque \textit{et al.}~\cite{cinque2010assessing} - Proposes a software fault injection approach to assess the effectiveness of logs in the recording of software faults in the deployed environment.
& Three open-source systems: \textit{Apache Web Server}, \textit{TAO Open Data Distribution System}, and \textit{MySQL Database Management System}.
& Approximately, only 40\% of the injected faults are covered by the existing logging statements in the three studies systems. 
& Faults are intentionally injected into the experimented software systems to determine the most common failure sequences and identify logging deficiencies and improve them.
&The faults are synthetic (might not necessarily match real faults) and it requires access to the source code of the software.\\

Chuah \textit{et al.}~\cite{chuah2010diagnosing} - Presents an approach to reconstruct event order and establish correlations among log events to discover the root causes of a given failure.
& \textit{Syslogs} of \textit{Ranger} and \textit{Turing} supercomputers, and \textit{BlueGene/L RAS} logs.
&The authors received positive feedback from system admins that they have found the tool analysis useful in facilitating their diagnosing efforts. 
& Introduces a \textit{Fault Diagnostics} tool \textit{FDiag}, to discover faults, which comprises three components: a \textit{Message
Template Extractor} (MTE), a \textit{Statistical Event Correlator} (SEC), and an \textit{Episode Constructor}.
&The approach depends on the availability of event-specific keywords as domain knowledge for correlation, and does not provide causality.\\

Yuan \textit{et al.}~\cite{yuan2010sherlog} - Proposes \textit{SherLog}, a tool that leverages runtime log information and source code analysis to infer the probable execution paths during a failed production run.
& Evaluated on eight real-world software failures collected from different application such as \textit{rmdir}, \textit{Squid}, and \textit{ln}.
& The experiments show the information inferred by SherLog is useful to assist developers in failure diagnosis.
&By accepting the execution log of a failed run and the source code, SherLog aims to identify what must or may have happened along the execution path.
&SherLog relies on the amount of information available in log messages to perform its analysis. As such, log messages that lack the necessary debugging information will significantly limit Sherlog's effectiveness.\\

Pecchia \textit{et al.}~\cite{pecchia2012detection} - Conducts an experimental study to examine factors from event logs that help with the detection of failures.  
& Performs experiments on a set of 17,387 instances of injecting faults into three systems: \textit{Apache Web Server}, \textit{TAO Open DDS}, and \textit{MySQL DBMS}.
& Features such as system architecture, placement of the logging statements, and support provided by the execution environment can have an impact on the accuracy and effectiveness of the logs at runtime.

& This research additionally investigates the logging improvement that can potentially increase the usefulness of the execution logs.
& The approach requires access to the source code and is only tested on open-source projects.\\

Nagaraj \textit{et al.}~\cite{nagaraj2012structured} - Presents \textit{DISTALYZER}, a tool which utilizes log data to assist developers in diagnosing performance problems.
& Case studies on three systems: \textit{TritonSort}, \textit{HBase}, and \textit{BitTorrent}.
& Results show that \textit{DISTALYZER} is able to uncover undiagnosed performance issues for the experimented systems.
& DISTALYZER uses machine learning techniques (\textit{i.e.}, \textit{Welch's t-test~\cite{welch1947generalization} and \textit{dependency networks}~\cite{heckerman2000dependency}}) to compare log files with acceptable and unacceptable performance.

&DISTALYZER leverages \textit{ad-hoc} approaches (\textit{e.g.}, thread id) to group log messages, which limits its application for less-structured logs.\\

Fronza \textit{et al.}~\cite{fronza2013failure} - Proposes an approach to perform log-based prediction by applying \textit{Random Indexing (RI)} and \textit{Support Vector Machines (SVMs)}.
& Experimented on log files of a large European manufacturing company (anonymous).
& According to the findings, weighted SVMs achieve the best performance by slightly shrinking specificity (true negative rate) scores to improve sensitivity or recall, and specificity stays greater than 0.8 in the majority of the experimented applications.
& It applies weighted SVM, which utilizes cost-sensitive learning to achieve balanced TPR and TNR values, and makes the method more reliable in classifying both failures and non-failures.
& SVM classification performs well in classifying non-failure instances, but poor in identifying failures, \textit{i.e.}, low \textit{true positive rate} or \textit{recall}.\\

Syer \textit{et al.}~\cite{syer2013leveraging} - Proposes an approach that combines performance counters and execution logs to diagnose memory-related issues in load tests.
& A case study of \textit{WordCount} application on \textit{Hadoop}. 
& The approach flags less than 0.1\% of the execution logs with a high precision of ($\geqslant$80\%).
& After clustering the events, authors apply scoring techniques to identify clusters that are abnormal and can be associated with a performance issue.
&The approach has limited applications to memory performance issues, such as memory leaks, spikes, and bloats.\\

Zhao \textit{et al.}~\cite{zhao2014lprof} - Proposes \textit{lprof}, a \textit{log profiling} tool that recreates the execution flow of distributed applications.
& Evaluated on four distributed systems: \textit{HDFS}, \textit{Yarn}, \textit{Cassandra}, and \textit{HBase}.
& lprof's reaches 88\% accuracy in attributing log messages to requests, and 65\% of the diagnostics are helpful for the operators.
& lprof performs control-flow (CF) and data-flow (DF) analyses, and infers if log messages are causally related and what variables are unmodified between multiple log printing statements, and then groups the logs and use them for diagnosing performance issues.
&The lprof's static analysis is limited to a single software component and needs to be readjusted for different languages (bytecode), which is cumbersome in practice.\\

Xu \textit{et al.}~\cite{xu2013detecting,xu2014pod} - Utilizes system logs to provision rolling updates in a cloud environment for process oriented dependability (POD) analysis.
& Experiments with rolling upgrade on AWS with injecting 8 different types of faults into the cloud-based clusters. Faults include \textit{machine image (MI) change during upgrade}, \textit{key pair management fault}, and \textit{security group configuration fault}.
& The evaluation results show acceptable performance (90+\%) in precision, recall, and accuracy scores in diagnosing the injected sporadic faults.
& It creates a process model of the desired provisioning activities through log data with added annotation and checkpoints. 
The deployment logs are checked and assertions are raised in case there has been a deployment violation. 
&The approach heavily relies on the specific information in the logs and the absence of this information severely impacts the performance of error detection. Following research~\cite{farshchi2018metric} aims to combine logs with system metrics for a more robust analysis.\\

\bottomrule
\end{tabular}
\end{center}
\vspace{-3mm}
\caption{Performance, fault, and failure diagnosis research - Topic (J).}
\label{log_performance_failure_table}
\end{table*}
}

\afterpage{\renewcommand{\arraystretch}{1.2}
\newcommand\rownumber{\stepcounter{magicrownumbers}\arabic{magicrownumbers}}
\rowcolors[]{2}{gray!15}{white}
\begin{table*}
\ContinuedFloat
\scriptsize	
\begin{center}
 \begin{tabular}{p{3cm} p{3cm} p{3.5cm} p{3.5cm} p{3cm}} 
 \toprule
 
\rowcolor{blue!20} \textbf{Reference - Aim} &  \textbf{Experiments}& \textbf{Results} & \textbf{Pro}& \textbf{Con} \\ [0.5ex] 
 \midrule

Russo \textit{et al.}~\cite{russo2015mining} - Proposes an approach to mine and learn error predictors from system logs, and then applies it to a real telemetry system for failure prediction.
& Experimented on log sequences of 25 different applications of a software system for telemetry and performance of cars.
& The evaluation achieves 78\% \textit{recall}, and 95\% \textit{precision}.
& Uses three popular support vector machines (SVMs): multilayer perceptron, radial basis function, and linear kernels - to learn and predict \textit{defective} (\textit{i.e.}, faulty) log sequences.
& The study is performed on the logs of a single system. 
Thus, the result of the applicability of the proposed approach to other systems and other software domains remains unknown.\\

Yu \textit{et al.}~\cite{yu2016cloudseer} - Introduces \textit{CloudSeer}, a lightweight and non-intrusive approach that works on interleaved logs for cloud workflow monitoring.
& CloudSeer is prototyped and evaluated on an open-source platform, \textit{i.e.}, multi-user \textit{OpenStack} logs. 
& The approach is accurate enough to check and infer workflows for most interleaved log sequences. Cloudseer reaches 92+\% accuracy in checking interleaved logs for six experimented groups with a satisfactory checking efficiency (\textit{i.e.}, computation time).
& CloudSeer constructs \textit{automatons} for the workflow of
management tasks based on their normal execution scenarios, and later checks log messages against these \textit{automatons} to detect workflow discrepancies and divergences in an online approach.
&The performance problems are only detected for log entries with \textit{`ERROR'} log verbosity level, and model creation requires multiple executions of a single task. For messages which do not accompany an error, finding a timeout is not trivial and requires further discussion.\\

Gurumdimma \textit{et al.}~\cite{gurumdimma2016crude} - Introduces \textit{CRUDE}, which combines console logs with \textit{resource usage} data to improve the error detection accuracy in distributed systems. 
& Experimented on \textit{Rationalized logs} (ratlogs) from the Ranger Supercomputer containing four weeks worth of data: resource usage data (32GB) and rationalized logs (1.2GB).
& The approach is able to identify 80\% of errors leading to failures, and achieves f-measure over 70\%. 
& The approach has three main steps: it clusters nodes with similar behavior, then uses an anomaly detection algorithm to detect jobs with anomalous resource usage, and finally, links anomalous jobs with erroneous nodes.
&The proposed approach does not model temporal relationships to improve fault identification~\cite{sorkunlu2017tracking}.\\

Zhao \textit{et al.}~\cite{zhao2016non} - Introduces \textit{Stitch}, a distributed and end-to-end performance profiler by flow reconstruction. 
& Evaluated both through a controlled user study and lab experiments on \textit{Hive}, \textit{Spark}, and \textit{OpenStack}. 
& On average, Stitch achieves 96\% and 95\% accuracy for object and edge detection, respectively, for workflow reconstruction.
& Stitch aims to construct the system model and the hierarchical relationship of objects in a distributed software stack without requiring domain-specific knowledge. 
&Although Stitch can establish correlations between different software's objects and modules, it cannot accurately infer causal relationships among them.\\

Zou \textit{et al.}~\cite{zou2016uilog} - Proposes \textit{UiLog}, which is a fault analysis tool, to collect logs and their statistics from various components and diagnose the detected faults.
& Performs experiments on logs of components (\textit{e.g.}, disk, I/O, memory) of a cloud environment, \textit{StrongCloud}, collected over a year period.
& Twelve categories of faults are detected, and fault detection precision is maxing out at 88\% when the length of logs is more than 200 words.
&The approach classifies logs by the fault type in real-time and performs fault correlation analysis to help administrators and locate the faults' root causes.
&Requires domain knowledge, and the precision of fault analysis is dependent on the size of the logs.\\

Zhang \textit{et al.}~\cite{zhang2017pensieve} - Introduces \textit{Pensieve}, a flow reconstruction tool for performance failure reproduction through system logs and bytecode.
& Evaluated on 18 randomly sampled real failures on four systems: \textit{HDFS}, \textit{HBase}, \textit{ZooKeeper}, and \textit{Cassandra}.
& Pensieve is able to reproduce 72\% of the sampled failures within ten minutes of analysis time.
& Pensieve leverages event chaining, and extrapolates a chain of causally dependent events leading to the failure while using \textit{partial trace observation} technique, which significantly limits the execution paths to observe. 
&Some domain-specific knowledge or a developer familiar with the system is needed to actually describe and diagnose the failure, and make sense of the chain of the events.\\

\bottomrule
\end{tabular}
\end{center}
\vspace{-3mm}
\caption{Performance, fault, and failure diagnosis research - Topic (J) (continued).}
\label{log_performance_failure_table}
\end{table*}
}

\subsubsection{Category J: Performance and Failure Diagnosis}\label{pfd}
In many cases, log messages are one of the most important clues and often the only available resource for the system's failure diagnosis and performance degradation, as it might be difficult and undeterministic to reproduce a failed scenario by replaying (\textit{i.e.}, rerunning). 
Developers often have to diagnose a production run performance degradation or failure based on logs collected in the field and returned by customers without having access to the infield user's inputs. Although some of the research is shared with \textit{'anomaly detection'}, but for performance and failure diagnosis, prior research usually aims to diagnosis a set of real-word known failures. Similarly, cloud-based distributed systems and supercomputers are the main categories that prior work has opted in for their evaluation. 
Table~\ref{log_performance_failure_table_OW} provides an overview for the research on mining of log files for performance and failure diagnosis, and Table~\ref{log_performance_failure_table} compares and summarizes the research in this category with additional details. 

\begin{tcolorbox}[breakable, enhanced]
\small \textbf{Finding} \textbf{14.} \textit{
In sum, performance and failure diagnosis approaches include:
\begin{enumerate*}[label=\protect\circled{\arabic*}]
\item probable program execution paths investigation with logs~\cite{yuan2010sherlog}, 
\item machine learning to compare and classify logs with good and bad performance~\cite{nagaraj2012structured}, 
\item correlating performance counters (\textit{e.g.}, CPU/memory usage) and logs~\cite{syer2013leveraging}, 
\item automaton-based workflow modeling~\cite{yu2016cloudseer}, \item process-oriented dependability analysis~\cite{xu2014pod}, \item control and data flow analyses to extrapolate causal relations among longs~\cite{zhao2014lprof,zhao2016non,zhang2017pensieve}, and \item fault diagnosis with logs~\cite{zou2016uilog}.
\end{enumerate*} 
}
\end{tcolorbox}

\afterpage{\renewcommand{\arraystretch}{1.2}
\newcommand\rownumber{\stepcounter{magicrownumbers}\arabic{magicrownumbers}}
\rowcolors[]{2}{gray!15}{white}
\begin{table*}
\scriptsize	
\begin{center}
 \begin{tabular}{p{3cm} p{3cm} p{4cm} p{3cm} p{3cm}} 
 \toprule
 
\rowcolor{blue!20} \textbf{Reference - Aim} &  \textbf{Experiments}& \textbf{Results} & \textbf{Pro}& \textbf{Con} \\ [0.5ex] 
 \midrule
 
Lee \textit{et al.}~\cite{lee2012unified} - Employs a unified logging infrastructure in Twitter to perform analysis on the user statistics by the use of log data. 
& \textit{``Client events''} within Twitter logging framework.
& Discusses a variety of applications for the proposed approach, such as \textit{summary statistics}, \textit{event counting}, \textit{funnel analytics}, and \textit{user modeling}. 
&The research applies techniques from natural language processing (NLP) to process the user's behavior on the website; the user's behavior right now is strongly influenced by immediately preceding actions.
& Currently, user session sequences only capture event names and do not provide enough details for more sophisticated types of analyses.\\

Lim \textit{et al.}~\cite{lim2014identifying} - Addresses the problem of automated identification of recurrent and unknown performance issues.
& Evaluated on two datasets: Transaction Processing Performance Council BenchmarkTM W (TPC-W)~\cite{urltpc}, and System X, a real production system (confidential).
& The results show the approach achieves higher \textit{AUC} when compared to approaches such as \textit{Fingerprint}, \textit{Signature}, \textit{K-means}, and \textit{Hierarchical} approaches in recurrent and unknown issues identification. 
&The approach works based on the mining and metric extraction of historical log records, and utilizes a \textit{\textbf{H}idden \textbf{M}arkov \textbf{R}andom \textbf{F}ield (HMRF)} based approach for the clustering of recurrent issues.  
&The proposed statistical analysis only works if a large amount of monitoring data over a long period of time is available.\\

Oprea \textit{et al.}~\cite{oprea2015detection} - Proposes a framework that analyzes log data collected at the enterprise network borders on a regular basis (\textit{e.g.}, daily). 
& Experimented with DNS logs released by \textit{Los Alamos National Lab (LANL)}, and \textit{AC} dataset of web proxies logs generated at the border of a large enterprise network (confidential).
&The approach can detect malicious web domains with high accuracy and low false-negative rates. The work also detects new malicious domains that are not previously reported/detected by other tools in the enterprise. 
&Creates a bipartite graph $G = (V,E)$, such that hosts and domains are vertices on each side of the graph. There will be an edge between a host and a domain if the host connects with the domain.
&The approach cannot detect regular connections to malicious domains which happen in the training phase.\\

Barik \textit{et al.}~\cite{barik2016bones} - Performs a case study of log utilization in Microsoft for business decisions and analytics.
& Performed interviews with 28 engineers at Microsoft, and followed that up with a survey of 1,823 respondents to confirm their findings.
& Use of log event data is pervasive within the organization and the usage primarily falls into eight categories, among them \textit{engineering the data pipeline}, \textit{instrumenting for event data}, \textit{troubleshooting problems}, and \textit{making business decisions}. 
& This research highlights that event log data surely plays an important role in the company's decision-making process as the industry makes a transition towards a data-driven decision-making paradigm. 
& The study is performed on a single software company and the findings may not generalize to other software and institutions.\\

Chen \textit{et al.}~\cite{chen2018automated} - Presents an approach, called LogCoCo (\textit{i.e.}, \textbf{Log}-based \textbf{Co}de \textbf{Co}verage), to estimate and measure the source code coverage using the readily-available execution logs. 
& Five commercial proprietary projects from \textit{Baidu} and one open-source project, \textit{i.e.}, \textit{HBase}.
& Measures the accuracy and usefulness of LogCoCo, and it achieves high accuracy for different types of code coverage (\textit{Must} and \textit{May} have been executed). 
Additionally, the tool's results are useful to evaluate and improve the test suites for code coverage.
&Using program analysis techniques, LogCoCo matches the execution logs with their corresponding code paths and estimates three different code coverage criteria: 1) \textit{method coverage}, 2) \textit{statement coverage,} and 3) \textit{branch coverage}. 
&The approach cannot accurately infer whether a \textit{May} executed code region is actually covered in a test.\\
\bottomrule
\end{tabular}
\end{center}
\vspace{-3mm}
\caption{User, business, security, and code coverage research - Topic (K).}
\label{log_user_business_table}
\end{table*}
}

\subsubsection{Category K: User, Business, Security, and Code Coverage}
\begin{figure}[h]
\centering
\includegraphics[scale=.67]{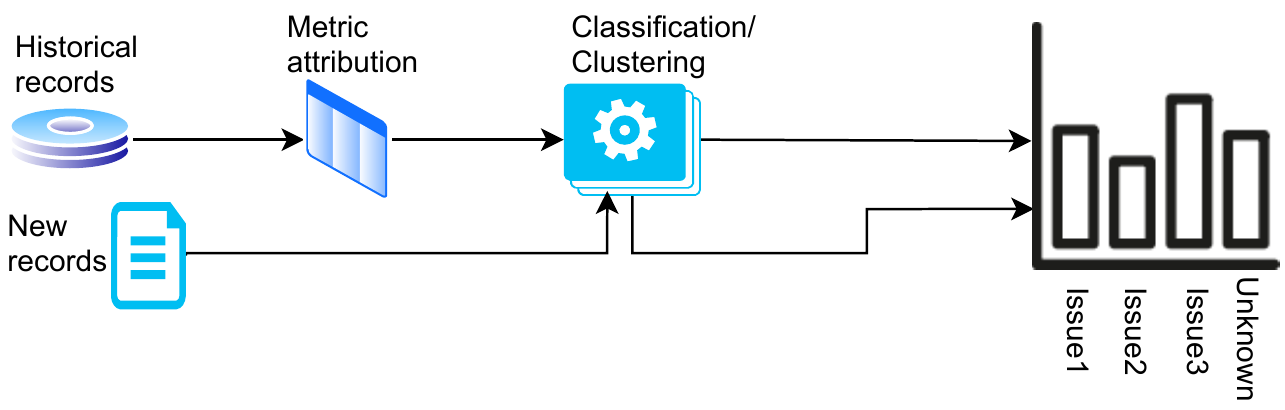}
\caption{Duplicate and recurrent issue detection tool.}
\label{recurrent_issues}
\vspace{-3mm}
\end{figure}

Other applications of log file mining include: \textbf{analyzing user statistics and behavior}~\cite{lee2012unified}, \textbf{application security}~\cite{oprea2015detection}, \textbf{duplicate issue identification}~\cite{lim2014identifying}, \textbf{code coverage with logging statements}~\cite{chen2018automated}, and \textbf{business analytics}~\cite{barik2016bones}. 
For example, Figure~\ref{recurrent_issues} shows how recurrent issues are first classified through historical log records, and once a new record is available, it is analyzed and compared against the historical issues~\cite{lim2014identifying}. 
Table~\ref{log_user_business_table_OW} provides a comparative summary of the research in this category, and
Table~\ref{log_user_business_table} further explains each research effort.    
\begin{tcolorbox}[breakable, enhanced]
\small \textbf{Finding} \textbf{15.} \textit{
In sum, because the logs are readily available, prior studies have expanded the usage of logs to adjacent domains, such as \textbf{code coverage analysis} and \textbf{business decision making}. 
We presume the list of possible applications of the logs will continue to grow, as logs have proven to be a rich source of information, which is non-intrusive and readily available.   
}
\end{tcolorbox}

\subsubsection{Implementation and Evaluation}
It is worth mentioning that many of the concepts with regards to feature selection, machine learning implementation, and evaluation metrics that are utilized for designing and assessing the log statement automation approaches (Sections~\ref{log_auto_theory}-\ref{metrics_section}) are also leveraged for designing and evaluating log file mining tasks. 
For example, \textit{DeepLog}~\cite{du2017deeplog}, proposes a machine learning algorithm for log file mining. 
LogRobust~\cite{zhang2019robust} extracts feature vectors from the log files and implements an LSTM deep learning approach, and evaluates its anomaly detection approach with \textit{Precision}, \textit{Recall}, and \textit{F-Measure}. 
\textit{Drain}~\cite{he2017drain}, a log parsing tool, evaluates its performance with \textit{F-Measure}.

\afterpage{\renewcommand{\arraystretch}{1.2}
\newcommand\rownumber{\stepcounter{magicrownumbers}\arabic{magicrownumbers}}
\rowcolors[]{2}{gray!15}{white}
\begin{table*}
\scriptsize	
\begin{center}
 \begin{tabular}{p{3cm} p{3cm} p{4cm} p{3cm} p{3cm}} 
 \toprule
 
\rowcolor{blue!20} \textbf{Reference - Aim} &  \textbf{Experiments}& \textbf{Results} & \textbf{Pro}& \textbf{Con} \\ [0.5ex] 
 \midrule

Miranskyy \textit{et al.}~\cite{miranskyy2016operational} - Discusses the challenges of event log analysis for big data systems (BDS), as the logs generated by a BDS can be big data themselves. 
&Categorizes seven challenges of log analysis for big data systems. 
& Highlights currently available solutions to each challenge and discusses unanswered questions, based on the authors' and industrial experience.
&The authors categorize the challenges of big data log processing into seven classes, including \textbf{scarce storage}, \textbf{unscalable log analysis}, \textbf{inaccurate capture \& replay}, and \textbf{ inadequate tools for instrumenting BDS source code}.
&Accurate mapping and case studies and examples of challenges in real-world big-data software can further illustrate the current issues.\\

Salman \textit{et al.}~\cite{salman2017designing} - Proposes \textit{PhelkStat}, a tool for analysis of system event logs of large-scale data centers on Apache Spark's big data platform. 
& A set of public (\textit{e.g.}, \textit{Spirit}, \textit{Thunderbird}, and \textit{Liberty}) and private (\textit{e.g.}, Cray and dartmouth/campus) log data. 
& Performs evaluation on a set of log analysis tasks such as \textit{arrival rate distribution}, \textit{anomaly content}, and \textit{runtime analysis}.
& The authors utilized a set of attributes, \textit{i.e.}, temporal and spatial metrics such as arrival rate and byte count, to characterize system event logs and then correlate the metrics with the runtime performance of the system. 
&Log analysis tasks are partially correlated with system events, but are not analyzed meaningfully to draw actionable steps for admins and users of the system.\\

Mavridis \textit{et al.}~\cite{mavridis2017performance} - Evaluates various log file analysis tasks with two cloud computational frameworks, \textit{Apache Hadoop} and \textit{Apache Spark}. 
& Experiments on log files of an \textit{Apache HTTP server}, and implemented useful log analysis tasks such as checking for denial of the service (DoS) attacks from the available logs.
& Compared performance of Hadoop and Spark by evaluating their execution time, scalability, resource utilization, and cost and power consumption
& The research showed the potential of utilizing distributed big-data platforms for facilitating log analysis. 
&The analysis needs to be expanded to designing tools that can leverage the parallelism of distributed big-data platforms to accomplish a faster and scalable analysis of logs.\\

Chowdhury \textit{et al.}~\cite{chowdhury2018exploratory} - Performs an exploratory study to investigate the energy impact of logging in Android applications using GreenMiner~\cite{hindle2014greenminer}, an automated energy test-bed for mobile applications. 
& Studies approximately a thousand versions of 24 Android applications (\textit{e.g.}, CALCULATOR, FEEDEX, FIREFOX, and VLC) with logging enabled and disabled, accompanied by a controlled experiment on a synthetic application.
& There is little to no energy impact when logging is enabled for \textbf{most} versions of the studied applications. However, about 79\% (19/24) of the studied applications have at least one version that exhibits a noticeable impact on energy consumption.
& The authors found that \textbf{the rate of logging}, \textbf{the size of messages}, and \textbf{the number of log buffer flushes} are significant factors of energy consumption attributable to logging on mobile devices.
&More accurate models that correlate mobile log events with the amount of energy consumption are required.\\

He \textit{et al.}~\cite{he2018characterizing} - Characterizes natural language attributes of log statements' descriptions.
& Experiments on ten Java and seven C++ open-source projects and answers four research questions. 
& Findings confirm the natural characteristics of logs, such as endemic and specific. 
& Proposes an automated approach for log description prediction based on source code similarity and edit distance. 
& The dynamic part of the log statements is left out, and the approach for log description automation is limited to cases that a similar code snippet is found.\\

Zeng \textit{et al.}~\cite{zeng2019studying} - Replicates the work of Yuan \textit{et al}.~\cite{yuan2012characterizing} and investigated the logging practices in Android applications. 
& Performs a case study on 1,444 open-source Android applications in the F-Droid repository.
& Although mobile app logging is less pervasive than server and desktop applications, logging is leveraged in almost all studied mobile apps, and there are noticeable differences between the logging practices applied in mobile applications versus the ones in server and desktop applications, as observed by prior studies~\cite{chen2017characterizing2,yuan2012characterizing}. 
& The majority of the logging statements in mobile apps are in \textit{debug} and \textit{error} verbosity levels, while \textit{info} level logging statements are the prevailing level in server and desktop applications. 
&The research can be expanded by providing developers with guidelines for mobile apps.\\

Gholamian and Ward~\cite{gholamian2021naturalness} - Performs an experimental study on natural and local characteristics of log files. 
& Experiments on eight system logs (\textit{e.g.}, \textit{HDFS} and \textit{Spark}), and two natural language language data (\textit{e.g.}, \textit{Gutenberg} and \textit{Wiki}). 
& Six findings confirm that log messages are natural and local, even more or so than common English text.
& Applies the findings and proposes an NLP-based anomaly detection approach from log files, which utilizes n-gram models.
&More advanced NLP models (\textit{e.g.}, deep learning and BERT) need to be investigated to improve the anomaly detection task.\\

\bottomrule
\end{tabular}
\end{center}
\vspace{-3mm}
\caption{Emerging log research - Topic (L).}
\label{log_emerging_research_table}
\end{table*}
}

\subsection{Category L: Emerging Applications of Logs}\label{log_apps}
Thus far, we discussed different logging practices and log applications, mainly in large-scale software systems. 
However, most recently, there has been a special interest in applications of logs in other domains such as mobile devices~\cite{chowdhury2018exploratory,zeng2019studying}, embedded~\cite{dong2015dynamic,gomez2017efficient,wang2018extracting}, and big data~\cite{miranskyy2016operational,salman2017designing,mavridis2017performance}. 
We summarize the key findings here:
 
\begin{itemize}
\item Prior studies have proposed the application of logs for emerging areas such as \textbf{mobile} and \textbf{big data systems}.  
\item For \textbf{mobile}, developers should be aware of different logging practices that might apply to alternative platforms with different design criteria and requirements. 
For example, because mobile devices operate on battery with limited storage space, the cost and overhead of logging (\textit{e.g.}, continuously flushing logs) become more exorbitant and unfavorable than software that is running on a workstation. 
\item Developers of log analysis tools have considered \textbf{big-data platforms} to scale and speed up log analysis.
\item \textbf{Natural language attributes} of logs~\cite{he2018characterizing,gholamian2021naturalness} open up a new avenue for log statement automation, \textit{e.g.}, log statement description, and automated log analysis of logs, \textit{e.g.}, anomaly detection. 
\end{itemize}

Table~\ref{log_emerging_research_table_OW} provides an overview of emerging applications of logs, and~\ref{log_emerging_research_table} provides additional details on research in \textbf{Category L}.

\begin{tcolorbox}[breakable, enhanced]
\small \textbf{Finding} \textbf{16.} \textit{Researchers have contributed to logging research in various categories, and research continues to progress in the existing categories and also grows to the emerging domains. 
Emerging application of logs include \textbf{mobile} and \textbf{big data},  and \textbf{natural language processing of logs}. 
}
\end{tcolorbox}
\vspace*{-3mm}

\section{RQ4: Challenges and Opportunities for Future Work}\label{opportunities}
In the previous sections, we reviewed and discussed the state-of-the-art logging research by providing an introduction to log messages and log files (\ref{log_file}), costs and benefits associated with logging (\ref{log_cost}), mining logging statements (\ref{log_mining}), automated logging approaches and their evaluation metrics (\ref{automated_logging}), mining of log files (\ref{mine_log_files}), and finally, the emerging areas of log application (\ref{log_apps}). 
In this section, as we revisit each section, we put emphasis on answering RQ4 and specify future directions and opportunities for each category. 
We point out the missing pieces of the puzzle for each area, followed by our intuitive approaches for tackling those issues, which are inspired by the collective knowledge of the prior work. In the following, we include opportunities for future work based on the research categories. 

\subsection{Category A: Logging Cost}
\subsubsection{Adaptive and Constraint-Based Logging}
An imperative trend that we foresee as future research is the need for adaptive logging~\cite{mizouchi2019padla}. 
On the one hand, continuously logging in details (\textit{e.g.}, in trace verbosity level) can infer performance overhead, and on the other hand, logging very little might degrade the effectiveness of the logs. 
Therefore, we anticipate further research that will work on dynamically adjusting the amount of logged data from the least verbose to the most verbose level in order to help with detailed postmortem analysis, if the system is in a detected anomaly state, and on the other side, minimize the performance overhead of logging while the system operates normally and as expected. 

\subsubsection{Whether to Log?}
We mentioned that the prior research has explored various challenges such as \textit{`what to log?'}, \textit{`where to log?'}, and \textit{`how to log?'}. 
We see potential for further research on all challenges of logging and with more emphasis on \textbf{`whether or not to log?'}. 
With the emergence of adaptive logging and logging less when not needed and log more details whenever necessary, the idea of whether or not to ultimately print an existing logging statement becomes of importance.   
We foresee future research that explores different scenarios of whether logging statements are eventually printed or filtered based on the goal of the logging analysis tasks, \textit{i.e.}, performance evaluation or failure diagnosis, and the operating state of the system, \textit{i.e.}, normal state vs. when a system anomaly is detected.

\subsection{Categories B, C, D: Logging Practices, progression, and Issues}
\subsubsection{Improved Logging Practices}
Although logging practices in the software development process have been reviewed and improved over the past decade~\cite{yuan2012characterizing,chen2017characterizing}, there is still room for betterment~\cite{hassani2018studying}. 
Additional tools that can automatically detect log-related issues are required. 
Moreover, because the majority of current logging practices and decisions are \textit{ad-hoc} and decided by developers on the spot, the introduction of systematic logging practices that can provide suggestions to developers while composing the code can ensure a higher quality of logs. 
Also, further research that can provide directives and insights for developers with regards to good versus poor logging practices, and hence help to improve their logging practices and make better use of logging, is of interest.
For example, more effective logging can enable the customers of the software systems to solve problems themselves using the logs without relying on developers or avoid unnecessary logging costs, such as exposing users' sensitive information in the logs~\cite{li2020qualitative}. 
Another angle for logging practices improvement includes studies that investigate cost-aware logging, which can help developers to estimate and optimize the cost of logging while benefiting from the logs. 
Although efforts such as Log20~\cite{zhao2017log20} have aimed to address this issue, there is still a sizable room to improve upon. 
This avenue of future research can be also expanded to other platforms such as mobile devices as the logging practices can be different depending on the applications and the system requirements~\cite{zeng2019studying}. 
Prior research has shown other areas such as mobile systems, that are not in the research community's spotlight, are even more in dire need of systematic guidance and automated support tool for assisting in logging practices~\cite{zeng2019studying}.

\subsubsection{Representation of Log Files}
It is a safe assumption that the log analysis methods require, or at the very least, perform better on logs with good quality to conduct meaningful analyses. 
Therefore, we foresee future research in improving the formatting and defining universal structure for log messages, which will directly help in achieving more symmetric organization of log files, and consequently, more effective log analysis with higher \textit{Precision} and \textit{Recall} values. This goal is also partially realized with proper selection and improvement of logging libraries and utilities, \textit{e.g.}, Log4j, SLF4J, and Logback. 

\subsubsection{Logging Libraries and Utilities}
Logging libraries and utilities (LLU) provide additional functionality, structure, and flexibility in logging for developers such as log verbosity levels and thread-safety~\cite{chen2020studying}.  
Although LLUs facilitate logging, there has been insufficient research on this topic. 
Further research that aims to improve the performance of logging libraries by performing some of the logging tasks during the compile time is necessary~\cite{yang2018nanolog}.
Furthermore, the development of application-specific logging libraries will provide higher logging flexibility and better API for developers in a specific domain, similar to \textit{Log++}~\cite{marron2018log++} for cloud logging, to perform workload-related logging. 
For example, in a cloud deployment and provisioning process, users are further interested in logging the machine image initialization and termination steps in more details to enable better debugging in case of failures.
Additionally, LLUs can improve to bring in new configurability, such as supporting different log verbosity levels for separate parts of a logging statement~\cite{li2020qualitative}. 
Another angle that LLUs can improve is to provide checks on the format of the developer's provided logging text and ensure that the provided content passes a minimum set of standards in order to make logs more useful and organized. 
Additionally, logging libraries can help to reduce the overhead of logging. 
The way that some of the LLU work is that, during the runtime, all of the logging statements are executed but based on the verbosity level of the logging statements, some of the logs are filtered from being written to the log file. 
This approach can still introduce a considerable overhead if logging statements make calls to other methods and variables. 
To cope with this situation, developers include log statement guarding (\textit{e.g.}, putting the log statement inside an \textit{if-clause}) to avoid the logging statement being executed based on whether or not the level is enabled.
Therefore, this type of log guarding improvement would be beneficial to be implemented inside the LLUs~\cite{hassani2018studying}.

\subsubsection{Application Specific Logging}
As logging messages can provide valuable information with regards to the different aspects of the running software, it is also evident that different tasks and applications that rely on logs require different types of information from the log files. 
Therefore, we anticipate application-specific logging research in the future to grow. 
That is to say, depending on the application, we need to log different categories of runtime information.  
For example, for a security log, certain values need to be printed while not compromising sensitive information; however, this might not be an issue in postmortem analysis of logs~\cite{rivera2019towards}. 
Additionally, developers of other platforms, such as mobile apps, should be aware of the differences between desktop/server and mobile practices as it comes to logging, as for mobile, there is energy overhead concern that should be taken into account~\cite{zeng2019studying}. 
Therefore, different platforms also might end up logging different information with varied frequencies of outputting logging statements.

\subsubsection{Maintenance of the Logging Code}
As the software systems continue to grow, maintaining the logging code becomes more challenging. 
Previous studies have observed that the logging code is not maintained as well as the feature code, as there is no straightforward way to test the correctness of the logging code~\cite{chen2019extracting}. 
Therefore, we emphasize that further research should consider the systematic maintenance and testing of logging code alongside the feature code evolution. 
Additionally, there are interesting opportunities for developing automated tools that can read the context of the feature code changes and suggest logging code maintenance and updates concerning the feature code updates while the new feature code is being checked in.\looseness=-1

\subsection{Category E: Log Printing Statement Automation}

\subsubsection{Automatic LPS Generation}
In contrast to developer-inserted logs, LPS automation aims to auto-generate or suggest new logging statements or enhance the quality of currently available logs inside the source code based on various source code and application criteria. 
Although this topic has been of interest recently~\cite{yuan2012improving,yuan2012conservative,zhu2015learning,zhao2017log20,jia2018smartlog,gholamian2020logging,li2021deeplv}, considering the continuous advancement and birth of new AI and learning methods, we anticipate future research in the development of machine learning methods to implement and automate logging, with statistical modeling, supervised, unsupervised, and deep learning approaches will continue to foster. 
These methods should consider automating different aspects of the logging statements, such as the \textit{\textbf{location}}, \textit{\textbf{content}}, and \textit{\textbf{verbosity level}}. 
The automated methods can also consider different criteria for automation, such as diagnosability versus cost-awareness~\cite{li2020qualitative}. 
The ultimate goal is to achieve an automated approach that can introduce high-quality log suggestions or enhancements for various applications. 
Subsequently, assuming the development of different approaches, a comparative study of different approaches and the areas that each one performs better becomes necessary, similar to the comparison of 
different log parsing techniques in~\cite{zhu2019tools}.

\subsubsection{Constraint-based Logging}
The majority of the log automation tools have aimed to mimic developers' logging habits~\cite{gholamian2020logging,jia2018smartlog,zhu2015learning}. 
In other words, the log learning approaches work to learn developers' logging habits to decide if a new unlogged code snippet requires a logging statement. 
However, prior work~\cite{hassani2018studying} has also shown that developers make mistakes, and in some places, they even forget to log in the first place. 
Thus, one remaining important challenge is to develop constraint-based automated logging approaches to guarantee a particular logging goal, \textit{e.g.}, at minimum, one iteration out of 100 iterations of method \textit{MtdM()} is logged, or a particular execution path is fully disambiguated with logging, \textit{i.e.}, we can accurately determine which code segments \textit{`must'} have been executed. 
Another example can be ensuring the beginning and the end of all methods of interest are logged. 
By doing so, we can guarantee that at least a minimum quality of logs is granted. 

\subsubsection{Golden Quality LPSs for Benchmarking}
To the best of our knowledge, there is no prior work that quantitatively measures the quality of logging statements in each software project. 
Many of the prior work consider the developers' inserted logging statements as ground truth to evaluate their automated logging approach~\cite{zhu2015learning,jia2018smartlog,gholamian2020logging}. 
However, prior research has shown that there is no general guideline for logging and developers mostly rely on their intuitions and insert logging statements in an \textit{ad-hoc} manner~\cite{liu2019variables}. 
As such, defining a set of quantitative metrics that can be applied to evaluate the quality of logs on various software projects and give them scores can be highly beneficial. 
This allows to find projects with high-quality logs, learn from them, and use them as a \textit{golden benchmark} for comparison with auto-generated logs.  

\subsection{Category F: Log Maintenance and Management}
\textbf{Log collections and compressors.} Log collections, similar to LogHub~\cite{he2020loghub}, are useful for evaluating various log analyses. 
Additionally, datasets that are labeled and differentiate normal against abnormal log records are well sought after, as they enable the application of supervised and deep learning approaches for log analysis~\cite{zhang2019robust,liang2007failure,he2017towards}. 
As such, we see value in further research to collect log data from various software and application domains, and develop automatic and accurate probabilistic methods~\cite{yang2021semi} to label the data to facilitate log analysis and logging practices research. 
For log compression, because logs generally benefit from higher repetition than natural text, future research can benefit from designing and evaluating log-tuned compressors, which not only can result in more effective compression but also more efficient and streamlined decompression, for later auditing and analysis.

\subsection{Categories G, H, I, J, K: Automated Log Analysis Applications}
\subsubsection{Log Parsing}
With the recent advancements of NLP models~\cite{vaswani2017attention,devlin2018bert}, we foresee the development of transformer-based log parsers that can potentially improve the performance of log parsing, and consequently, the downstream log mining tasks. 
In addition, pre-trained language models~\cite{liu2019roberta} can potentially be fine-tuned on log files to enable higher performance for log parsing.  

\subsubsection{Log Analysis and Tools}
Prior research has proposed plenty of log analysis methods and tools for different applications, such as \textbf{anomaly and problem detection}~\cite{xu2009detecting,fu2009execution}, \textbf{performance and failure diagnosis}~\cite{zhao2014lprof,zhao2016non,yu2016cloudseer,zhang2017pensieve}, \textbf{system's runtime behavior}~\cite{hassan2008industrial,he2018identifying,shang2013assisting}, \textbf{system profile building}~\cite{oprea2015detection}, \textbf{code quality assessment}~\cite{shang2015studying}, and \textbf{code coverage}~\cite{chen2018automated}. 
Log analysis, starting with log parsing, plays an essential role in extracting useful information from the log files.  
As logs can be viewed in different ways, such as events, time series, and feature vectors, this enables different types of analyses. 
Complementary to the available research, because logs are non-intrusive and readily available, we anticipate new methods of log analysis or improvement of the current methods will be sought after for different applications. 
The quality of log analysis can directly impact the amount of actionable information that we can extract from the log files. 
Therefore, we expect new logging analysis approaches will emerge that utilize and combine a variety of algorithms to achieve a more accurate analysis. The approaches might also assume a specific format of logs, \textit{e.g.}, log messages following a specific template within the log files, to achieve a more personalized analysis. 
For example, the research can benefit from considering multiple factors, such as the content of each log message, the frequency, and the sequencing of log messages in log analysis tasks, \textit{e.g.}, anomaly detection, in order to achieve a deeper understanding of what happens in the logs. 
Lastly, we foresee that future research will benefit from utilizing AI approaches in understanding and leveraging the hidden semantics of the log messages, rather than solely focusing on learning log patterns and templates. 
This will enable a more sophisticated log analysis.

\subsubsection{Scalable and Online Log Processing}
In order to keep pace with the massive amount of growing logs in size (at the rate of approximately multiple terabytes per day~\cite{he2017towards}) and various formats, which is the by-product of the software growth as well as the number of software users' growth, we anticipate further research will be conducted to develop and update the current logging processing tools and platforms. 
Thus, future research should consider leveraging distributed and parallel processing platforms (\textit{e.g.}, Apache Spark) in conjunction with efficient machine learning approaches to implement scalable log analysis tools for all stages of the process, \textit{i.e.}, real-time collection, processing, and storage of voluminous logs~\cite{candido2019contemporary}. 
In addition, as many of the enterprise software platforms require 24/7 up-time and availability, the need for online tools that can perform the log analysis simultaneously as the system generates logs becomes more apparent. 
We require the tools to be efficient enough to perform analysis at the same speed or faster than the log generation rate.

{\renewcommand{\arraystretch}{1.2}
\newcounter{magicrownumbers2}
\newcommand\rownumber{\stepcounter{magicrownumbers2}\arabic{magicrownumbers2}}
\rowcolors[]{2}{gray!15}{white}
\begin{table*}[h]
\scriptsize
\begin{tabular}{P{.2cm}|p{3cm}|p{10cm}|p{3cm}} \toprule
   \rowcolor{blue!20} \hspace*{-2mm}\textbf{No.} & \textbf{Avenue} & \textbf{Rationale} & \textbf{Selected research} \\ \midrule
   \rownumber. & \textbf{Adaptive logging} &Dynamic adjustment of the logging level from the least to the most verbose level helps with detailed postmortem analysis.&\cite{mizouchi2019padla}\\
    \rownumber.& \textbf{Whether to log?}  & Different scenarios of whether LPSs are printed or filtered based on the goal of the logging analysis tasks should be studied.   & \cite{ding2015log2}   \\   
    \rownumber. & \textbf{Logging practices} & Future research can improve on logging practices in the software development process and reduce the \textit{ad hoc} and forgetful developers' logging habits.& \cite{yuan2012characterizing,chen2017characterizing,hassani2018studying}\\ 
  \rownumber.& \textbf{Representation of the log files}& Further research can improve the formatting and standardization of log messages, which directly results in more organized log files and more accurate automated analysis. & \cite{salfner2004comprehensive,ogle2004canonical}  \\
    \rownumber.& \textbf{Logging libraries and utilities} & Logging libraries and utilities (LLU) can provide additional functionality, structure, and flexibility in logging for developers.   &\cite{chen2020studying,yang2018nanolog,marron2018log++} \\
 \rownumber.& \textbf{Application-specific logging}  & Research can investigate and ensure that how different tasks and applications that rely on logs can record application-specific information (\textit{i.e.}, based on the application needs) into the log files.& ~\cite{rivera2019towards,zeng2019studying}   \\
    \rownumber.&\textbf{Maintenance of logging code} & As maintenance of the logging code becomes more challenging, future research requires to develop automated approaches to ensure up-to-date and issue-free logging code.&  \cite{chen2019extracting,li2019dlfinder,chen2017characterizing} \\    
    \rownumber.& \textbf{Automated and constraint-based log generation}& Research requires to improve the quality of auto-generated LPSs and, also, enhance the quality of the developer-inserted \textit{ad-hoc} logs by adding additional variables, \textit{etc}. & \cite{candido2021exploratory,li2021deeplv,liu2019variables,li2020shall,yuan2012improving,yuan2012conservative,zhu2015learning,zhao2017log20,jia2018smartlog,gholamian2020logging}   \\    
\rownumber.& \textbf{Golden quality log statements} & High quality logs are required to learn from, and use them as a \textit{golden benchmark} for comparison with auto-generated logs.&  \cite{liu2019variables,zhu2015learning,jia2018smartlog,gholamian2020logging} \\    
\rownumber.& \textbf{Log collections} & There is a need for labeled log data collections and development of automated log labeling approaches to facilitate automated log analysis.  &  \cite{he2020loghub,urlopenstackdata,cotroneo2019bad} \\
  \rownumber.  & \textbf{Log compression} & Compressors which are designed for logs are needed to improve the compression/decompression efficiency and enable efficient long-term storage and backup of logs.&  \cite{liu2019logzip,yao2020study,hassan2008industrial} \\

\rownumber.&\textbf{Log analysis for various objectives} & Research will actively continue to propose and improve approaches for more accurate log analysis for different log mining tasks and postmortem debugging. 
& \cite{yang2021semi,xu2009detecting,fu2009execution,zhao2014lprof,zhao2016non,yu2016cloudseer,zhang2017pensieve}  \\
\rownumber.& \textbf{Scalable and online log processing} &  Scalable and real-time log processing is required to keep pace with the massive amount of growing logs in size, \textit{e.g.}, multiple terabytes per day.  &  \cite{he2017towards} \\

\rownumber.& \textbf{Natural language processing of logs} & Leveraging NLP characteristics of logs will benefit automated log generating (\textit{e.g.}, log description generation) and NLP processing of log files.   & \cite{gholamian2020logging,hindle2012naturalness,tu2014localness,he2018characterizing}  \\
\rownumber.& \textbf{Log summarization \& visualization} & Development of approaches that aim to elicit and condense big-picture insights from logs with visualizing and summarization are in demand, and this will enable practitioner to only focus on significantly smaller but most important portion of logs.  & \cite{rabkin2010graphical,cao2020knowledge,ehrlinger2016towards}  \\
    
 \bottomrule
\end{tabular}
\caption{Summary of avenues for future work in logging research.}
\label{summary_future}
\vspace*{-5mm}
\end{table*}
}

\subsection{Category L - Emerging Logging Research}
\subsubsection{Natural Language Processing of Logs}
Prior work~\cite{hindle2012naturalness,tu2014localness} in software engineering has utilized natural language processing (NLP) for software tasks such as source code next token suggestion.
Recently, there has been a thread of research on analyzing logging statements as natural language sequences. 
He \textit{et al.}~\cite{he2018characterizing} characterized the NLP characteristics of LPS descriptions in Java and C\# projects, and Gholamian and Ward~\cite{gholamian2021naturalness} showed software execution logs are natural and local, and these features can be leveraged for automated log analysis, such as anomaly detection. 
We hypothesize that further research is required to confirm the NLP characteristics of software logs, and eventually, leveraging NLP characteristics of logs will further benefit automated log generating and analysis of log files. 
Moreover, the recent advancements in NLP models, \textit{e.g.}, BERT models~\cite{devlin2018bert}, calls for further investigation and application of them in improving the performance of log mining tasks. 
The intuition is that these models can embed and learn a higher degree of log semantics, and thus, can better enable actionable diagnosis from logs.   

\subsubsection{Log Summarization and Visualization}
Prior works~\cite{rabkin2010graphical,cao2020knowledge,locke2021logassist} have proposed approaches to summarize and visualize console and security logs. 
Log summarization and visualization is a natural response to the ever-growing scale of logs to gain high-level insight into the logs. 
In large-scale distributed software systems, as the scale of logs continue to grow, and various subsystems continue to generate logs in heterogeneous formats and rates, we foresee the development of approaches and solutions, both in academia and industry, that aim to make high-level sense of logs and to gain big-picture insight with visualizing logs. 
In addition, log summarization will help developers and practitioners to focus their troubleshooting efforts on a smaller set of relevant logs. 
Knowledge graph representation is a potential candidate for this aim~\cite{ehrlinger2016towards}. 
Lastly, we provide a digest of the avenues for future of logging research in Table~\ref{summary_future}. 
\begin{tcolorbox}[breakable, enhanced]
\small \textbf{Finding} \textbf{17.} \textit{Outstanding problems exist for each category of logging research. 
Future research can consider and tackle these challenges to improve the quality of log statements and log files, and thus enable more effective log analysis tasks.}
\end{tcolorbox}
\vspace*{-3mm}
\section{Conclusions}\label{conclusions}
Logging statements and log files are the inevitable pieces of the puzzle in analyzing and ensuing various aspects of correct functionality of software systems, such as debugging, maintaining, and diagnosability. 
The valuable information gained from logs has motivated the research and development of a plethora of logging practices, logging applications, and log automation and analysis tools.

In this survey, we initially started with the basics of log statements and log files and the involving challenges in extracting useful information from them in Sections~\ref{introduction_section} and~\ref{log_file}. 
As we conduct the survey, we aim to answer four crucial research questions related to software logging: \textbf{(RQ1)} categories of logging research, \textbf{(RQ2)} publication trends based on topics, years, and venues, \textbf{(RQ3)} available research in each category, and finally \textbf{(RQ4)} challenges and opportunities for future logging research. 
We next reviewed the costs and benefits associated with logging in Section~\ref{log_cost} and followed that up with research that mines logging statements to derive logging practices in Section~\ref{log_mining}. 
In Section~\ref{automated_logging}, we reviewed the proposed methods for automated logging, and we mentioned evaluation methods and metrics for auto-generated logs and learning-to-log platforms in Section~\ref{metrics_section}. 
In section~\ref{mine_log_files}, we reviewed log file mining and log analysis research which aims to expedite and scale up the log processing, and apply logs for different system maintenance tasks such as anomaly and failure detection/diagnosis, performance issues, and code quality assessment. 
We also reviewed the emerging domains and applications for logging, such as in NLP, mobile, and big data in Section~\ref{log_apps}. 
Finally, we discussed the opportunities for future research in different aspects of logging statements and log files, their practices, and their analyses in Section~\ref{opportunities}. 
Overall, we reviewed \textbf{112} primary studies published between 2010 and  2021 (inclusive), and we included our findings for each subsection of the survey, which tallied up to \textbf{17 succinct and quick-grasping findings} in total. 

\balance
Although current research advances have made logs more useful and effective, there are still multiple remaining challenges and avenues for future work and improvement. 
Categories of challenges remain in various aspects of \textbf{automated log analysis}, \textbf{LPS auto-generation}, \textbf{scalable logging analysis and infrastructure}, \textbf{cost-aware logging}, \textbf{log maintenance and management}, and \textbf{improved logging practices}. 
We foresee future research in multiple directions for logging as follows: 
\begin{itemize}
\item As the size of computer systems increases, we anticipate the voluminousness and heterogeneity of logs, which turns it into a big-data problem, will demand further quantitative cost analysis for collecting, processing, and storing of logs, as logging can infer computation, storage, and network overhead. 
Additionally, due to the voluminousness and heterogeneity of generated logs, and in some cases, the need for real-time processing of logs, we anticipate the development of efficient, scalable, and real-time log analysis tools~\cite{miranskyy2016operational}. 
\item We anticipate continued research on current logging practices and log-statement-related issues, as this will enable improvements of future practices and help to create guidelines for developers when making logging decisions. 
We also predict the evolvement of learning and AI-based log recommender tools and IDE plugins, which utilize the readily available code repositories of open-source projects to provide just-in-time logging practice suggestions to developers~\cite{li2017towards,gholamian2020logging}. 
Additionally, we expect further work on logging libraries to collaborate with emerging logging practices and bring in the development of application-specific logging practices~\cite{rivera2019towards}. 
\item We foresee automated log file analysis techniques continue to evolve and become more effective and sophisticated (with machine learning and AI-based techniques) in their information extraction from log files and log statements. 
We also see an emerging trend of new applications that utilize log analysis recently, such as log analysis for code coverage~\cite{chen2018automated}. Moreover, we predict further research will be performed in enabling the analysis of logs for other platforms, such as mobile systems and big-data applications. 
Log collections will also continue to grow to help with log analysis. 
\item With regards to log statement prediction, we anticipate that future research on supervised, unsupervised, and deep learning techniques will continue to benefit logs and their analyses. 
\end{itemize}
In this study, we aim to systematically summarize, discuss, and critique the state-of-the-art knowledge in the logging field for experienced researchers, and simultaneously, help new researchers to get a quick and critical grasp of the available research in this area. 
Additionally, we envision the uncovered research opportunities in this survey serve as a beacon for advancing the logging research. 
Lastly, we provide a link to the data used in this survey, available at: \url{https://github.com/sgholamian/comprehensive-software-logging-survey/}~\cite{url_comprehensive_survey_git}.

\onecolumn
\section{List of Papers}\label{paper_full_list}
Table~\ref{reference_per_topic} provides the list of papers per each category of logging research.
\newcommand\rownumber{\stepcounter{magicrownumbers}\arabic{magicrownumbers}}
\begin{table}[h]
\scriptsize	
\begin{center}
 \begin{tabular}{p{2.5cm} p{11cm} |p{.7cm} p{1.3cm} p{1.2cm}} 
 \toprule
 \rowcolor{blue!20} \textbf{Topic} & \textbf{Paper title} & \textbf{Year} & \textbf{Venue}& \textbf{Subtopic(s)} \\ [0.5ex] 
 \midrule
 \multirow{6}{2.5cm}{\textbf{(A) Costs and benefits of logging (6)}}
&Be conservative: Enhancing failure diagnosis with proactive logging~\cite{yuan2012conservative}.& 2012& OSDI (J)&(E)\\

&Linux auditing: Overhead and adaptation~\cite{zeng2015linux}.& 2015& ICC (C)& \\
      
&Log2: A cost-aware logging mechanism for performance diagnosis~\cite{ding2015log2}.& 2015&ATC (C)& (E)\\ 

&A qualitative study of the benefits and costs of logging from developers' perspectives~\cite{li2020qualitative}.& 2020&TSE (J)&\\

&Log4Perf: Suggesting and updating logging locations for web-based systems' performance monitoring~\cite{yao2020log4perf}.& 2020& EMSE (J)& (E)\\

&What distributed systems say: A study of seven spark application logs~\cite{gholamian2021distributed}.& 2021& SRDS (C)& (H)\\

\midrule

 \multirow{7}{2.5cm}{\textbf{(B) Logging practices (7)}}
 &Characterizing logging practices in open-source software~\cite{yuan2012characterizing}.& 2012&ICSE (C)& (C)\\ 
  &Where do developers log? An empirical study on logging practices in industry~\cite{fu2014developers}.&2014& ICSE (C)& (A), (E)\\  
  &Industry practices and event logging: Assessment of a critical software development process~\cite{pecchia2015industry}.&2015&ICSE (C)&\\   
  &Studying the relationship between logging characteristics and the code quality of platform software~\cite{shang2015studying}.&2015& EMSE (J)& (D)\\  
&Characterizing logging practices in Java-based open source software projects - A replication study in Apache Software Foundation~\cite{chen2017characterizing2}.&2017&EMSE (J)& (C)\\  
&An exploratory study of logging configuration practice in Java~\cite{zhi2019exploratory}.&2019&ICSME (C)& (C)\\

&Studying the use of Java logging utilities in the wild~\cite{chen2020studying}.&2020& ICSE (C)& (C), (D)\\

\midrule

 \multirow{5}{2.5cm}{\textbf{(C) Logging progression (5)}}
 &An exploratory study of the evolution of communicated information about the execution of large software systems~\cite{shang2014exploratory}.& 2014&JSS (J)& (D)\\ 
    &Logging library migrations: A case study for the Apache software foundation projects~\cite{kabinna2016logging}.& 2016&MSR (C)& (B)\\  
 &Examining the stability of logging statements~\cite{kabinna2018examining}.& 2018&EMSE (J)& (B), (D)\\  
  &Guiding log revisions by learning from software evolution history~\cite{li2019guiding}.& 2019&EMSE (J)& (D)\\  
&Can you capture information as you intend to? A case study on logging practice in industry
 ~\cite{rong2020can}.& 2020&ICSME (C)& (B)\\  
  
\midrule 
 
 \multirow{7}{2.5cm}{\textbf{(D) Log-related issues (7)}} 
&Simple testing can prevent most critical failures: An analysis of production failures in distributed
data-intensive systems~\cite{yuan2014simple}.& 2014&OSDI (C)&(B)\\  

&Understanding log lines using development knowledge~\cite{shang2014understanding}.& 2014&ICSME (C)& (B)\\  

&Characterizing and detecting anti-patterns in the logging code~\cite{chen2017characterizing}.& 2017&ICSE (C)&\\  

&Studying and detecting log-related issues~\cite{hassani2018studying}.& 2018&EMSE (J)& (E)\\     

&Studying duplicate logging statements and their relationships with code clones~\cite{li2021studying}.& 2021&TSE (J)&(C)\\

&An exploratory semantic analysis of logging questions~\cite{gujral2021exploratory}.& 2021&SP\&E (J)&()\\

&Demystifying the challenges and benefits of analyzing user-reported logs in bug reports~\cite{chen2021demystifying}.& 2021& EMSE (J)&()\\  

\midrule

 \multirow{15}{2.5cm}{\textbf{(E) Log printing statement automation (15)}}&AutoLog: Facing log redundancy and insufficiency~\cite{zhang2011autolog}.& 2011&APSys (C)& (A)\\
   
   &Improving software diagnosability via log enhancement~\cite{yuan2012improving}.& 2012&TOCS (J)& (J)\\

&Learning to log: Helping developers make informed logging decisions~\cite{zhu2015learning}.& 2015&ICSE (C)& (B)\\

&LogOptPlus: Learning to optimize logging in catch and if programming constructs
~\cite{lal2016logoptplus}.& 2016&{\fontsize{5}{8}\selectfont COMPSAC (C)}&\\ 

&Log20: Fully automated optimal placement of log printing statements under specified overhead threshold
~\cite{zhao2017log20}.& 2017&SOSP (C)& (A)\\

&Towards just-in-time suggestions for log changes~\cite{li2017towards}.& 2017&EMSE (J)& (C), (D)\\

&Which log level should developers choose for a new logging statement?
~\cite{li2017log}.& 2017&EMSE (J)& (B)\\

&SMARTLOG: Place error log statement by deep understanding of log intention
~\cite{jia2018smartlog}.& 2018&{\fontsize{6}{9}\selectfont SANER (C)}& (A)\\

&An approach to recommendation of verbosity log levels based on logging intention
~\cite{anu2019approach}.& 2019&ICSME (C)&\\

&Which variables should I log?~\cite{liu2019variables}.&2019&TSE (J)&\\

&Automatic recommendation to appropriate log levels~\cite{kim2020automatic}.&2020&SP\&E (J)&\\

&Logging statements’ prediction based on source code clones~\cite{gholamian2020logging}.& 2020&SAC (C)& (B)\\

&Where shall we log? Studying and suggesting logging locations in code blocks
~\cite{li2020shall}.& 2020&ASE (C)& (B)\\ 

&An exploratory study of log placement recommendation in an enterprise system  
~\cite{candido2021exploratory}.& 2021&MSR (C)& (B)\\  

&DeepLV: Suggesting log levels using ordinal based neural networks
~\cite{li2021deeplv}.& 2021&ICSE (C)&(B)\\

\midrule

 \multirow{13}{2.5cm}{\textbf{(F) Log maintenance and management (13)}}
&An integrated data-driven framework for computing system management
~\cite{li2009integrated}.&2010&TSMCA (J)& (I)\\

&Cloud application logging for forensics
~\cite{marty2011cloud}.& 2011& SAC (C)&\\

&FLAP: An end-to-end event log analysis platform for system management  
~\cite{li2017flap}.&2017&KDD (C)& (I), (J)\\

&Using finite-state models for log differencing  
~\cite{amar2018using}.&2018&{\fontsize{5.5}{7}\selectfont ESEC/FSE (C)}&\\  

&Statistical log differencing 
~\cite{bao2019statistical}.&2019&{ASE (C)}&\\  

&Logzip: Extracting hidden structures via iterative clustering for log compression
~\cite{liu2019logzip}.&2019&ASE (C)& (G)\\

 &A study of the performance of general compressors on log files
~\cite{yao2020study}.& 2020&EMSE (J)&\\

&Effective removal of operational log messages: an application to model inference{\normalsize *}
~\cite{shin2020effective}.&2020&arXiv (A)& (I)\\ 

&Loghub: A large collection of system log datasets towards automated log analytics{\normalsize *}
~\cite{he2020loghub}.&2020&arXiv (A)& (H)\\ 

&A survey of software log instrumentation
~\cite{chen2021survey}.&2021&CSUR (J)& (B), (C)\\ 

&LogAssist: Assisting log analysis through log summarization
~\cite{locke2021logassist}.&2021&TSE (J)&\\ 

&Improving state-of-the-art compression techniques for log management tools
~\cite{yao2021improving}.&2021&TSE (J)&\\ 

&Would you like a quick peek? Providing logging support to monitor data processing in big data applications~\cite{wang2021would}.&2021&{\fontsize{5.5}{7}\selectfont ESEC/FSE (C)}&\\

\bottomrule

\multicolumn{5}{l}{\vspace*{1mm}{\normalsize *}\footnotesize{Snowballing}}

\end{tabular}
\end{center}
\vspace*{-3mm}
\caption{\footnotesize A full list of reviewed publications. \textit{`Subtopic'} column shows what other topics are discussed in the research, if applicable.}
\label{reference_per_topic}
\end{table}

\begin{table}
\ContinuedFloat
\scriptsize	
\begin{center}
 \begin{tabular}{p{2.5cm} p{11cm} |p{.7cm} p{1.3cm} p{1.2cm}} 
 \toprule
 \rowcolor{blue!20} \textbf{Topic} & \textbf{Paper title} & \textbf{Year} & \textbf{Venue}& \textbf{Subtopic(s)} \\ [0.5ex] 
 \midrule

\multirow{8}{2.5cm}{\textbf{(G) Log parsing (8)}}
&What happened in my network: Mining network events from router syslogs 
~\cite{qiu2010happened}.&2010&IMC (C)& (J)\\
 
&Baler: Deterministic, lossless log message clustering tool
~\cite{taerat2011baler}.&2011&CSRD (J)&\\ 
 
 &LogSig: Generating system events from raw textual logs 
~\cite{tang2011logsig}.&2011&CIKM (C)&\\

 &LogCluster - A data clustering and pattern mining algorithm for event logs 
~\cite{vaarandi2015logcluster}.&2015&CNSM (C)&\\

&Spell: Streaming parsing of system event logs 
~\cite{du2016spell}.&2016&ICDM (C)&\\

&Drain: An online log parsing approach with fixed depth tree
~\cite{he2017drain}.&2017&ICWS (C)&\\

&A search-based approach for accurate identification of log message formats
~\cite{messaoudi2018search}.&2018&ICPC (C)&\\
   
&Logram: Efficient log parsing using n-gram dictionaries
~\cite{dai2020logram}.& 2020&TSE (J)& (L)\\

\midrule
   
\multirow{16}{2.5cm}{\textbf{(H) Anomaly detection (16)}}
&Detecting large-scale system problems by mining console logs{\normalsize *}
~\cite{xu2009detecting}.&2009&SOSP (C)& (J)\\ 
 
&Execution anomaly detection in distributed systems through unstructured log analysis
{\normalsize *}~\cite{fu2009execution}.&2009&ICDM (C)&  (G)\\ 
 
&Mining invariants from console logs for system problem detection{\normalsize *}~\cite{lou2010mining}.&2010& ATC (C) & (G)\\ 

&Linking resource usage anomalies with system failures from cluster log data
~\cite{chuah2013linking}.&2013&SRDS (C)& (J)\\  
 
&DeepLog: Anomaly detection and diagnosis from system logs through deep learning
~\cite{du2017deeplog}.&2017&CCS (C)& (G), (I), (J)\\

&Experience report: Log mining using natural language processing and application to anomaly detection
~\cite{bertero2017experience}.&2017&ISSRE (C)& (J)\\

&Execution anomaly detection in large-scale systems through console log analysis{\normalsize *}
~\cite{bao2018execution}.&2018& JSS (J) & (J)\\

&Metric selection and anomaly detection for cloud operations using log and metric correlation analysis{\normalsize *}
~\cite{farshchi2018metric}.&2018& JSS (J) & (J)\\

&LogAnomaly: Unsupervised detection of sequential and quantitative anomalies in unstructured logs~\cite{meng2019loganomaly}.&2019&  IJCAI (C) & (G)\\

&Robust log-based anomaly detection on unstable log data
~\cite{zhang2019robust}.&2019& {\fontsize{5.5}{7}\selectfont ESEC/FSE (C)} & (C)\\

&Anomaly detection via mining numerical workflow relations from logs
~\cite{zhang2020anomaly}.&2020& SRDS (C) &\\

&HitAnomaly: Hierarchical transformers for anomaly detection in system log
~\cite{huang2020hitanomaly}.&2020& TNSM (J) &\\

&LogSayer: Log pattern-driven cloud component anomaly diagnosis with machine learning
\cite{zhou2020logsayer}.&2020& IWQoS (C) & (J)\\

&LogTransfer: Cross-system log anomaly detection for software systems with transfer learning
\cite{chen2020logtransfer}.&2020& ISSRE (C) &\\

&Semi-supervised log-based anomaly detection via probabilistic label estimation
~\cite{yang2021semi}.&2021&ICSE (C)&\\

&Log-based anomaly detection without log parsing
~\cite{le2021log}.&2021&ASE (C)&\\

 \midrule

 \multirow{7}{2.5cm}{\textbf{(I) Runtime behavior (8)}}&An approach for mining web service composition patterns from execution logs 
~\cite{tang2010approach}.& 2010& WSE (C)&\\

&Online detection of multi-component interactions in production systems
~\cite{oliner2011online}.& 2011& DSN (C)& (H)\\

&Logmaster: Mining event correlations in logs of large-scale cluster systems
~\cite{fu2012logmaster}.& 2012& SRDS (C)&(J)\\

&Assisting developers of big data analytics applications when deploying on Hadoop clouds
~\cite{shang2013assisting}.& 2013&ICSE (C)&(J)\\

&Behavioral log analysis with statistical guarantees
~\cite{busany2016behavioral}.& 2016&{ICSE (C)}& \\

&Performance model derivation of operational systems through log analysis
~\cite{awad2016performance}.& 2016&{\fontsize{5.5}{7}\selectfont MASCOTS (C)}& (J)\\

&Exploring properties and correlations of fatal events in a large-scale hpc system
~\cite{di2018exploring}.& 2018& TPDS (J)& (J)\\

&Identifying impactful service system problems via log analysis
~\cite{he2018identifying}.& 2018&{\fontsize{5.5}{7}\selectfont ESEC/FSE (C)}&(J)\\

\midrule

 \multirow{15}{2.5cm}{\textbf{(J) Performance, fault, and failure diagnosis (15)}}&Assessing and improving the effectiveness of logs for the analysis of software faults
~\cite{cinque2010assessing}.& 2010& DSN (C)&(I)\\  
 
 &Diagnosing the root-causes of failures from cluster log files
~\cite{chuah2010diagnosing}.& 2010& HiPC (C)& \\ 
 
 &SherLog: Error diagnosis by connecting clues from run-time logs
~\cite{yuan2010sherlog}.& 2010&{\fontsize{6}{7}\selectfont  ASPLOS (C)} & (I)\\

&Detection of software failures through event logs: An experimental study
~\cite{pecchia2012detection}.& 2012& ISSRE (C)& (B)\\

&Structured comparative analysis of systems logs to diagnose performance problems{\normalsize *}
~\cite{nagaraj2012structured}.& 2012& NSDI (C)&\\

&Failure prediction based on log files using random indexing and support vector machines
~\cite{fronza2013failure}.& 2013& JSS (J)& (I)\\

&Leveraging performance counters and execution logs to diagnose memory-related performance issues
~\cite{syer2013leveraging}.& 2013& ICSE (C)&\\

&lprof: A non-intrusive request flow profiler for distributed systems{\normalsize *}
~\cite{zhao2014lprof}.& 2014& OSDI (C)& (G), (H)\\

&POD-Diagnosis: Error diagnosis of sporadic operations on cloud applications{\normalsize *}
~\cite{xu2014pod}.& 2014& DSN (C)&\\

&Mining system logs to learn error predictors: A case study of a telemetry system
~\cite{russo2015mining}.& 2015& EMSE (J)&\\

&Cloudseer: Workflow monitoring of cloud infrastructures via interleaved logs{\normalsize *}
~\cite{yu2016cloudseer}.& 2016& {\fontsize{6}{7}\selectfont  ASPLOS (C)}& (I)\\

&CRUDE: Combining resource usage data and error logs for accurate error detection in large-scale distributed systems
~\cite{gurumdimma2016crude}.& 2016& SRDS (C)& (H)\\

&Non-Intrusive performance profiling for entire software stacks based on the flow reconstruction principle
~\cite{zhao2016non}.& 2016& OSDI (C)& (I)\\

&Uilog: Improving log-based fault diagnosis by log analysis
~\cite{zou2016uilog}.& 2016& JCST (J)&\\

&Pensieve: Non-intrusive failure reproduction for distributed systems using the event chaining approach{\normalsize *}
~\cite{zhang2017pensieve}.& 2017& SOSP (C)&\\

\midrule

 \multirow{5}{2.5cm}{\textbf{(K) User, business, security, and code
coverage (5)}}&The unified logging infrastructure for data analytics at Twitter
~\cite{lee2012unified}.& 2012& VLDB (C)&\\

&Identifying recurrent and unknown performance issues
~\cite{lim2014identifying}.& 2014& ICDM (C)& (J)\\

&Detection of early-stage enterprise infection by mining large-scale log data
~\cite{oprea2015detection}.& 2015& DSN (C)& (H)\\

&The bones of the system: A case study of logging and telemetry at Microsoft
~\cite{barik2016bones}.& 2016& ICSE (C)&(B), (I)\\

&An automated approach to estimating code coverage measures via execution logs
~\cite{chen2018automated}.& 2018& ASE (C)&\\

\midrule

\multirow{7}{2.5cm}{\textbf{(L) Emerging applications (7)}}&Operational-log analysis for big data systems: Challenges and solutions
~\cite{miranskyy2016operational}.& 2016&{\fontsize{6}{7}\selectfont IEEE Softw (J)}& (A), (B), (I)\\

&Designing PhelkStat: Big data analytics for system event logs
~\cite{salman2017designing}.& 2017& HICSS (C)& (H)\\

&Performance evaluation of cloud-based log file analysis with Apache Hadoop and Apache Spark
~\cite{mavridis2017performance}.& 2017& JSS (J)& (A)\\

&An exploratory study on assessing the energy impact of logging on android applications
~\cite{chowdhury2018exploratory}.& 2018& EMSE (J)& (A)\\

&Characterizing the natural language descriptions in software logging statements
~\cite{he2018characterizing}.& 2018& ASE (C)& (E)\\

&Studying the characteristics of logging practices in mobile apps: a case study on F-Droid
~\cite{zeng2019studying}.& 2019& EMSE (J)&(A), (B)\\

&On the naturalness and localness of software logs
~\cite{gholamian2021naturalness}.& 2021& MSR (C)& (H)\\
\midrule

\textbf{Total (112)}&&&\\

\bottomrule

\multicolumn{5}{l}{\vspace*{1mm}{\normalsize *}\footnotesize{Snowballing}}

\end{tabular}
\end{center}
\vspace*{-3mm}
\caption{\footnotesize A full list of reviewed publications (continued). \textit{`Subtopic'} column shows what other topics are discussed in the research, if applicable.}
\label{reference_per_topic}
\end{table}

\twocolumn


\ifCLASSOPTIONcaptionsoff
  \newpage
\fi

\nobalance
\bibliographystyle{ieeetran}

\bibliography{bare_adv}

\begin{IEEEbiography}
[{\includegraphics[width=1in,height=1.1in,clip,keepaspectratio]{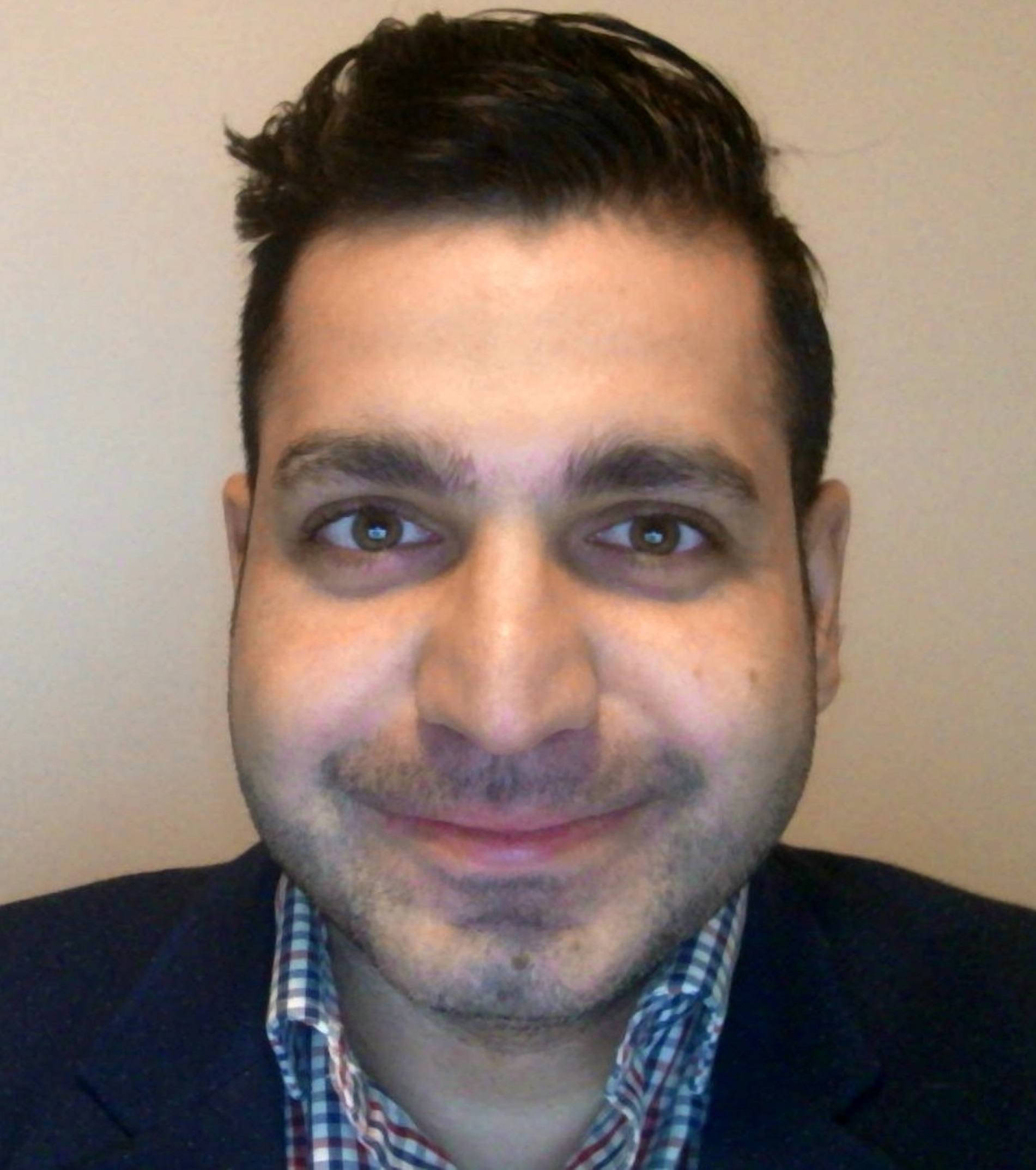}}]
{Sina Gholamian} is a Ph.D. student in software engineering at the University of Waterloo, Canada. His research interests include application of machine learning and data mining approaches for designing and improving automated software systems. His prior work has been published at premier venues such as ASE, MSR, SRDS, and ACM SAC.
\end{IEEEbiography}

\begin{IEEEbiography}
[{\includegraphics[width=1in,height=1.2in,clip,keepaspectratio]{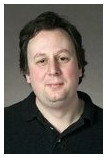}}]
{Paul A. S. Ward} is an associate professor at the University of Waterloo, Canada. He is also a faculty fellow at the IBM Centre for Advanced Studies. His expertise lies in the area of distributed systems and computer networks. 
\end{IEEEbiography}

\end{document}